\documentclass[11pt,twoside]{article}
\usepackage[ansinew]{inputenc}
\usepackage[hidelinks]{hyperref}
\usepackage{multirow}

\usepackage{ulem}
\hypersetup{ colorlinks=false }
\usepackage{amsthm}
\usepackage{amssymb,bm}
\usepackage{amsmath}
\usepackage{verbatim}

\usepackage[T1]{fontenc}
\usepackage{selinput}
\SelectInputMappings{%
  aacute={á},
  ntilde={ñ},
  Euro={€}
}

\usepackage{array}
\usepackage{amsfonts}
\usepackage{latexsym}
\usepackage{longtable}
\usepackage{schemata}
\usepackage{lscape}
\usepackage[skip=1ex]{caption}
\usepackage{natbib}
\setcitestyle{authoryear,open={(},close={)}} 

\numberwithin{figure}{section}
\numberwithin{table}{section}

\usepackage[pdftex]{color,graphicx}
\usepackage[usenames,dvipsnames,table]{xcolor}
\usepackage{float} 
\usepackage{subfigure} 
\usepackage{wrapfig} 
\usepackage{graphicx}
\usepackage{subcaption}
\usepackage[rflt]{floatflt} 
\graphicspath{{./FIGURES/}}

\usepackage{adjustbox}
\usepackage{multirow}
\date{}
\numberwithin{equation}{section}
\numberwithin{figure}{section}

    {\par\noindent{\bf Proof\/.}}

\title{An statistical analysis of COVID-19 intensive care unit bed occupancy data\footnote{This research is part of the grant PID2020-113578RB-I00, funded by MCIN/AEI/10.13039/501100011033/. It has also been supported by the Spanish grant PID2022-136878NB-I00, the Valencian grant Prometeo/2021/063, by the Xunta de Galicia (Grupos de Referencia Competitiva ED431C-2020/14) and by CITIC that is supported by Xunta de Galicia, convenio de colaboraci\'{o}n entre la Conseller\'{i}a de Cultura, Educaci\'{o}n, Formaci\'{o}n Profesional e Universidades y las universidades gallegas para el refuerzo de los centros de investigaci\'{o}n del Sistema Universitario de Galicia (CIGUS). The first author was also sponsored by the Spanish Grant for Predoctoral Research Trainees RD 103/2019 being this work part of grant PRE2021-100857, funded by MCIN/AEI/10.13039/501100011033/ and ESF+. In addition, we thank the Centro de Supercomputaci\'{o}n de Galicia (CESGA) for providing their services for part of the simulations in this work.}}
\author{Naomi Diz-Rosales$^{1}$, Mar\'{i}a Jos\'{e} Lombard\'{i}a$^{2}$, Domingo Morales$^{3}$
\vspace{0.1 cm}\\
{\small  $^{1}$naomi.diz.rosales@udc.es}\\
{\small $^{2}$maria.jose.lombardia@udc.es}\\
{\small $^{3}$d.morales@umh.es}\\
{\small  $^{1, 2}$Universidade da Coru\~{n}a, CITIC, Spain.}\\
{\small $^{3}$Universidad Miguel Hern\'{a}ndez de Elche, IUICIO, Spain.}\\
{\small 28-04-2024}}
\begin{document}
\maketitle

\begin{abstract}
\noindent
The COVID-19 pandemic has had far-reaching consequences, highlighting the urgency for explanatory and predictive tools to track infection rates and burden of care over time and space. 
However, the scarcity and inhomogeneity of data is a challenge. In this research we develop a robust framework for estimating and predicting the occupied beds of Intensive Care Units by presenting an innovative Small Area Estimation methodology based on the definition of mixed models with random regression coefficients. We applied it to estimate and predict the daily occupancy of Intensive Care Unit beds by COVID-19 in health areas of Castilla y León, from November 2020 to March 2022. \\[.2cm]

\end{abstract}

\section{Introduction}\label{main.intro}
COVID-19 disease, caused by the acute respiratory syndrome coronavirus type 2 or SARS-CoV-2, has had a major impact on all areas of human welfare.
In fact, since the first cases were reported in December 2019 in Hubei province in the People's Republic of China as of 31/12/2023, confirmed cases have surpassed the 770,000,000 mark, with a total of 773,449,299 reported infections, while official deaths total 6,991,842 \citep{WHO2024}.

Among the multiple consequences, the collapse of the healthcare system stands out, with Intensive Care Units (ICU) as the main bottleneck for survival.
Fortunately, with vaccines, the severity of the disease has been greatly reduced.
However, COVID-19 has been a turning point that has prompted countries to start planning strategies to deal with possible epidemics that cannot be ruled out in the near future \citep{WHO2023}, as well as to assess the behaviour of seasonal diseases that put high pressure on health systems, such as influenza.
A good example of this is the situation experienced in December 2023, in countries such as Spain, where a high circulation of influenza is being recorded, in coexistence with hospital admissions due to COVID-19 and other respiratory viruses, such as the Respiratory Syncytial Virus (RSV) \citep{ISCIII2023}.
Therefore, there is a clear need to model such outbreaks in order to develop predictive tools for spatial and temporal monitoring of hospital saturation capacity.

The scientific community has been fully committed to data analysis. 
Trying to outline the main fields of research in statistics and operational research, we find that COVID-19 has been approached from different objectives, mainly involved in: inference on the dynamics of virus transmission; estimating the basic ($R_{0}$) and dynamic ($R_{t}$) reproduction number; inference on other epidemiological parameters such as the incidence rate, deaths, or length of stay (\textit{LoS}), both on the ward and in ICU; as well as evaluating the impact of non-pharmacological intervention measures on the evolution of the pandemic \citep{Gnanvi2021}.

However, despite these magnificent contributions, there are still major challenges to overcome in estimating and predicting hospital saturation. As the authors themselves point out, reliance on existing data and their quality is key, being a handicap during this pandemic, where delays in reporting, changes in registration criteria and a marked spatial-temporal heterogeneity of the disease have stood out notably. In this respect, it should be emphasised that COVID-19 has not affected the different geographical areas of each country in the same way, neither in intensity nor in speed, due to a wide variety of factors that have a different influence on its spread across different territories. In fact, the Instituto de Salud Carlos III (ISCIII) has funded the COV20-00881 project, precisely to identify these factors and be able to prevent future outbreaks and their intensity. Thus, as preliminary results, they obtained that there are different clusters of provinces depending on their degree of affectation and the most influential factors \citep{CNE2022}.

In order to incorporate this heterogeneity in the modelling and to obtain indicators for planning hospital resources and intervention measures, it is necessary to carry out analyses at a lower level of aggregation than the country, community or region. Thus, in Spain, health areas (H.A., from now on) have been one of the potentially targeted spatial units during the phase of de-escalation and the beginning of the new normality. Defined as the fundamental structure of the health system for the management of health centres in their autonomous community, and characterised by geographical, socioeconomic, demographic, epidemiological, cultural, labour, climatological and communication route factors \citep{SG20}, these organisational demarcations have been of great importance in the control of intervention measures by planning specific restrictions according to the results on risk level.

Nevertheless, this poses a challenge for obtaining precise estimates, which can be taken up by Small Area Estimation (SAE), a field of statistics that aims to obtain precise estimates of complex indicators in areas or domains with a small or even zero number of observations. Its development began with the study by \cite{FayHerriot1979}, and since then, especially in recent years under the Sustainable Development Goals (SDGs), methodological research has not ceased, with a succession of publications on inference on continuous and discrete variables. For a detailed review of this methodology, we recommend consulting the works of \cite{FayHerriot1979}, \cite{Battese1988}, \cite{PrasadRao1990} and \cite{JiangLahiri2001}, founders of the field, as well as consulting articles such as \cite{Pfeffermann2013} and monographs such as \cite{RaoMolina2015} and \cite{Morales2021}, which provide an exhaustive and clear follow-up of the development of SAE.

In fact, leading researchers in disease mapping highlight SAE techniques for their great potential to address one of the main problems in this area of research, the size of the areas under study. Disease mapping studies aim to estimate epidemiological indicators for administrative divisions of a region of interest. However, there are corresponding challenges that arise when monitoring rare diseases in the region of interest or diseases that are difficult to record \citep{MartinezBeneitoBotellaRocamora2019}.

Despite its possibilities, few studies to date have applied the SAE methodology to the study of the COVID-19 pandemic, most notably that of \cite{MartinezBeneito22}, which aims to make inferences about the incidence of COVID-19 cases using mixed models. In this research, following the line of mixed models with spatio-temporal dependence of random effects, they try to achieve greater flexibility also by incorporating random slopes, being, to date, to the best of our knowledge, the only study of COVID-19 and SAE with this type of models. Simulation analysis and application with real data from Comunidad Valenciana and Castilla y León demonstrated their potential usefulness in detecting different patrons in the evolution depending on their geographical area, facilitating the most appropriate decision making in each territory in terms of epidemic containment.

Indeed, incorporating random slopes can be key to providing the necessary flexibility in modelling, especially when the relationship between the target variable and the auxiliary variables is not constant over all areas and time. Thus, the first mixed models with random regression coefficients were introduced by \cite{Dempster1981}. However, despite an early initial development phase, with works such as those by \cite{PrasadRao1990} to derive the Empirical Best linear unbiased predictor (EBLUP), and the analytical estimator of the Mean squared error (MSE), research has been limited, with the work of \cite{Hobza2013} and \cite{Morales2021}, always in the context of linear mixed models (LMMs), being noteworthy. Recently, \cite{DizRosales2024a} have studied random slopes in the context of GLMMs by defining a Poisson area model with random regression coefficients (ARRCP model), emphasising the optimal results obtained both in simulation and in application to real data, over other models in which the regression coefficients are fixed.

Therefore, against this background, in this paper, for the first time, we derive predictors of the count of ICU beds occupied by COVID-19 patients, as well as occupancy rates, under the ARRCP model. All this under the main objective of estimating and predicting the counts and proportions of ICU beds occupied by COVID-19 patients by H.A. and day, applying it to data from Castilla y León, between November 2020 and March 2022, to support health system planning.

The paper is organised as follows. Section \ref{main.dataset} describes the COVID-19 dataset, specifying the choice of the temporal and spatial range under study, as well as motivating the suitability of the random slope models. Section \ref{main.app} presents the results of the application of the ARRCP model to real COVID-19 data from Castilla y León, with the aim of estimating and predicting the count and occupancy rate in ICU by COVID-19 patients by H.A. and day. Finally, some conclusions are presented in Section \ref{main.conc}.

In addition, the document contains supplementary material with some appendices. Appendix \ref{app.ext.descr.data} extends the descriptive information on the data, presented in Section \ref{main.dataset} of the main article, for H.A. and day of the study period. Finally, Appendix \ref{app.further.ap.data} extends the results from Section \ref{main.app} of the main article to assess the fit and forward predictive ability under the ARRCP model for H.A. and target day.

\section{COVID-19 dataset}\label{main.dataset}

It should be noted that, in this research, we have created the database from scratch. Thus, we first proceeded to search for and compile figures from official sources. Subsequently, based on the study guidelines of the Ministry of Health and specialised research articles, we constructed a series of auxiliary variables that could potentially explain ICU occupation.
Finally, with the database assembled, we created the aggregated version according to the target spatio-temporal domains or areas, in this case the crossing of H.A. with day in Castilla y León.

Importantly, we have made publicly available a GitHub repository, with all the original files, the aggregated files and the final database, accompanied by a comprehensive description of each step of the construction of the dataset, available at \cite{DizRosales2024b}.

\subsection{Spatial areas}\label{main.dataset.sa}

\begin{figure}[ht!]
  \begin{minipage}{0.45\textwidth}
    \centering
    \includegraphics[width=\linewidth]{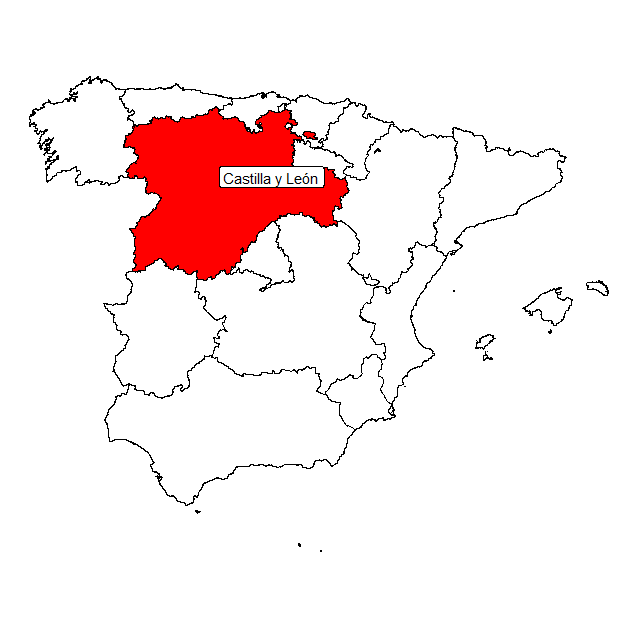}
  \end{minipage}%
  \begin{minipage}{0.45\textwidth}
    \centering
    \includegraphics[width=\linewidth]{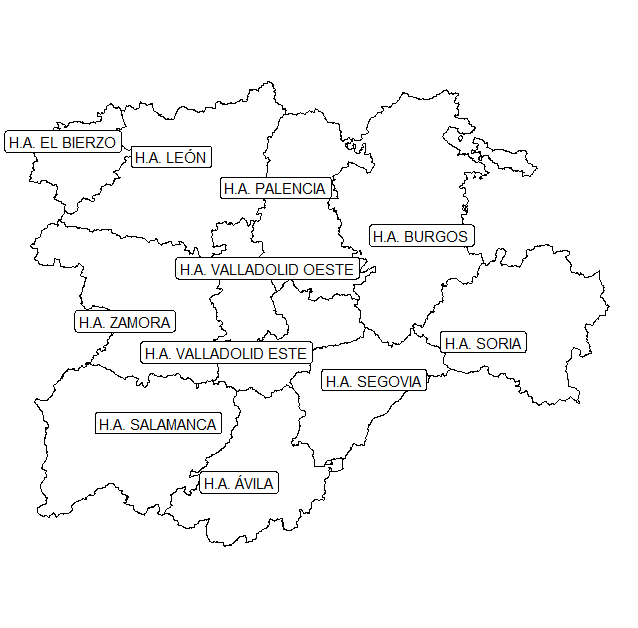}
  \end{minipage}
  \caption{Geographical location (left) and spatial layout of each H.A. in Castilla León (rigth).}\label{fig:main.dataset.saMAPHA}
\end{figure}

The domains or spatial areas of interest correspond to the H.A. of Castilla y León, an autonomous community in Spain, located, as can be seen in Figure \ref{fig:main.dataset.saMAPHA}, in the north-western fringe of the peninsula. 

Castilla y León, with its 2,385,223 inhabitants in 2021, is the sixth most populated autonomous community in Spain \citep{INE2023}, and the largest in the country, with a surface area of 94,227 $km^{2}$ \citep{IGN24}. It has a population density of 25.31 inhabitants per $km^{2}$, the lowest of all the Spanish regions. However, most of the population is concentrated around the urban centres, with high levels of depopulation in a large part of its rural area, which is aggravated by the notable ageing of its inhabitants \citep{Lopez2021}.
In fact, approximately 26.31\% of the population of the region was aged 65 or over, a figure significantly higher than the national proportion of around 20.20\% \citep{MinisterioSanidad2022a}. 

In the field of health, Castilla y León is the third autonomous community in Spain with the highest life expectancy in years at birth, with an average of 84.25 years, above the national level.

However, since one of the risk factors for more severe COVID-19 is comorbidities with diseases that may increase the likelihood of hospital admission, this analysis should be further explored. Thus, analysing the 2021 data on hospital morbidity rates per 100,000 inhabitants, in both communities, as is the case at national level, the main diseases or disorders requiring hospitalisation are: 1) Diseases of the circulatory system, 2) Diseases of the digestive system, 3) Neoplasms and 4) Diseases of the respiratory system \citep{INE2021}.

It is noteworthy that, within these four categories, Castilla y León has higher prevalences than the national average of diseases or disorders that cause more severe COVID-19.  Of note for their association with a higher risk of ICU admission and mortality due to COVID-19 are heart failure, high blood pressure or respiratory disorders such as COPD or Chronic Obstructive Pulmonary Disease \citep{IGN2021}.

In short, there are challenges that require a health system that sustains the care of citizens. In this respect, Castilla y León has a higher concentration of health centres in more populated areas.
In addition, in 2020, Castilla y León was in the top 10 autonomous communities with the highest ratios of functioning beds per 1,000 inhabitants \citep{MinisterioSanidad2020}.

Despite this, Castilla y León has been one of the most affected communities, with the highest case fatality rates \citep{Lopez2021}.

Given the heterogeneity of conditions, the partitioning of Castilla y León into units that allow more detailed information to be obtained on the dynamics of the virus is potentially useful. Therefore, in this study, the target spatial domain or area is defined as H.A, a key unit for determining the risk of health collapse during the process of implementing virus containment measures. In Figure \ref{fig:main.dataset.saMAPHA}, each of the health areas that subdivide Castilla y León can be identified and geographically located. 

As is the case at the community and provincial level, each H.A. has its own socio-demographic characteristics that are different from the rest of the areas.
For example, most of the population and health centres are concentrated in urban centres, and therefore, the health areas of Burgos, Salamanca, León and Valladolid together exceed the third quartile of total population size. However, the highest proportions of the population over 65 are to be found in the health areas of Zamora, León and El Bierzo, with the lowest values in Valladolid Este, Valladolid Oeste, Burgos and Segovia.

In the following subsections, the auxiliary and objective variables will be examined in depth, disaggregated by H.A, making a comparison based on their characteristics. 

\subsection{Temporal Areas}\label{main.dataset.ta}

The dataset incorporates observations corresponding to 490 consecutive days for each H.A., collected from 2 November 2020 to 6 March 2022. Of this total, it should be specified that the 182 days between 2 November 2020 and 2 May 2021 are used to assess the adjustment capacity of the ARRCP model. On the other hand, the remaining 308 days, from 3 May 2021 to 6 March 2022, are used to assess the quality of the forward forecast with data that have not been used in the model fit.

The choice of this time frame corresponds to the following reasons. Firstly, the availability of data. The selection of 2 November 2020 as the starting date is due to the fact that there are only records, for both communities, of the occupancy of ward and ICU beds by patients with COVID-19 since 7 October 2020. Therefore, in order to be able to incorporate into the study the dynamics of the evolution of the disease, as well as the delays in the reporting of pandemic data, it is necessary to be able to incorporate not only the occupancy recorded on the target day, but also its delayed versions. Consequently, we leave a time window from 7 October to the start date of the study in order to create the lagged variables using the available information. In this way, we also aim to incorporate the time lag between infection, symptom development, possible admission to ward and/or ICU, and unfortunately, death.

Secondly, the wide range of days under study aims to reproduce the highly variable scenario experienced during the pandemic, with periods of both high and low pressure, as well as incorporating the effect of the emergence of new variants.
Specifically, according to the National Epidemiology Centre (CNE), a total of five epidemic periods (EP) occurred during this time range \citep{ISCIII2022}.
\begin{itemize}
\item Second epidemic period (2nd EP): Between 22 June 2020 and 6 December 2020, at which point the 14-day CI (Cumulative Incidence) tipping point of COVID-19 cases occurs, leading to the third period.
\item Third period (3rd EP): Between 7 December 2020 and 14 March 2021, when the 14-day CI tipping point of COVID-19 cases occurs, leading to the fourth period.
\item Fourth period (4th EP): Between 15 March 2021 and 19 June 2021, when the 14-day CI turning point of COVID-19 cases occurs, leading to the fifth period.
\item Fifth period (5th EP): Between 20 June 2021 and 13 October 2021, when the 14-day CI tipping point for COVID-19 cases occurs, leading to the sixth period.
\item Sixth period (6th EP): Between 14 October 2021 and 27 March 2022, the last day before the implementation of a new epidemiological surveillance strategy.
\end{itemize}

This temporal structuring allows the assessment of the adjustment and predictive capacity at critical moments for any model, such as the transition between periods. Thus, errors of greater magnitude are generally expected since these changes are usually marked by significant variations in the dynamics of the disease, as shown by the benchmark of the turning point in the 14-day CI. Furthermore, during such periods and transitions, various events can and have occurred, such as changes in intervention measures, population behaviours, changes in virus variants and other factors that may affect the spread and impact of COVID-19, introducing uncertainty into future adjustment and prediction models.

In the same vein, the 490 days under study allow to take into consideration the impact of three main SARS-CoV-2 variants, which were considered by the European Centre for Disease Prevention and Control as Variants of Concern (VOC) for their significant epidemiological impact: Alpha (B.1.1.7), Delta (B.1.617.2) and Omicron (B.1.1.529). Their dominance was detected in wastewater analyses in the study regions from January 2021, July 2021 and December 2021, respectively \citep{MinisterioSanidad2022b,TrigoTasende2023}, and led to certain changes in epidemiological parameters.
Thus, Alpha led to both an increase in the incidence and severity of COVID-19, as it increased the reproduction number of SARS-CoV-2, as well as severity, measured as admissions to hospitalisation units or ICUs, or lethality, in deaths. Delta, on the other hand, caused a decrease in the incubation period of the disease, while its incidence increased due to an increase in the reproduction number. Finally, Omicron, whose circulation coincided with a period of high vaccination coverage, also led to a decrease in the incubation period, and an increase in incidence as the number of cases increased. However, there was also a decrease in disease duration, severity and lethality, probably due to the aforementioned vaccination campaign.

\begin{figure}[h!]
  \begin{minipage}{0.50\textwidth}
    \centering
    \includegraphics[width=\linewidth]{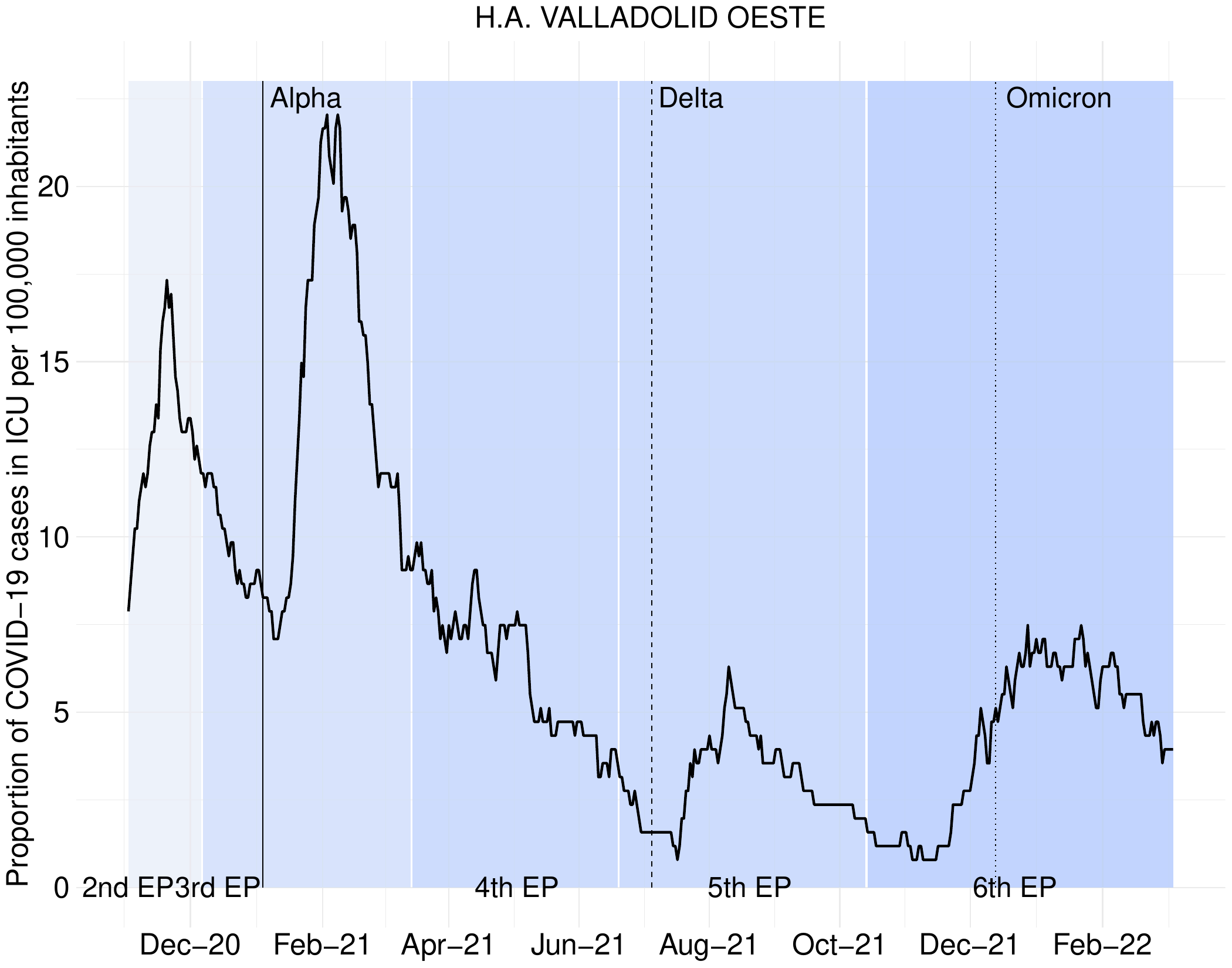}
  \end{minipage}%
  \begin{minipage}{0.50\textwidth}
    \centering
    \includegraphics[width=\linewidth]{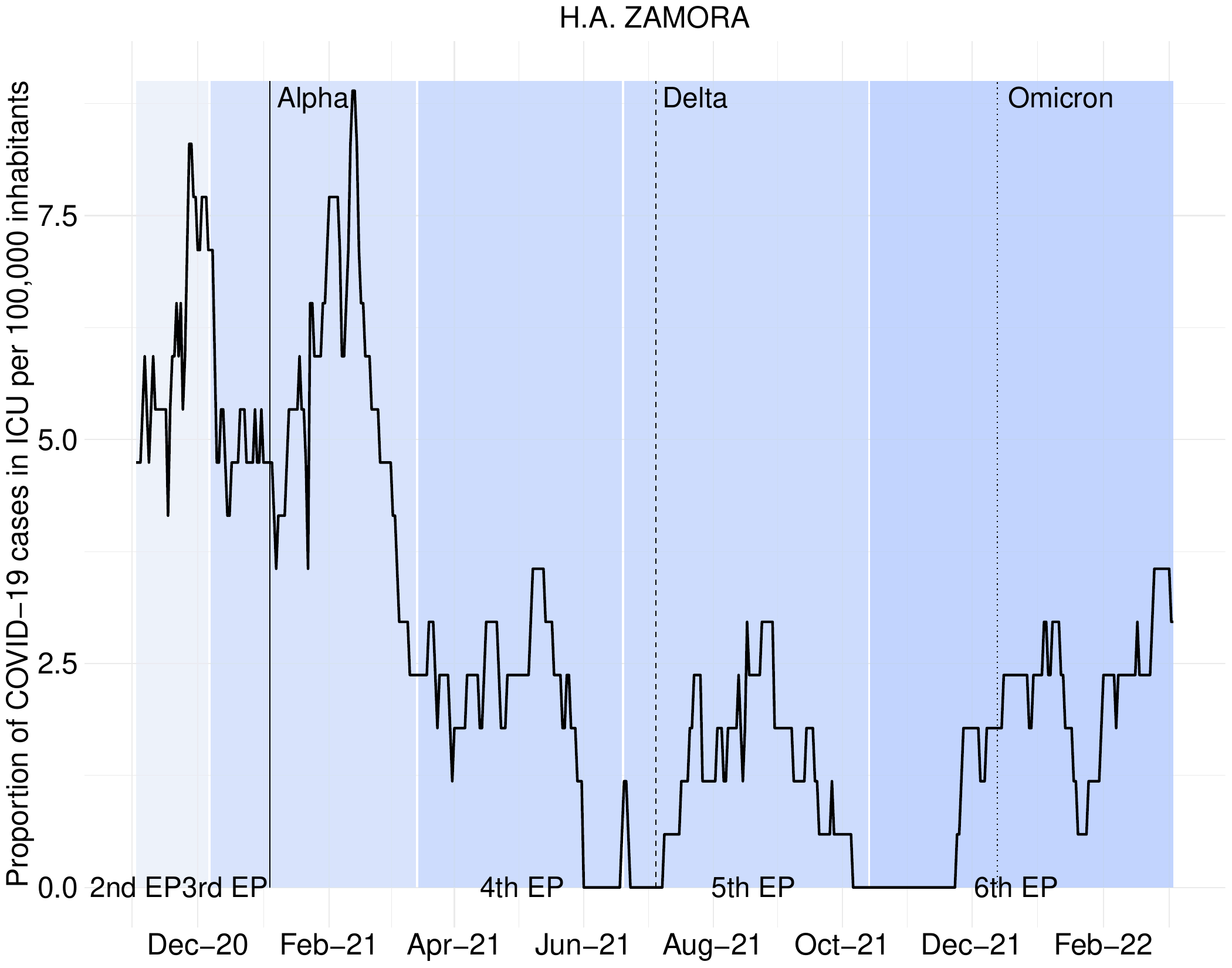}
  \end{minipage}
  \caption{Evolution of the ICU occupancy rate by COVID-19 per 100,000 inhabitants.}\label{fig:main.dataset.ta.ICUHA}
\end{figure}

The pattern of ICU admissions due to COVID-19 per 100,000 inhabitants is broken down below for each H.A. and day. 
Figure \ref{fig:main.dataset.ta.ICUHA} shows the health areas of Zamora and Valladolid Oeste in Castilla y León. In this way, the areas with the lowest and highest proportion of average ICU occupancy are represented for each community, respectively, while the areas with the lowest and highest number of inhabitants, Soria and Burgos, can be viewed in the Subsection \ref{app.ext.descr.data.ICU} of the Appendix \ref{app.ext.descr.data}. This choice is due to two reasons. As discussed in the Section \ref{main.intro}, a smaller number of available observations of the target variable may make estimation and/or prediction more difficult. Also, focusing on both extremes allows us to test the performance of our proposal in both high and low-pressure scenarios.

Several interesting conclusions can be drawn from the graphs. Firstly, the highest number of admissions, are concentrated mainly in the second and third epidemic period. When analysed by H.A., the heterogeneity of patterns between areas is noteworthy.

It can be seen that in the H.A. Zamora, two peaks of similar magnitude are reached at the end of the second epidemic period and at the end of the third, while in the H.A. Valladolid Oestese experiences a major peak in the third. In addition, both have hospital pressure at different times, with a certain delay in the H.A. Zamora, which is consistent with the epidemiological bases, since in densely populated and more connected areas, such as the H.A.Valladolid Oeste, the disease arrives earlier and spreads more quickly, subsequently leading to possible admissions. Along these lines, in the H.A. Zamora  sharper fluctuations are perceived, with more abrupt peaks and troughs in occupancy.

The behaviour described is consistent with the nature of the data. On the one hand, in the H.A. Zamora, the population size is substantially smaller than in the H.A. Valladolid Oeste. It is logical to expect sharper fluctuations in areas with smaller population size, as even a small number of additional cases can have a significant impact on the proportion of ICU occupancy, while in more populated areas, the variation in the number of cases can be cushioned by the larger population base. This is extensible to the rest of the areas, as these abrupt changes are also detected in areas such as the H.A. Soria. Especially in these areas, it is noteworthy that when occupancy peaks are reached, they sometimes exceed those in areas with higher overall ratios. This needs to be further studied, but could be due to the fact that when outbreaks occur, the spread may be faster, coupled with lower herd immunity, a more limited health care system or an older population. Finally, it should be noted that, as expected from their definition, the dominance of the Alpha and Delta variants has a clear impact on the increase in occupancy, as does the dominance of the Omicron variant, although not to the same extent, possibly due to vaccination coverage.

This visualisation allows us to detect both the heterogeneous behaviour between health areas and within the same H.A. as a function of time, with respect to the dynamics of occupancy in ICUs due to COVID-19. Therefore, the specification of random effects could be key to provide the flexibility that is desired in any modelling and prediction tool.
For this, an additional consideration must be made. Throughout the pandemic, the reporting of incidence data, but also of deaths, discharges and hospital occupancy, has experienced delays in notification. Particularly in the first months, this was due to the high pressure on hospitals, which resulted in a shortage of resources to carry out the process with the desired accuracy. However, once the most critical phases have been overcome, the counting of data is still affected by the so-called weekend effect. This is that, typically, data from Sunday, and even national or local holidays, experience a local minimum, with a sharp spike, usually at the beginning of the week. This has been widely reported in the research that has been carried out, and is usually dealt with by averaging the data on a weekly basis \citep{Area2021}. 

However, averaging has its drawbacks. \cite{MartinezBeneito22} point out that this leads to a danger of inducing time dependence, as well as making the estimates less sensitive to the latest changes that the data might show, decreasing the effectiveness for epidemiological surveillance. Therefore, in this research, temporal random effects are defined at the week level. By defining a random intercept that groups days into weeks we aim to have the following benefits. First, it can help to reduce variability in the data, which can improve the accuracy and reliability of the results. In addition, by grouping days into weeks, a better understanding of trends over time can be obtained and patterns can be identified that may not be evident if the data are analysed on a day-by-day basis. In addition, by using a random effect at the intercept level to group days into weeks, differences between weeks can be taken into account and more accurate estimates of model parameters can be obtained. Finally, the number of available observations is increased, which is beneficial for the optimal prediction of the modal predictors of the random effects.

\subsection{Variables}\label{main.dataset.var}

Based on the spatio-temporal areas defined, the aggregated dataset presents a total of 5390 rows, corresponding to 490 consecutive days of observations for a total of 11 health areas in Castilla y León.

For each H.A. and day, the objective variable of the ARRCP model is the count of people hospitalised with COVID-19 in ICU. Therefore, based on the guidelines indicated in the Epidemic Surveillance Strategies of the Ministry of Health \citep{MinisterioSanidad2022c} and the specialised literature, observations of potential auxiliary variables specified in Table \ref{tab:main.dataset.var} have been recompiled.

\begin{table}[ht!]
\caption{Definition of the main auxiliary variables considered.}\label{tab:main.dataset.var}

\centering
  \begin{adjustbox}{width=\textwidth}
\begin{tabular}{|lll|}
\hline
\multicolumn{1}{|l|}{\textbf{Variable}}       & \multicolumn{1}{l|}{\textbf{Description}}                                                                       & \textbf{Units}                                             \\ \hline
\multicolumn{3}{|l|}{Assessment of the level of transmission}                                                                                                                                                                \\ \hline
\multicolumn{1}{|l|}{\textit{CI14}}           & \multicolumn{1}{l|}{14-day cumulative incidence: number of confirmed 14-day COVID-19 cases}                     & Per 100,000 inhabitants (rescaled per 1,000 inhabitants)   \\
\multicolumn{1}{|l|}{\textit{CI7}}            & \multicolumn{1}{l|}{7-day cumulative incidence: number of confirmed 7-day COVID-19 cases}                       & Per 100,000 inhabitants (rescaled per 1,000 inhabitants)   \\
\multicolumn{1}{|l|}{\textit{positive.rate7}} & \multicolumn{1}{l|}{Weekly tests with a positive result between the number of tests performed in that period} & Percentage                                                (rescaled to base 1) \\ \hline
\multicolumn{3}{|l|}{Assessment of the level of utilisation of care services by COVID-19}                                                                                                                                    \\ \hline
\multicolumn{1}{|l|}{\textit{ward.rate}}      & \multicolumn{1}{l|}{Hospital beds occupied by COVID-19 cases}                                                   & Per 100,000 inhabitants (rescaled per 1,000 inhabitants)   \\
\multicolumn{1}{|l|}{\textit{disch.rate14}}   & \multicolumn{1}{l|}{COVID-19 discharges within 14 days}                                                         & Per 1,000 inhabitants    \\
\multicolumn{1}{|l|}{\textit{disch.rate7}}    & \multicolumn{1}{l|}{COVID-19 discharges within 7 days}                                                          & Per 1,000 inhabitants    \\ \hline
\multicolumn{3}{|l|}{Assessment of the level of severity}                                                                                                                                                                    \\ \hline
\multicolumn{1}{|l|}{\textit{acute.rate7}}    & \multicolumn{1}{l|}{Cases admitted to ICU among all COVID-19 inpatients within seven days}                      & Percentage                                               (rescaled to base 1) \\
\multicolumn{1}{|l|}{\textit{death.rate7}}    & \multicolumn{1}{l|}{7-day rate of COVID-19 deaths}                                                              & Per 1,000,000 inhabitants (rescaled per 1,000 inhabitants) \\
\multicolumn{1}{|l|}{\textit{death.rate14}}   & \multicolumn{1}{l|}{14-day rate of COVID-19 deaths}                                                             & Per 1,000,000 inhabitants (rescaled per 1,000 inhabitants) \\ \hline
\end{tabular}
\end{adjustbox}

\end{table}

For the construction of the target and auxiliary variables for Castilla y León, we have taken the data available on the Open Data Portal managed by the Junta de Castilla y León \citep{JCL2024}.

At this point, we would like to highlight the following. In Spain, health competences are transferred to its 17 autonomous communities and two autonomous cities. Thus, each community is responsible for the collection and reporting of COVID-19 data occurring in its territory, with criteria that have not always been homogeneous among them. At the same time, it should be borne in mind that there are also important differences in non-pharmacological intervention measures. Thus, as is also highlighted in the literature, data from different communities cannot be compared \citep{Area2021}.

It should also be noted that, due to the interest in knowing the population proportion of ICU beds occupied by COVID-19 patients, the model, defined in Section \ref{main.app}, has an offset, which in this case corresponds to the population living in each H.A. in 2021 according to the figures of the Population and Housing Census on 2021, published by the National Institute of Statistics \citep{INE2023}. Thus, the population figures for each municipality have been collected and aggregated at the H.A. level.

As a result of a model selection process, specified in Section \ref{main.app}, the auxiliary variables finally defined in the ARRCP model are: \textit{ward.rateL2}, \textit{disch.rateL14} and \textit{acute.rate7}, whose proportions vary depending on the epidemic period. For clarification purposes, it should be noted that the suffixes {\it L2} and {\it L3} in the auxiliary variables refer to a time lag of 2 and 3 days with respect to the target variable, which falls within the confidence band indicated in the literature on disease dynamics \citep{MinisterioSanidad2022c}. 
Descriptive statistics for each auxiliary variable, as well as their evolution by EP, H.A. and day, can be found in detail in the Subsections \ref{app.ext.descr.data.WARD}, \ref{app.ext.descr.data.RD} and \ref{app.ext.descr.data.ACUTE} of the Appendix \ref{app.ext.descr.data}.

\begin{figure}[h!]
  \begin{center}
    \includegraphics[width=0.4\linewidth]{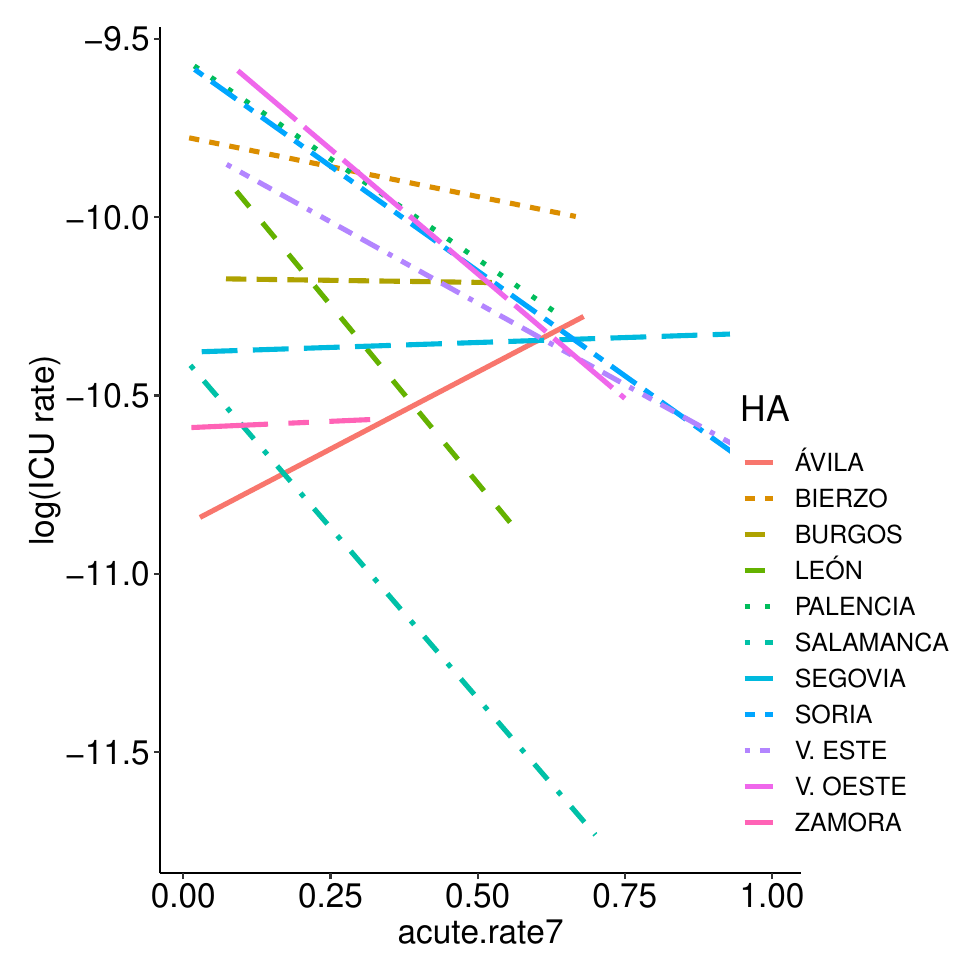}
  \caption{Log-ICU rate versus {\it acute.rate7} by Health Area (H.A.).}\label{fig:main.dataset.var.SLOPES}
  \end{center}
\end{figure}

Given the heterogeneity of patterns detected in the ICU occupation, a preliminary analysis is performed to check whether this diversity carries over to the relationship of the auxiliary variables with the target variable. In Figure \ref{fig:main.dataset.var.SLOPES}, we visualise the relationship of the variable {\it acute.rate7}, selected in Section \ref{main.app} as random slope, with the logarithm of the proportion of COVID-19 patients in ICU per 100,000 inhabitants, for Castilla y León. Note that the use of the logarithm is based on the model's equation \ref{ARRCPmodel}, which defines a logarithmic link, which is the natural one for the Poisson distribution. In this way it makes sense to represent the logarithm of the occupancy rate, equivalent to the natural parameter $p_{it}$ incorporated in that equation. For the rest of the auxiliary variables, the corresponding figures can be found in the Subection \ref{app.ext.descr.data.SLOPES} of the Appendix \ref{app.further.ap.data}.

We can observe that the slopes per H.A. are different, both in magnitude and direction. Thus, in H.A. Ávila, there is a slightly increasing trend. That is, as the proportion of {\it acute.rate7} increases, the proportion of admissions to the ICU increases, which indicates that the number of COVID-19 patients in the ICU increases in relation to the increase in the number of patients on the ward, either due to a greater severity of the disease, or due to a higher incidence that causes more admissions and therefore greater hospital saturation. On the other hand, the trend is practically horizontal in areas such as H.A. Burgos, H.A. Segovia and H.A. Valladolid Oeste. In the rest, a decreasing trend can be seen, since admission to the ICU may be accompanied by an equal or greater increase in hospitalisation on the ward. For all these reasons, the inclusion of spatial and temporal random effects is proposed in order to take a firm step towards the development of an explanatory and predictive model, with sufficient flexibility to maintain optimal precision in the face of the dynamics of the disease.

\section{Application to real data}\label{main.app}

During the model selection phase, we consider several criteria, highlighting that must be a model with easily accessible auxiliary information in epidemic contexts for its applicability to be viable. In addition, we considered: a) significance of model parameters and epidemiological interpretability; b) convergence of the ML-Laplace approximation algorithm; c) validity of model assumptions; and d) lower conditional AIC (for more information on model selection in SAE see \cite{Vaida2005} and \cite{Lombardia2017} among others).

Regarding the assessment of the significance of the model parameters, two studies are carried out, the estimation of the studentized confidence intervals and the p-value. For the convergence analysis, any possible failures or warnings during the estimation process are monitored, while applying the golden rule as stated by \cite{Bates2023}, the developers of the used package. That is, the model is fitted with all the optimisers available in the statistical software, and the coefficient of variation (CV) of the estimates of each of the parameters is calculated.

The validity of the model assumptions is checked with an analysis of the residuals, to detect possible problems of heteroscedasticity, non-normality, as well as to study the ability to fit.

Finally, for each model under study, the cAIC is estimated using the `cAIC4' package \citep{Saefken2021}, so that the model that meets the three preconditions and shows the lowest cAIC is the one finally selected.

In this way, the natural parameters of the selected ARRCP model are
\begin{equation}\label{ARRCPmodel}
\log\mu_{it}=\log \nu_{it}+\log p_{it}=\log \nu_{it}+\sum_{\ell=1}^4\beta_\ell x_{\ell,it}+\sigma u_{iw(t)}+\phi_1v_{1,i}x_{1,it},
\end{equation}

where $p_{it}$ corresponds to the domain proportion of people with COVID-19 in ICUs and $\nu_{it}=N_{it}$ is the population size in H.A. $i$ and day $t$. The model contains $p=4$ auxiliary variables: $x_0=$ intercept, $x_1=$ acute.rate7, $x_2=$ disch.rate14L3, $x_3=$ ward.rateL2,  with regression parameters
$\beta_0=-10.6206$, $\beta_1=1.7776$, $\beta_2=0.3551$, $\beta_3=1.1910$. Furthermore, we define the random effects at the intercept level in H.A. $i$ and week $w$ as $u_{iw(t)}\sim N(0,1)$ and the random slope $v_{1,i}\sim N(0,1)$ in H.A. $i$. 
The standard deviation parameters are
$\sigma=0.2075$, $\phi_0=0$, $\phi_1=1.1906$, $\phi_2=0$ and $\phi_3=0$.
Finally, the correlation parameters is  $\rho_{ab}=0$ for $a<b$ belonging to the set $\mathbb{I}=\{1,\ldots,11\}$.

First, the significance of the regression and variance parameters fitted under the ARRCP model for the real data is assessed. In addition, following the indications provided at the beginning of the section, the convergence of the model is evaluated, calculating the CV for each parameter.

Table \ref{tab:main.app.real.data.CL.CICV} presents the parameter estimates, the confidence intervals studied at $\alpha=5\%$ obtained using the double bootstrap approximation, with a total of 1000 replicates, as well as the CV.

\begin{table}[ht!]
\caption{t-percentile confidence intervals $\alpha$=5\% and CV of parameter estimates.}\label{tab:main.app.real.data.CL.CICV}
\renewcommand{\arraystretch}{1.0}
\centering
\begin{tabular}{|l|rrrrrr|}
\hline
&$\beta_0$& $\beta_1$ & $\beta_2$ & $\beta_3$ & $\sigma$ &$\phi_1$\\
\hline
2.5\%&-10.7467&0.8288&0.2509&1.0371&0.1887&0.8948  \\
Estimate &-10.6206&1.7776&0.3551&1.1910&0.2075&1.1906\\
97.5\% &-10.4857&2.6603&0.4442&1.3512&0.2324&2.1535\\
\hline
CV &-0.0036  & 0.2222   & 0.0169   & 0.0251   & 0.0160   & 0.0543\\
\hline
\end{tabular}
\end{table}

Firstly, all parameters are significant, since in no case do the confidence intervals contain zero. Likewise, the regression parameter $\beta_1$ and the variance parameter associated with the random slope, $\phi_1$, are the ones with the longest interval length. This is consistent, as it is a possible indication of the inherent variability between the groups or categories that led to the definition of the slope, and which increases the uncertainty in the estimation.

Likewise, for $\phi_1$ the $p-value=0$, corroborates the significance of the variance parameter associated with the random slope, contributing to increase the explanatory capacity of the COVID-19 occupation in ICU.
Once the significance of the parameters has been verified, it is essential to check their epidemiological interpretability.

Regarding the intercept, since the expected proportion of people with COVID-19 in the ICU cannot be negative, it has no direct interpretation in this context.
However, for the rest of the regression parameters, we find that they have a protective effect on ICU occupation due to COVID-19. This is consistent with what has been recorded and with the nature of the disease itself.
Broken down variable by variable, the positive value of $\beta_{1}$, associated with the variable $acute.rate7$, indicates that an increase in ICU bed occupancy relative to total ICU and ward occupancy is associated with an increase in the proportion of ICU occupancy due to COVID-19. This is consistent, since higher values of $acute.rate7$ may indicate that severe cases have increased relative to total inpatients, and these will eventually be referred to the ICU.

On the other hand, the positive value of $\beta_{2}$, associated with the variable $disch.rate14L3$, might seem contradictory, since the variable represents discharges. However, a higher number of discharges may imply that recovery from the disease is being successful, but the rate of admissions is maintained or increased, and/or that this is due to the fact that there has been an increase in the number of cases, so that both the proportion of cured and potential candidates for admission to the ICU increases.

Finally, the positive value of $\beta_{3}$, associated with the variable $ward.rateL2$, is possibly the most intuitive, since it indicates that an increase in the rate of hospitalisation on the ward leads to an increase in the rate of hospitalisation in the ICU. Although there is not always a direct relationship, since it depends on the evolution of the severity of COVID-19 and the intervention measures, it is consistent, since there is a greater possibility of leading to more severe symptomatology.

These results are consistent with reports published by health authorities, which use these variables as risk indicators to detect when there is an increase that may lead to action being taken to reduce the likelihood of healthcare collapse.
However, it is important to note that these interpetations are based on statistical associations and do not necessarily imply a direct causal relationship. In fact, the relationship between auxiliary variables and the target variable may be complex and depend on multiple factors. In the disease monitoring reports themselves, it is emphasised precisely that it is a multifactorial problem, so that the risk assessment must also consider the specific characteristics of the territorial unit, and other indicators of care capacity, epidemiological situation, public health capacity and characteristics of the susceptible population \citep{MinisterioSanidad2022c}.

Regarding the convergence assessment, the CV is of order $10^{-2}$, except for the parameter $\beta_{1}$, which is of order $10^{-1}$. This is consistent since the inclusion of a slope can introduce more variability in the parameter estimation, resulting in higher coefficients of variation. However, this order of variation shows consistent behaviour of all optimisers, with no outstanding problems.

As for the $cAIC$, it is again indicative of a good choice of model. Thus, the model with random slope has a $cAIC= 9683.9037$, lower than $cAIC=9729.2458$ for the model with only random intercept.

\begin{figure}[ht!]
\centering
  \begin{minipage}{0.4\textwidth}
    \centering
    \includegraphics[width=\linewidth]{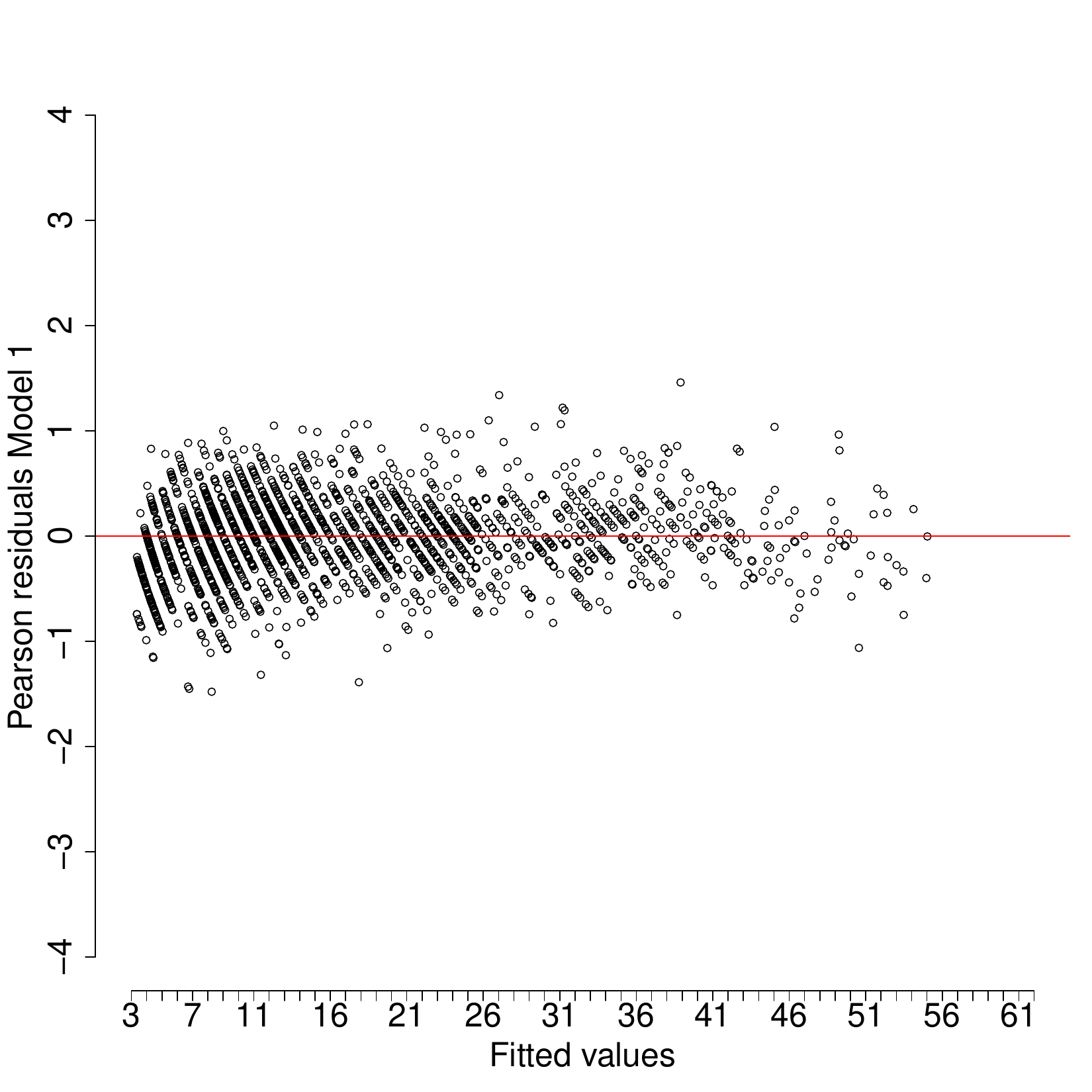}
  \end{minipage}%
  \begin{minipage}{0.4\textwidth}
    \centering
    \includegraphics[width=\linewidth]{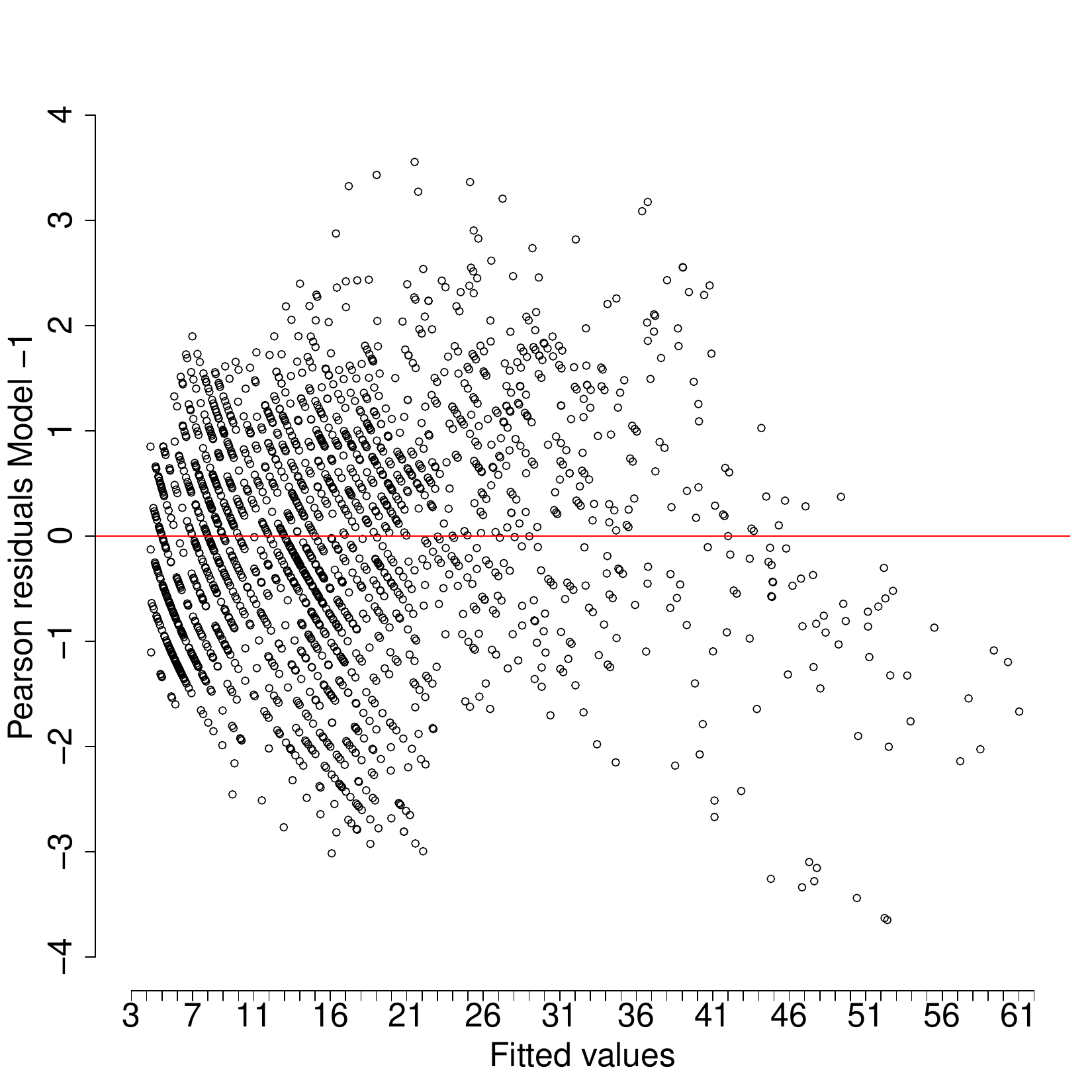}
  \end{minipage}
  \begin{minipage}{0.4\textwidth}
    \centering
    \includegraphics[width=\linewidth]{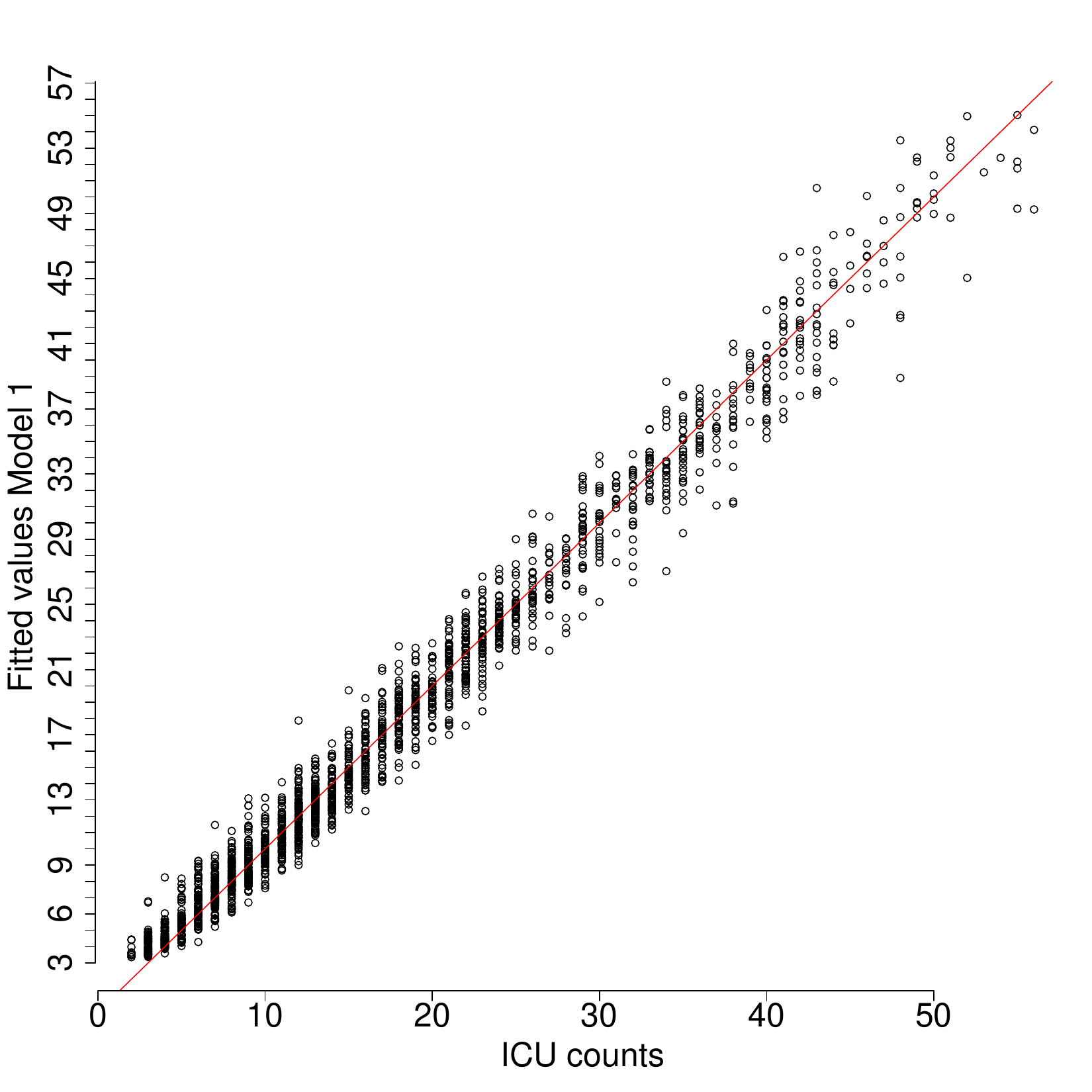}
  \end{minipage}%
  \begin{minipage}{0.4\textwidth}
    \centering
    \includegraphics[width=\linewidth]{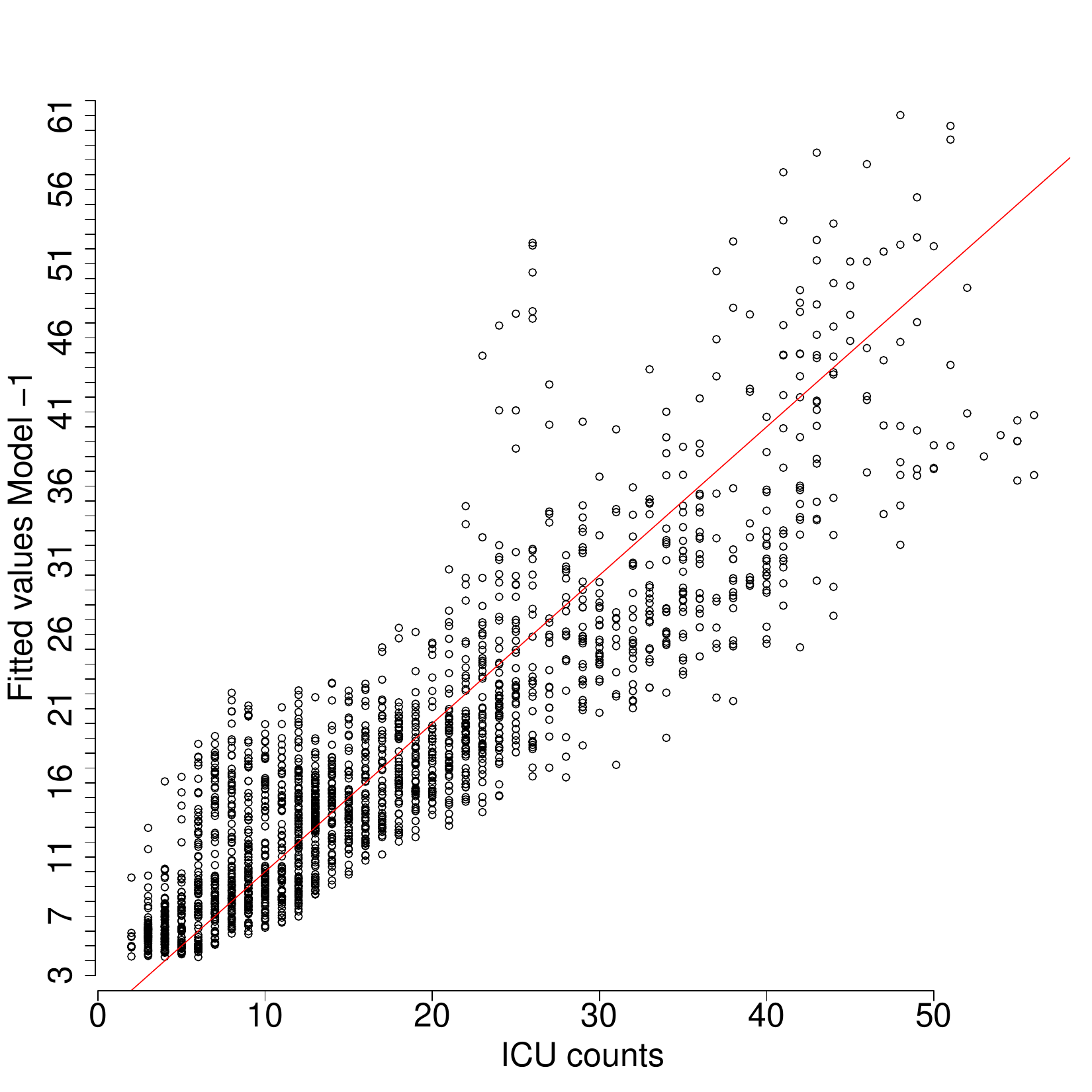}
  \end{minipage}
  \caption{Diagnosis of the Model 1 vs Model -1.}
\label{fig:main.app.real.data.CL.RES}
\end{figure}

The last step to finalise the assessment of the adequacy of the selected ARRCP model is to analyse the residuals. In this problem, we work with the Pearson residuals.
In Figure \ref{fig:main.app.real.data.CL.RES}, the goodness of fit of the model with random slope (Model 1) is assessed and compared to a model without random effects (Model -1) by analysing, on the one hand, the ratio of the residuals versus the fitted values (top row) and the ratio between the fitted values and the count of COVID-19 cases recorded in the ICU (bottom row).

Thus, in Model 1, the residuals are randomly centred around the $y=0$ line, mostly concentrated between -1.5 and 1.5, while when no random effects are included, patterns are detected and the dispersion increases significantly, with ranges between -4 and 4. On the other hand, Model 1 also represents a notable gain in goodness of fit compared to Model -1.

\begin{figure}[h!]
\centering
  \begin{minipage}{0.4\textwidth}
    \centering
    \includegraphics[width=\linewidth]{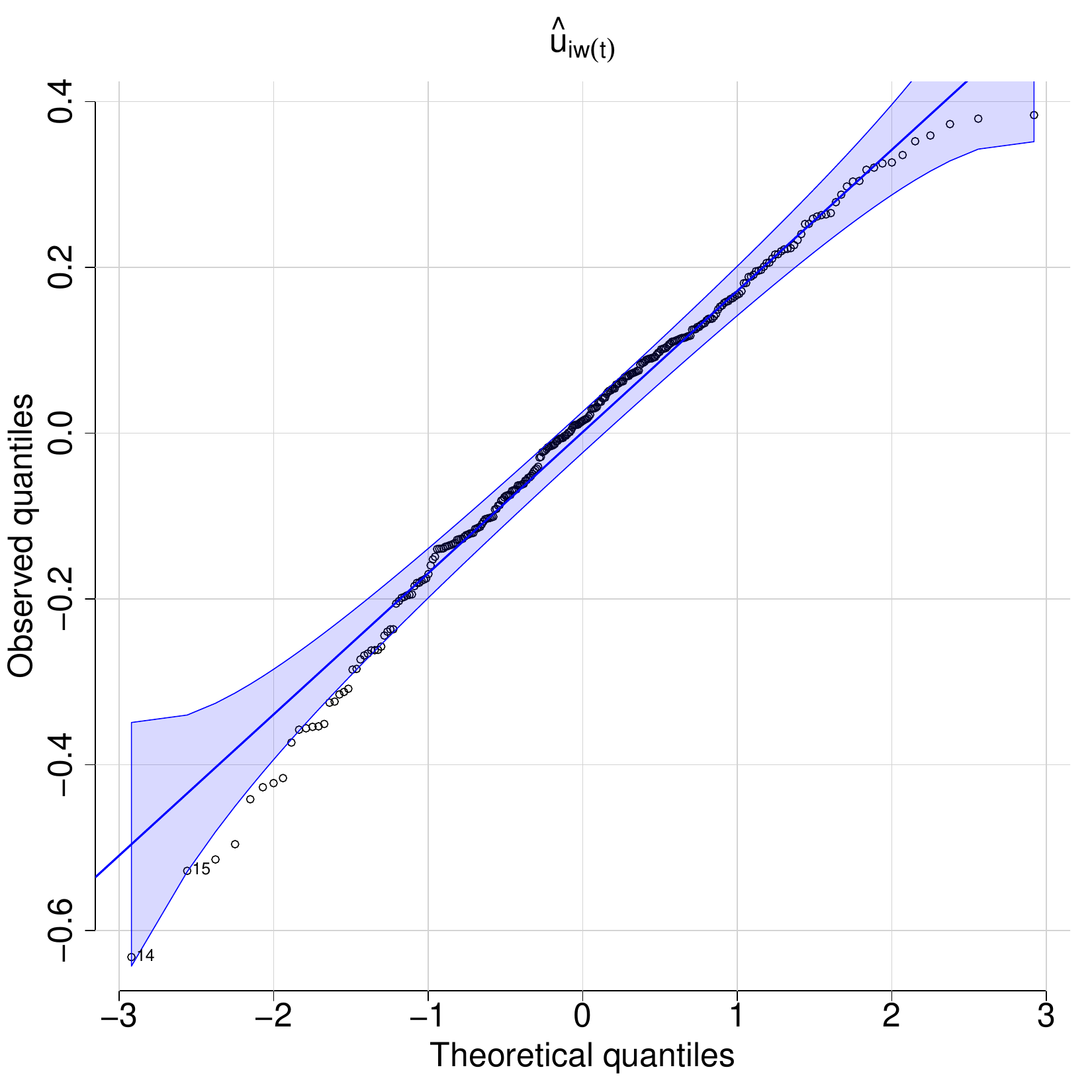}
  \end{minipage}%
  \begin{minipage}{0.4\textwidth}
    \centering
    \includegraphics[width=\linewidth]{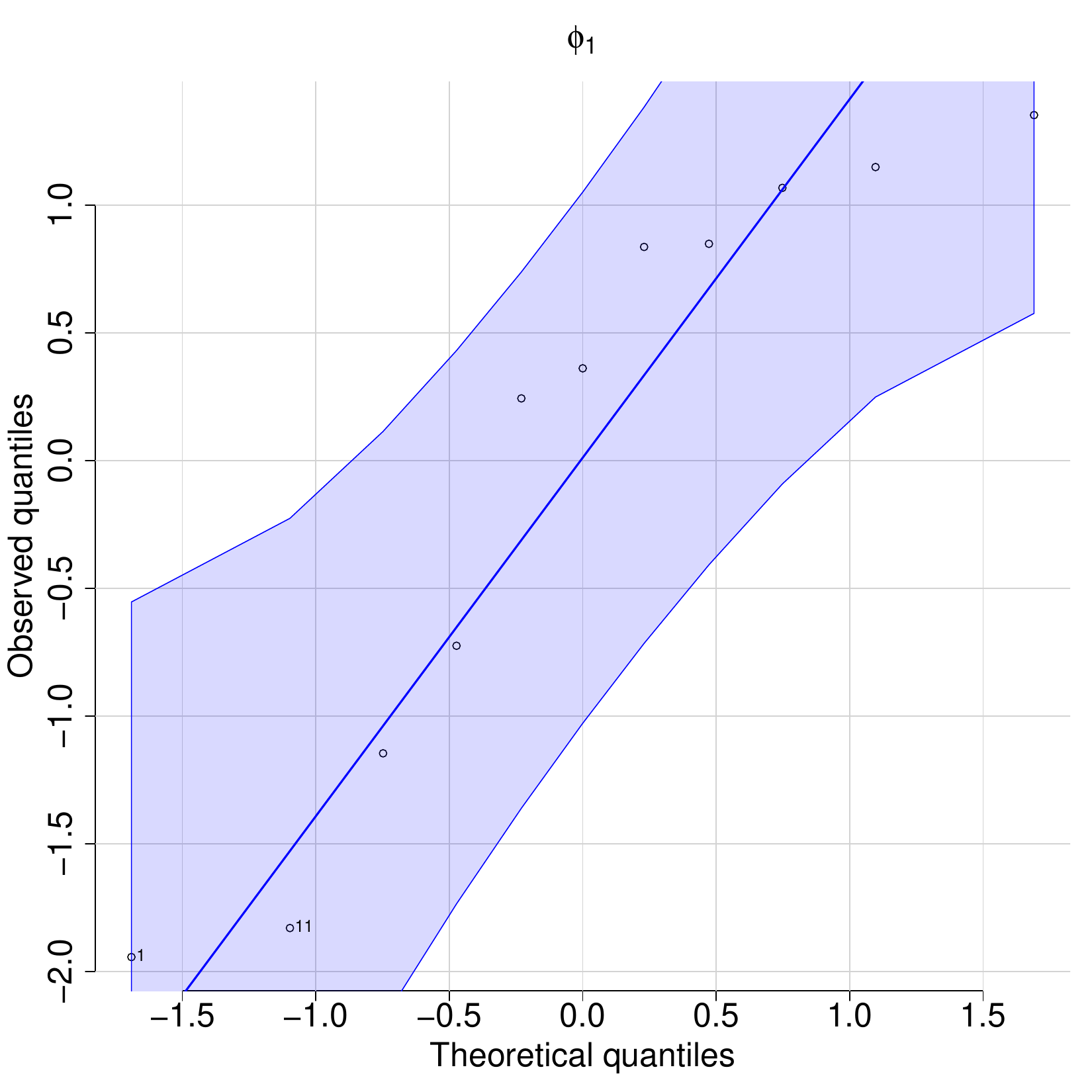}
  \end{minipage}
  \caption{Pearson residuals with respect to the normal values.}
\label{fig:main.app.real.data.CL.QQ}
\end{figure}

Regarding the assumption of normality for the random effects, a qqplot for the random intercept and for the random slope is plotted in Figure \ref{fig:main.app.real.data.CL.QQ}. For the random intercept, it is seen that although, with a deviation in the tails, the normality assumption is not an inconsistency. For the random slopes, only 11 points are available, so the assessment of normality is not very definitive, but again, in view of the results, there is nothing to suggest that the assumptions are wrong.

\begin{figure}[h!]
  \begin{minipage}{0.50\textwidth}
    \centering
    \includegraphics[width=\linewidth]{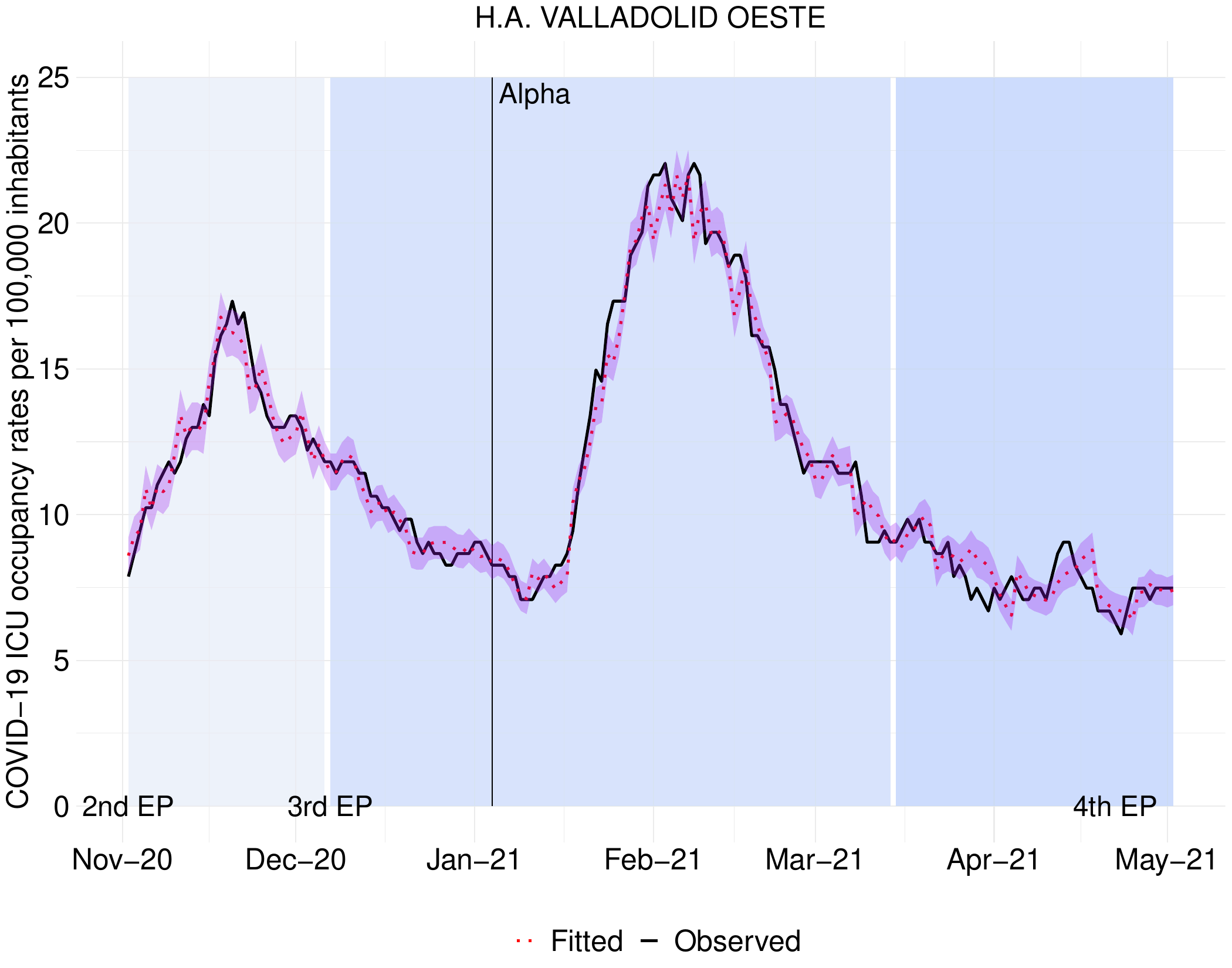}
  \end{minipage}%
  \begin{minipage}{0.50\textwidth}
    \centering
    \includegraphics[width=\linewidth]{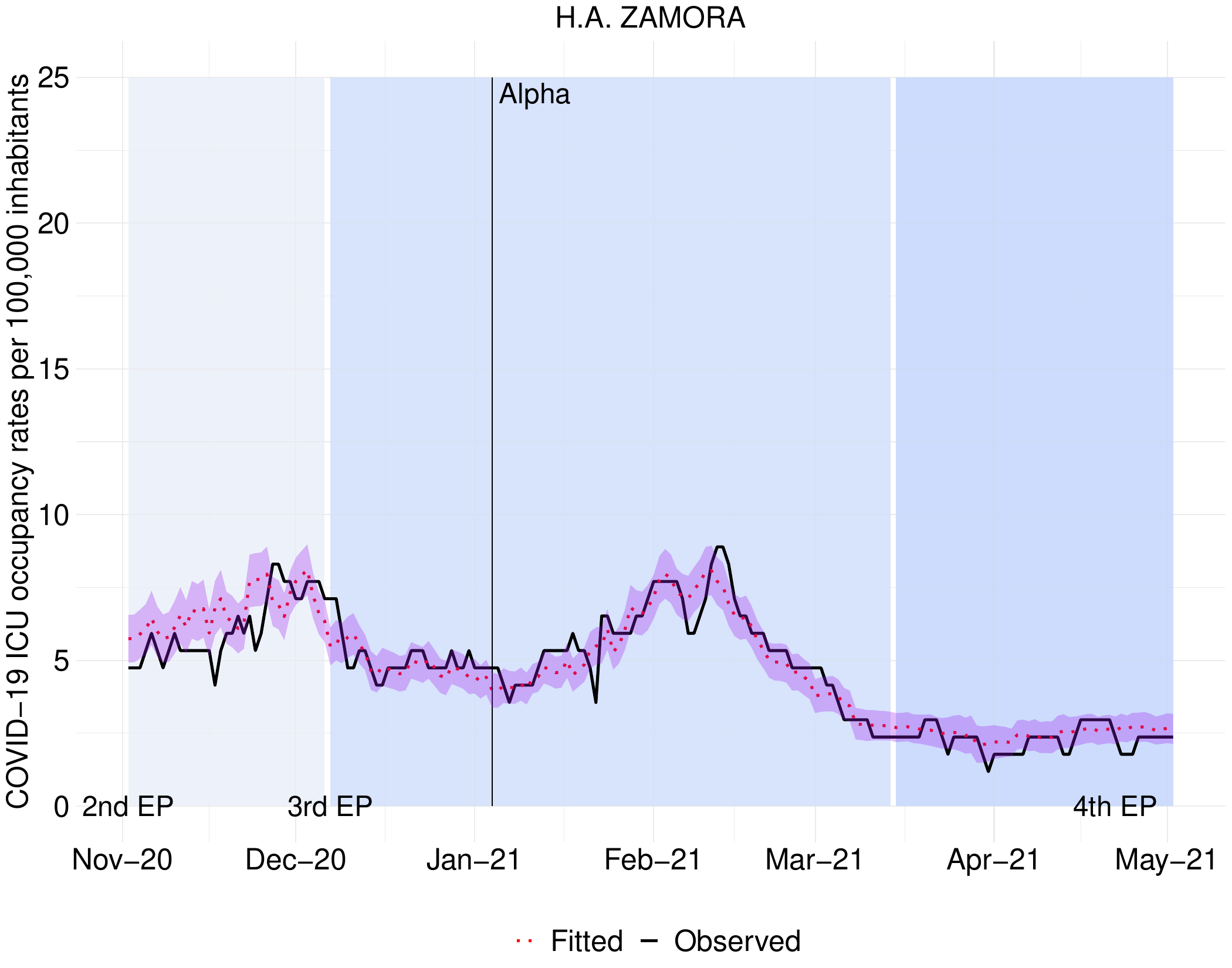}
  \end{minipage}
  \caption{Observed vs estimated ICU occupancy rate per 100,000 inhabitants.}
\label{fig:main.app.real.data.CL.FIT}
\end{figure}

Ultimately, given that the selected model meets all the key criteria, it is applied to obtain the adjustment of ICU beds occupied by COVID-19 patients. Figure \ref{fig:main.app.real.data.CL.FIT} plots the observed and adjusted values of ICU occupancy rates by COVID-19 per 100,000 inhabitants, with a confidence margin resulting from adding and subtracting the root mean squared error (RMSE) obtained by a parametric bootstrap estimator with 1000 bootstrap iterations. It should be specified that the results are displayed for the areas described in Section \ref{main.dataset}, with the highest and lowest mean proportion of COVID-19 patients in ICU, which at the same time are the health areas with the lowest RMSE and highest RMSE respectively. Additionally, all results can be seen in the Figure \ref{fig:app.further.ap.data.FIT.I} in the Subsection \ref{app.further.ap.data.FIT} of Appendix \ref{app.further.ap.data}.

It can be seen that in the H.A. Zamora the ICU occupancy ratio is notably lower than in the health area of H.A. Valladolid Oeste. In fact, if we look at the graph of the evolution of ICU occupancy in this period (Figure \ref{fig:main.dataset.ta.ICUHA}), the H.A. Zamora is one of the health areas with the fewest recorded cases. It is consistent to reason that, given a smaller number of available observations, there may be greater uncertainty in the estimate, and therefore the margin of error is greater. In fact, the following health areas with the highest percentage of error are the H.A. Ávila, the H.A. El Bierzo and the H.A. Segovia, which are those with the lowest proportion of ICU admissions.
Furthermore, it is also seen that the width of the error band increases in periods of particularly low occupancy, both in the H.A. Zamora and in the H.A. Valladolid Oeste. This behaviour is also detected in stages when there are sharp rises and falls in occupancy.

However, estimating the error in relative terms, as the ratio between the RMSE and the proportion in ICU estimated by the model, the maximum error obtained is around 30 \%, with an average error of around 10 \%, and a 75 \% between 6.626 \% and 12.308 \%. These results are really positive to support the capacity of the ARRCP model to explain the evolution of COVID-19 by H.A. in Castilla y León during this period.

Given the good results obtained in the adjustment, we proceed to evaluate the forecasting capacity. For this purpose, we adopt the rolling origin forecast technique, for forecast horizons of 1 to 7 days. It was initially described by \cite{Armstrong1972} in econometric studies. It consists of fitting the model with a set of data, and establishing the forecast origin on the first day after the model is fitted. The prediction is evaluated at different time horizons, and when the measurement is obtained, the process is restarted, moving the prediction origin one day, and so on. In this way, each displacement of the prediction origin leads to a readjustment of the model, which incorporates one day to the dataset, and therefore to an update in the prediction equation.  In addition, it should be specified that in order to carry out the forward prediction it is necessary to impute the random effects at the intercept level and the auxiliary variables. In order to make it generalisable at user level and to adapt to the changing dynamics and nature of the data, it has been carried out with the function {\it ets} of the {\it forecast} package \citep{Hyndman2008}.

\begin{figure}[h!]
  \begin{center}
    \includegraphics[scale=0.4]{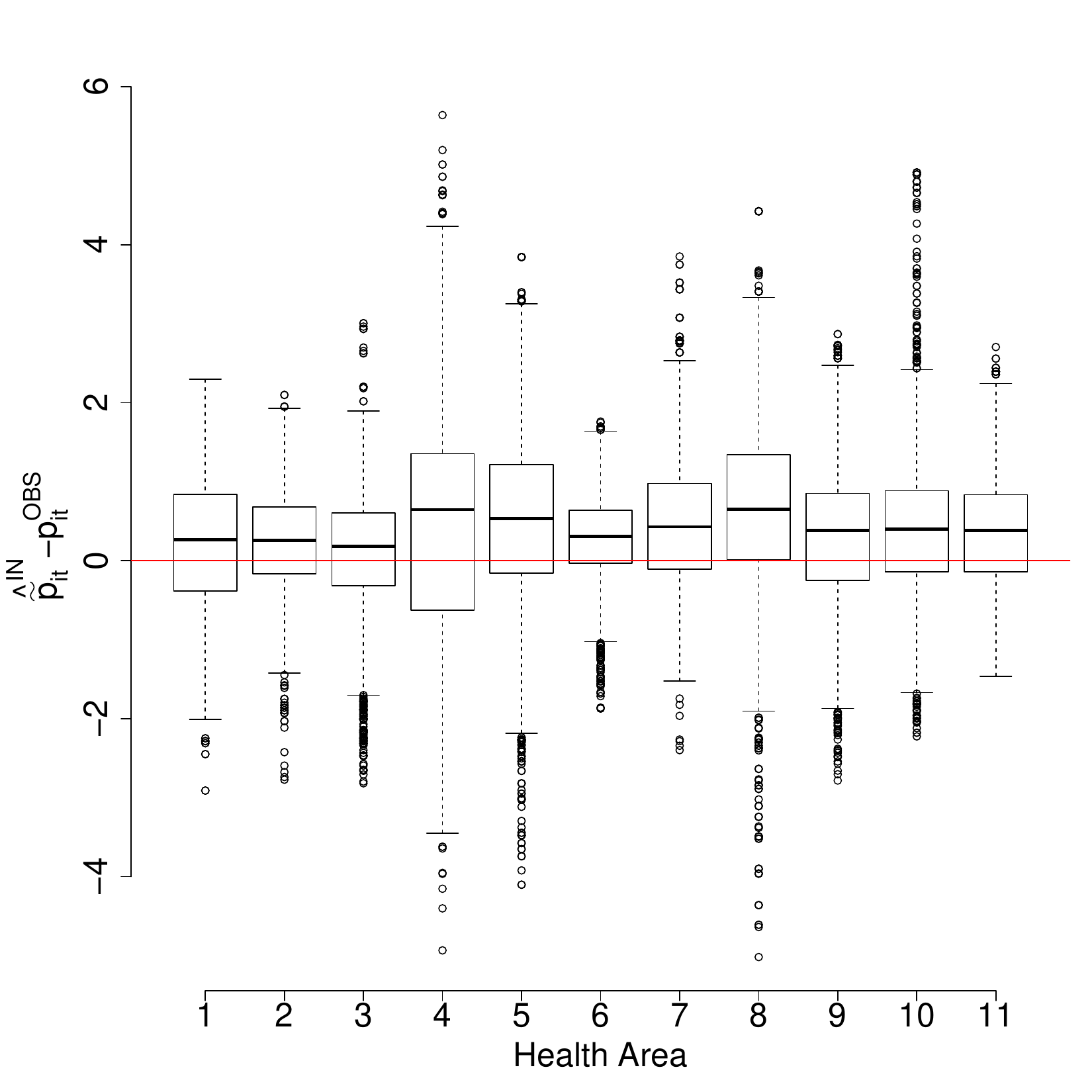}

  \caption{Differences between predicted and recorded ICU occupancy values by H.A.  For clarification purposes, consider the following equivalence: 1$=$H.A. Ávila, 2$=$H.A. Burgos, 3$=$H.A. León, 4$=$H.A. El Bierzo, 5$=$H.A. Palencia, 6$=$H.A. Salamanca, 7$=$H.A. Segovia, 8$=$H.A. Soria, 9$=$H.A. Valladolid Oeste, 10$=$H.A. Valladolid Este, 11$=$H.A. Zamora.}
\label{fig:main.app.real.data.CL.DIF}
  \end{center}
\end{figure}

Firstly, in Figure \ref{fig:main.app.real.data.CL.DIF}, the quality of the prediction is evaluated by analysing the difference between the predicted and observed values for each of the health areas.

The tendency is to predict more cases than actually exist, which in a context of epidemiological control is preferable when it comes to managing the possible health resources needed. Moreover, since the proportions are measured per 100,000 inhabitants, a margin of error between -4 and 6 cases per 100,000 inhabitants seems acceptable in this context, taking into account that the time range of the study is very wide. Thus, the measure of precision is influenced by all the changes in dynamics, non-pharmacological intervention measures such as local confinements and pharmacological measures such as vaccination coverage, and the evolution of virus variants themselves. The health areas of Burgos, León, Salamanca, Zamora and Ávila, which, on the other hand, was the area with the worst performance in the estimation of ICU occupancy, are particularly noteworthy for their good performance.

\begin{figure}[h!]
  \begin{minipage}{0.50\textwidth}
    \centering
    \includegraphics[width=\linewidth]{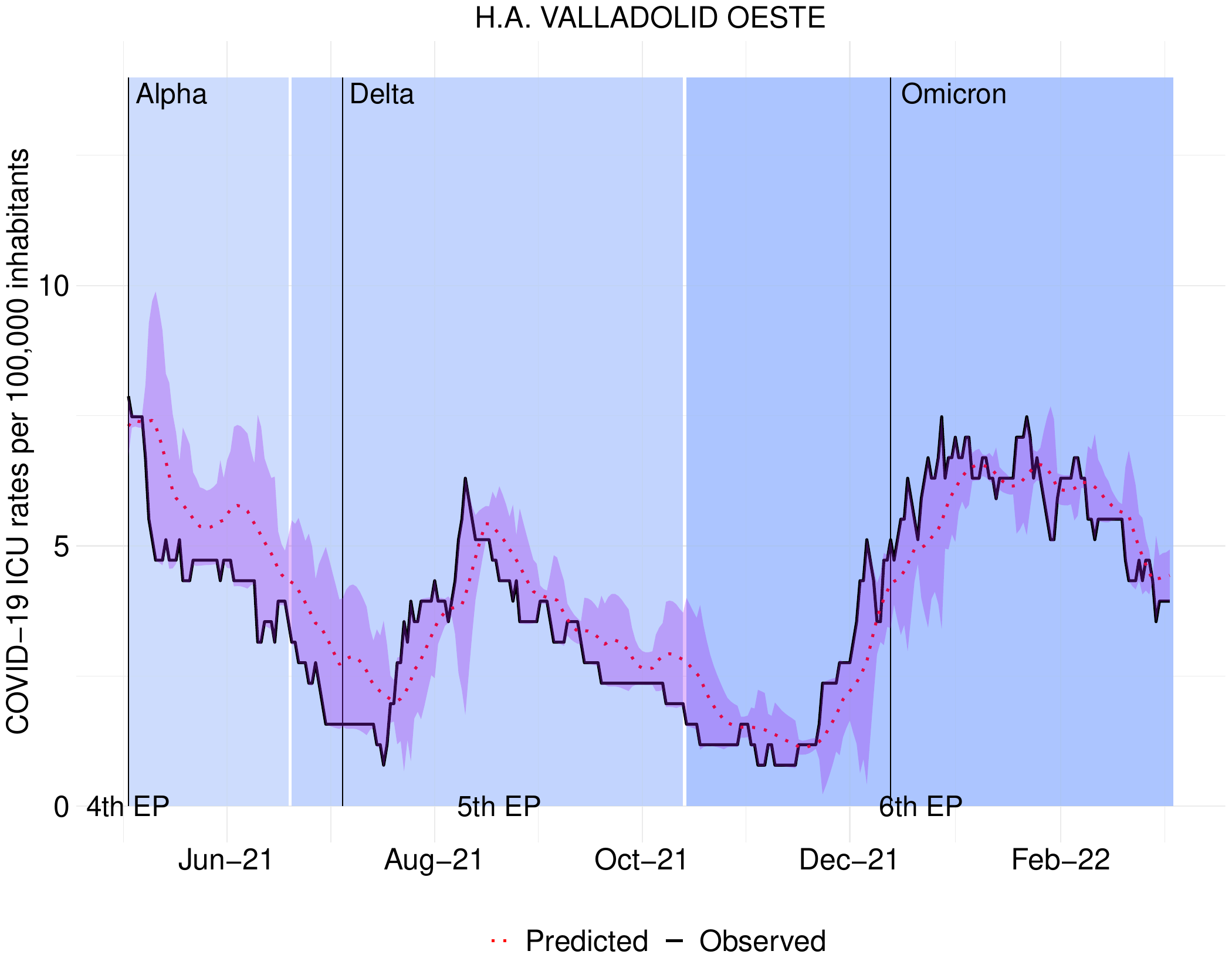}
  \end{minipage}%
  \begin{minipage}{0.50\textwidth}
    \centering
    \includegraphics[width=\linewidth]{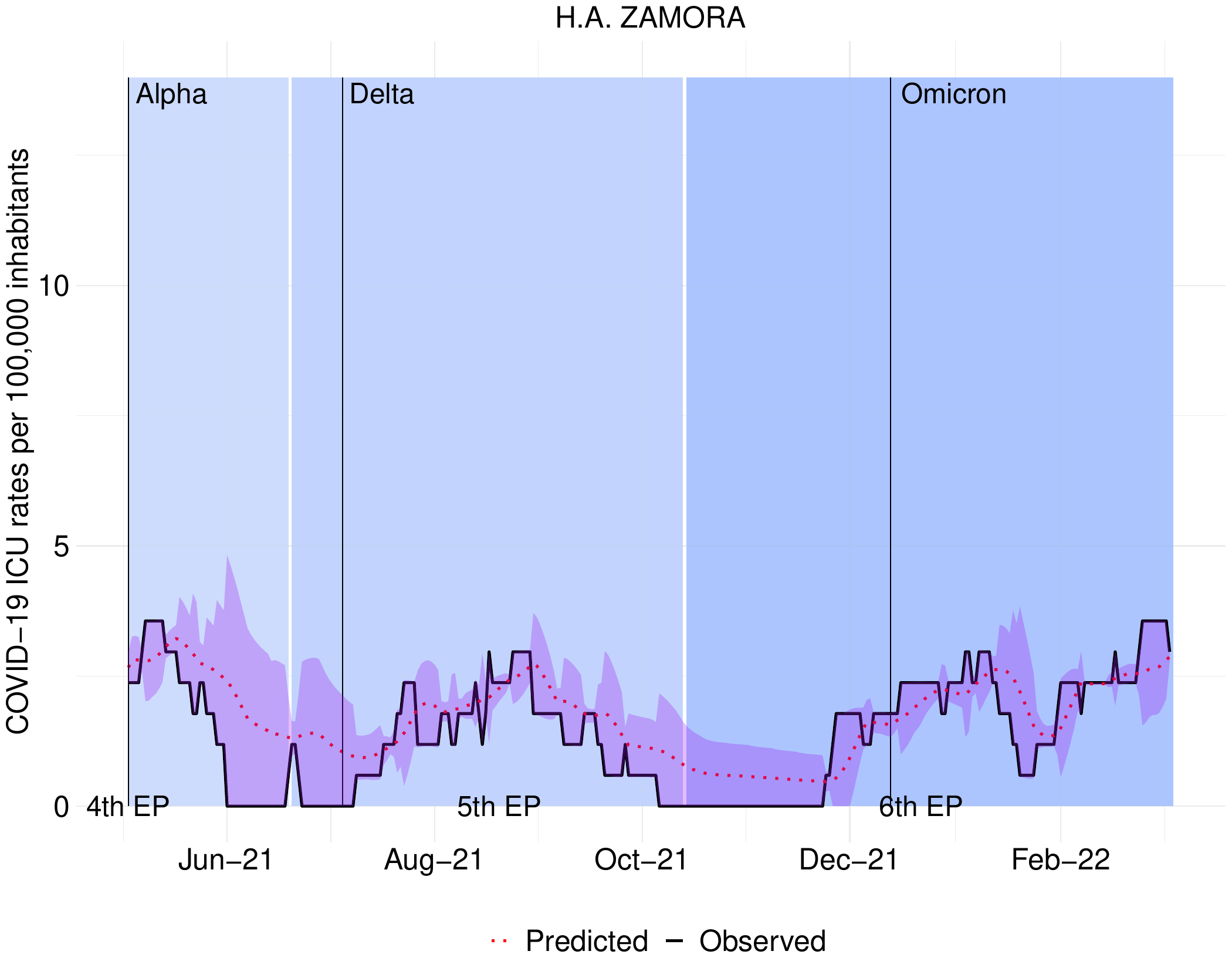}
  \end{minipage}
  \caption{Observed vs predicted ICU occupancy rate per 100,000 inhabitants.}
\label{fig:main.app.real.data.CL.PREDREAL}
\end{figure}

In order to deepen the analysis, Figure \ref{fig:main.app.real.data.CL.PREDREAL} shows the average predictions for each date obtained with the rolling forecast technique versus the value actually recorded, adding a margin of confidence resulting from the addition and subtraction to the average prediction of the RMSE between the predictions and the actual observations.
It is worth mentioning that on this occasion, the H.A. with the highest average error was the H.A. El Bierzo, while the one with the lowest error was the H.A. Salamanca.

In the Figure \ref{fig:main.app.real.data.CL.PREDREAL} the H.A. Zamora and the H.A. Valladolid Oeste are represented for comparative purposes with the adjustment capacity discussed in Figure \ref{fig:main.app.real.data.CL.FIT}, but the results for more health areas are fully available in the Subsection \ref{app.further.ap.data.PRED} of the Appendix \ref{app.further.ap.data}.

By definition, the margins of prediction confidence are wider than the margins of error contemplated in the analysis of the adjustment capacity, since in forward forecasting the inherent uncertainty is greater. Thus, in addition to the fact that the true auxiliary variables are not available, but rather an imputation of their possible future value, random effects at the intercept level must also be imputed.

However, the margin of error moves within acceptable ranges, with higher amplitudes being obtained in the period of lower pressure, as well as at times of sudden ups and downs. One possible explanation may be due to the fact that when working with the average of the past values of the auxiliary variables, when this type of peaks occur, the imputations are not fully reliable to the future behaviour, making the prediction of the target variable difficult. It is also worth highlighting the greater amplitude of the ranges in the H.A. Zamora, with a lower proportion of ICU cases.

It should be noted that, although these results provide key information to evaluate the correct performance of the prediction technique, in real time it would not be possible to provide this margin of confidence based on the RMSE, since we do not know the true values that will occur in our target prediction horizon.

Therefore, based on a RMSE estimator algorithm, these error margins are evaluated with 700 bootstrap iterations, in three time horizons, three days ahead, five days ahead and seven days ahead.

\begin{figure}[h!]
  \begin{minipage}{0.50\textwidth}
    \centering
    \includegraphics[width=\linewidth]{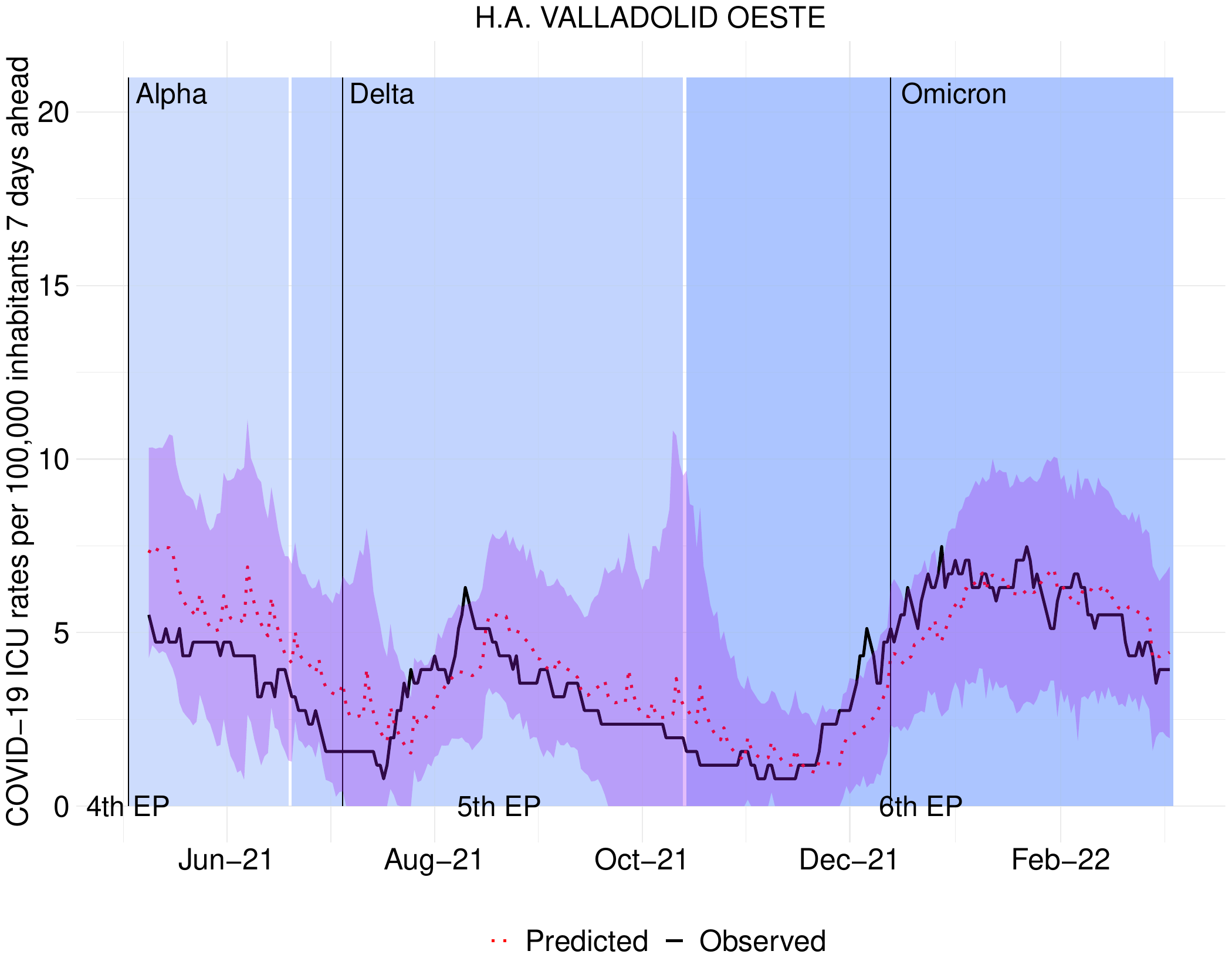}
  \end{minipage}%
  \begin{minipage}{0.50\textwidth}
    \centering
    \includegraphics[width=\linewidth]{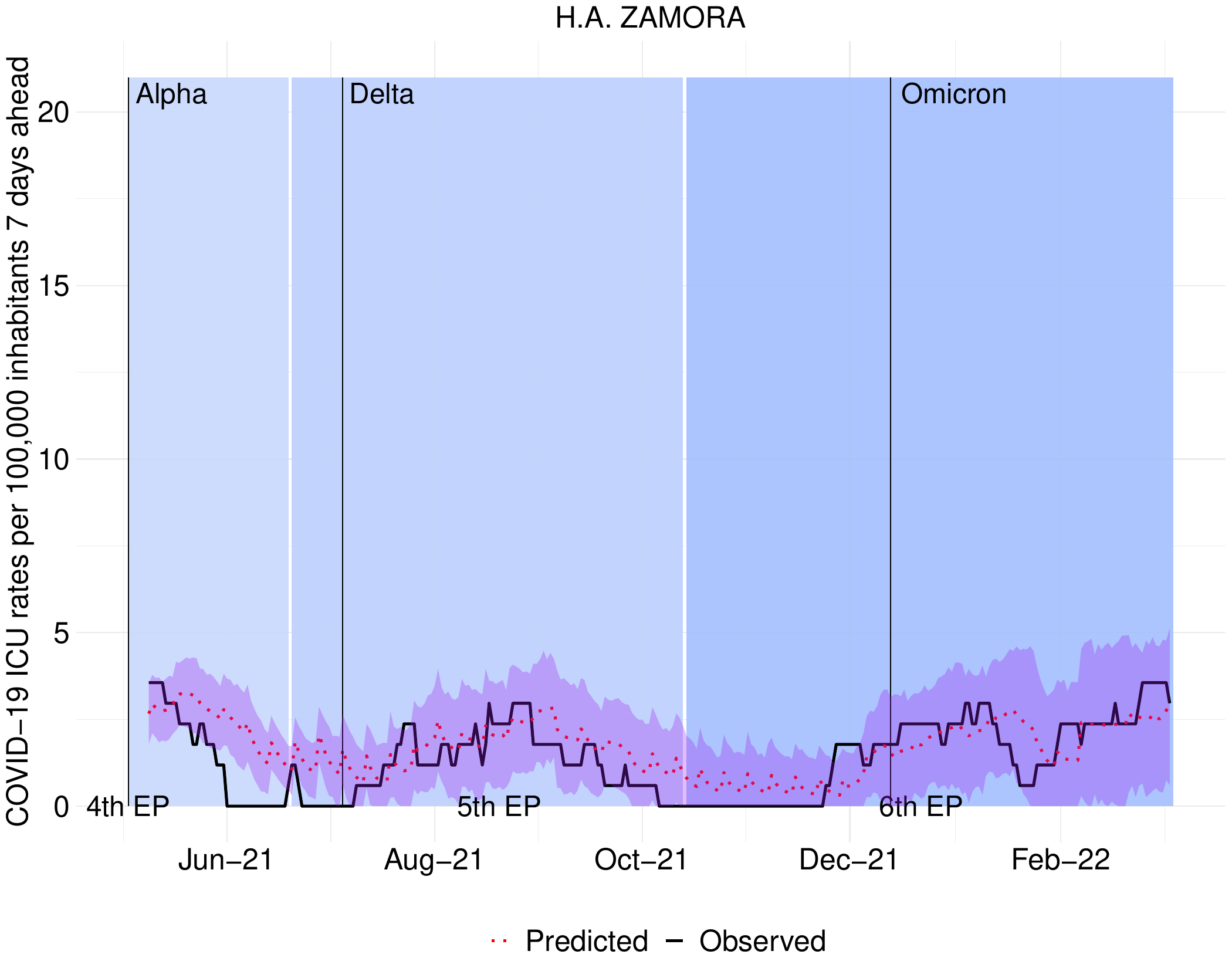}
  \end{minipage}
  \caption{Observed vs predicted ICU occupancy rate seven days ahead per 100,000 inhabitants.}
\label{fig:main.app.real.data.CL.PREDSIMH7}
\end{figure}

Figure \ref{fig:main.app.real.data.CL.PREDSIMH7} shows the worst-case forecast horizon, seven days ahead, for the Zamora and Valladolid Oeste health areas. The full results for each time horizon, H.A. and day are shown in Subsection \ref{app.further.ap.data.PREDSIM} of Appendix \ref{app.further.ap.data}.

For both health areas, the simulated error margin is wider than the actual error margin, which in this context is preferable in order not to underestimate the required health resources. In both areas, but especially in the case of Valladolid Oeste, between the fifth and sixth epidemic period, at the beginning of the prediction, generally wider bands are obtained. One possible explanation is that at this stage there are continuous peaks of descent and ascent over time, and when the imputation of the auxiliary variables is carried out, this information may be attenuated, with the corresponding attenuation in the prediction of the target variable. Likewise, in the H.A. Valladolid Oeste there is a higher degree of predicted error in the transition between the last two periods under study. Thus, for example, with the dominance of Omicron in the sixth epidemic period, the error range is again widened too. As previously mentioned, these turning points are usually marked by changes in the dynamics of the disease or intervention that can further increase uncertainty.

\subsection{ARRCP model for risk assessment}\label{main.app.risk}

Given the high performance obtained in the estimation of ICU occupancy due to COVID-19, the ARRCP model is a clear candidate to help understand the evolution of the healthcare burden by H.A. during the adjustment period. Likewise, the results obtained in prediction are equally promising. Therefore, we consider the generation of tools such as risk maps as potentially useful tools to help in the management of resources or the implementation or relaxation of non-pharmacological intervention measures.

The risk maps consist of representing each H.A. in a green to maroon colouring, according to the lowest to highest risk category assigned to it by the proportion of people with COVID-19 admitted to ICUs per 100,000 inhabitants following the guidelines published by the \cite{MinisterioSanidad2022c}.

\begin{figure}[h!]
  \begin{minipage}{0.50\textwidth}
    \centering
    \includegraphics[width=\linewidth]{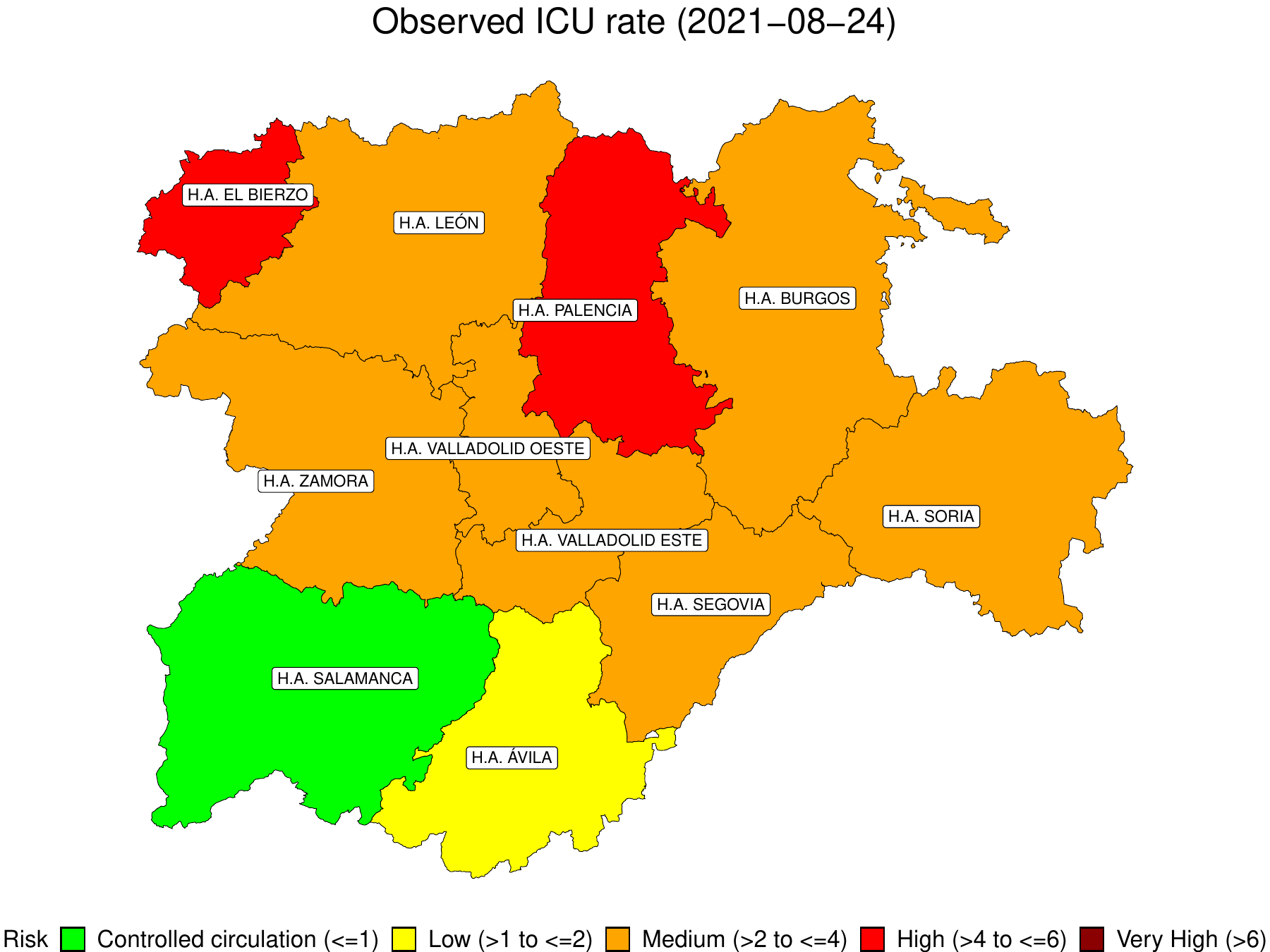}
  \end{minipage}%
  \begin{minipage}{0.50\textwidth}
    \centering
    \includegraphics[width=\linewidth]{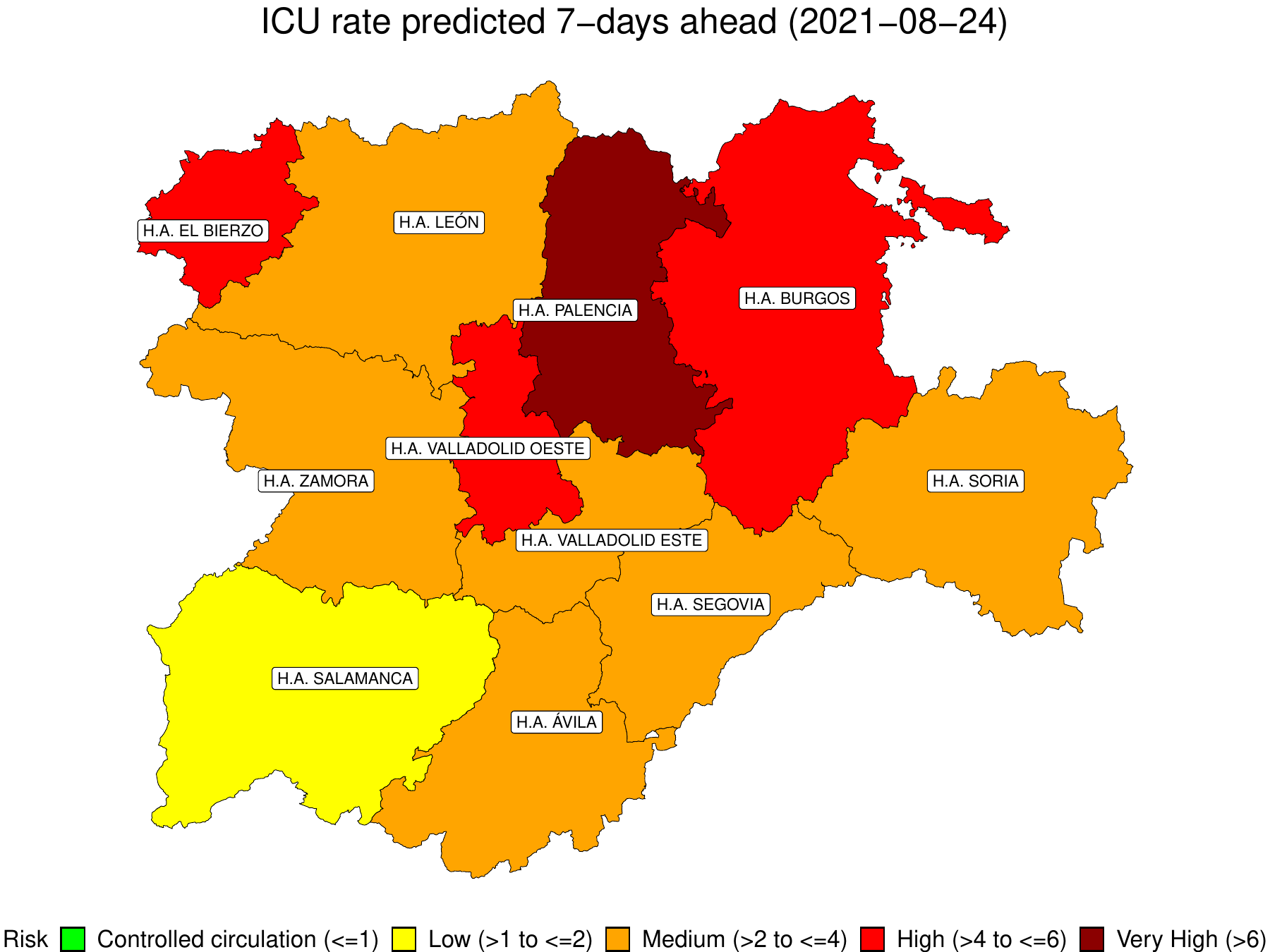}
  \end{minipage}
  \begin{minipage}{0.50\textwidth}
    \centering
    \includegraphics[width=\linewidth]{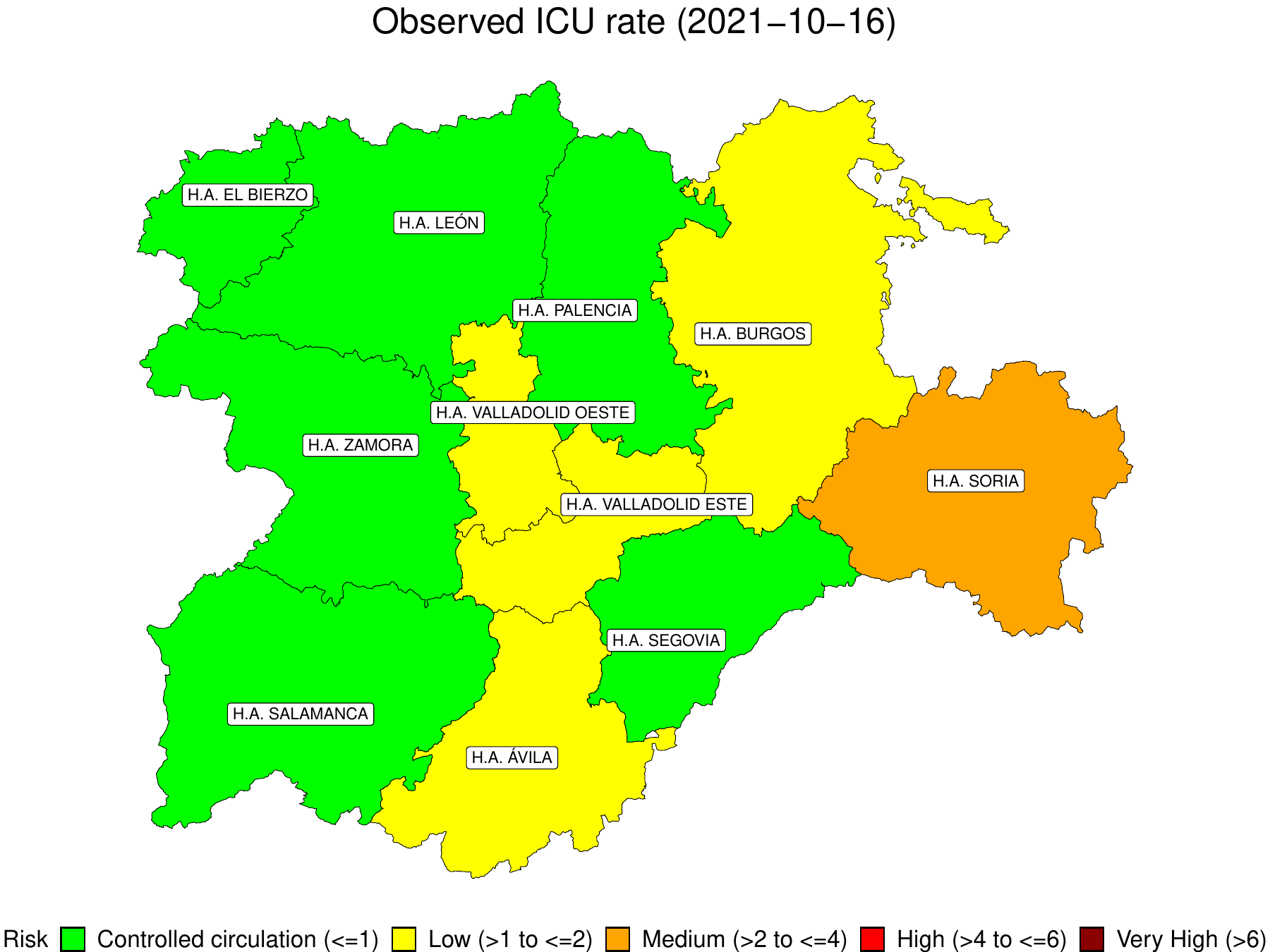}
  \end{minipage}%
  \begin{minipage}{0.50\textwidth}
    \centering
    \includegraphics[width=\linewidth]{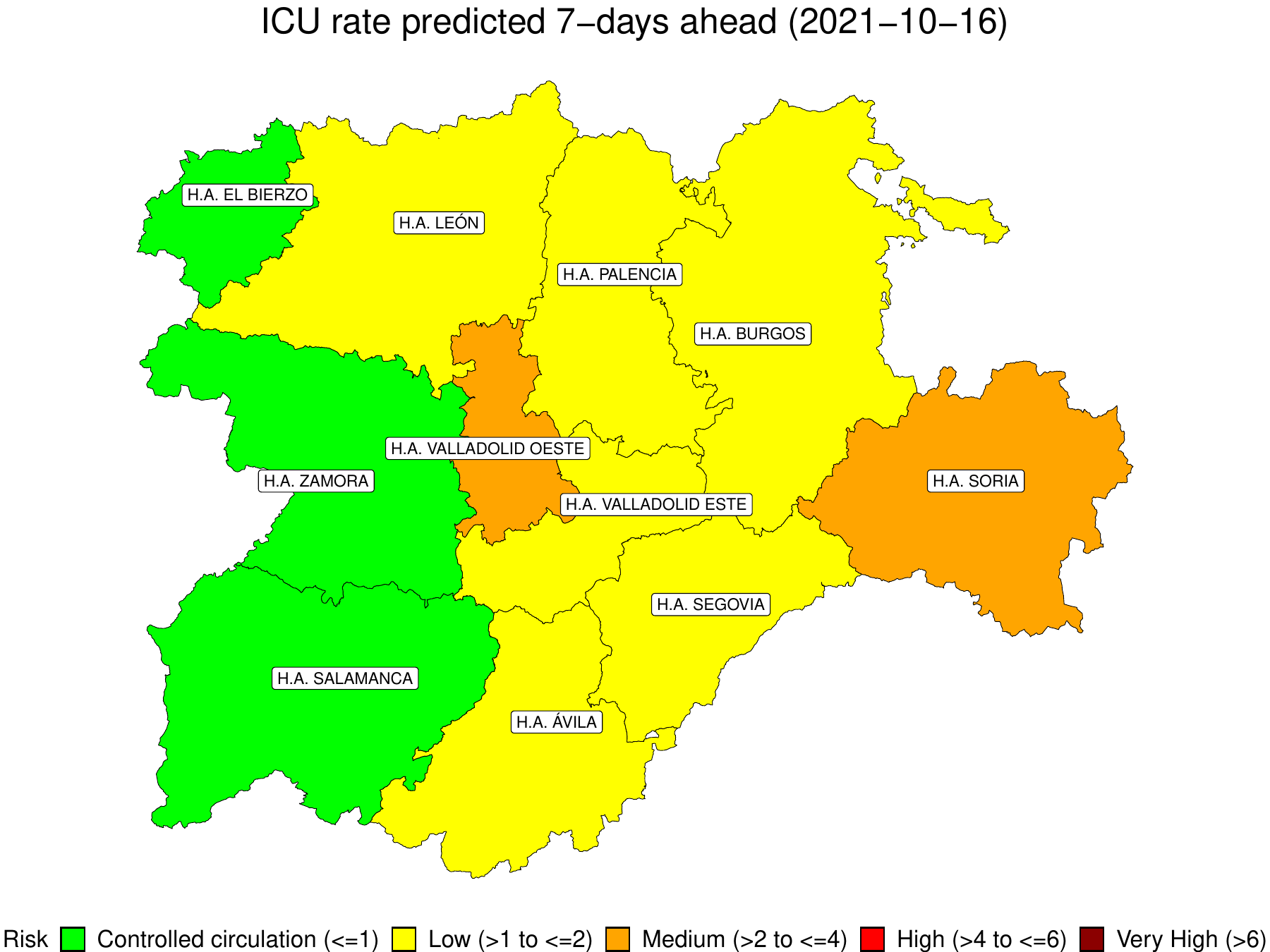}
  \end{minipage}
  \caption{Maps of risk of collapse in ICUs in Castilla y León.}
\label{fig:main.app.real.data.ARRCP.RISK.CL}
\end{figure}

Figure \ref{fig:main.app.real.data.ARRCP.RISK.CL} shows maps plotting the recorded and predicted ICU occupancy observations for two dates, 24-08-2021 and 16-10-2021. The 24-08-2021 falls within the fifth epidemic period, after the dominance of the Delta variant, coinciding with an upturn in the disease. On the other hand, 16-10-2021 falls three days after the transition from the fifth to the sixth epidemic period, in a period of low pressure. Thus, we intend to represent the behaviour of the predictors of the occupancy rate in two different scenarios, medium-high hospital saturation and no hospital saturation.

In Castilla y León, on 24-08-2021 a heterogeneous level of risk is recorded, with four discrepancies between the observed and the predicted seven days ahead. Specifically, the risk is overestimated by one level for H.A. Valladolid Oeste, H.A. Palencia, H.A. Salamanca and H.A. Ávila. This error results in categorising an area of medium risk as high risk, from high risk to very high risk and from controlled circulation to low risk. Therefore, for a prediction margin of 7 days, this error seems acceptable and preferable to underestimating the resources that would be necessary. For 16-10-2021, the seven-day observations and predictions coincide for a controlled risk.

Thus, in general, the large provincial cities have been the most affected, with H.A. Valladolid Este, H.A. Valladolid Oeste, H.A. Burgos, H.A. Palencia and H.A. León standing out, with the least severity located in areas such as H.A. Zamora, H.A. Salamanca and H.A. Ávila. 

While understanding the differences that caused the opposite impact in terms of severity in these areas is complex, as COVID-19 is a multifactorial disease, some studies provide initial explanations for this observed pattern. Thus, the study carried out by the \cite{CNE2022} on the main factors that influenced the incidence during the first wave highlights the main impact in Castilla y León was the ratio of residences for people over 70 years of age and, with a moderate-low influence, external mobility to areas such as the País Vasco and Comunidad de Madrid, bordering some of the most affected healthcare areas in this community. Due to the high ageing of the population, and the proximity of the large urban areas to these communities, the greatest impact could be coherent. 

In sum, in addition to their great usefulness for the preparation of resources needed in the future, risk maps are also very useful for understanding how the pandemic has evolved, studying possible past patterns and evaluating future patterns.

\section{Conclusions}\label{main.conc}

In this paper, we apply, for the first time, a Poisson-type area model with random regression coefficients to model the daily overload in ICUs as a consequence of COVID-19 disease in each of the eleven health areas of Castilla y León, assessing both modelling and forecast capacity.

In terms of accuracy in modelling the occupancy of ICUs, the predictors show excellent performance, both in simulations and in application to real data. Thus, in the period between 2nd November 2020 and 2nd May 2021 there is a heterogeneity in the impact of COVID-19 depending on each area, with the H.A. Soria and H.A. Valladollid Oeste, in Castilla y León, standing out for their higher rates. Furthermore, although the precision of the estimates is better in scenarios of high care pressure, it is worth noting the good performance in terms of relative RMSE in all types of scenarios.

Regarding forecasting the future, we started the evaluation of the performance of the predictors in the elaboration of future predictions in time horizons of 3, 5 and 7 days, for each of the health areas in the period between 3rd May 2021 and 6th March 2022. To our current knowledge, this is the first time that is maked for GLMMs with random slopes and random effects imputation. The results, in terms of RMSE obtained by parametric bootstrap, are encouraging in both low and high care pressure scenarios.
Thus, in the future, even in the most unfavourable prediction horizons, the actual margins of error show an optimal behaviour, with a better performance in times of high pressure and an overestimation of occupancy, preferable in epidemic contexts.  Therefore, we are currently developing a Shiny application to obtain these predictions at user level, e.g. by providing risk maps, to facilitate resource planning.

\newpage

\setcounter{page}{1}

\begin{center}
{\huge Supplementary Material}
\vspace{2mm}

\title{An statistical analysis of COVID-19 intensive care unit bed occupancy data\footnote{This research is part of the grant PID2020-113578RB-I00, funded by MCIN/AEI/10.13039/501100011033/. It has also been supported by the Spanish grant PID2022-136878NB-I00, the Valencian grant Prometeo/2021/063, by the Xunta de Galicia (Grupos de Referencia Competitiva ED431C-2020/14) and by CITIC that is supported by Xunta de Galicia, convenio de colaboraci\'{o}n entre la Conseller\'{i}a de Cultura, Educaci\'{o}n, Formaci\'{o}n Profesional e Universidades y las universidades gallegas para el refuerzo de los centros de investigaci\'{o}n del Sistema Universitario de Galicia (CIGUS). The first author was also sponsored by the Spanish Grant for Predoctoral Research Trainees RD 103/2019 being this work part of grant PRE2021-100857, funded by MCIN/AEI/10.13039/501100011033/ and ESF+. In addition, we thank the Centro de Supercomputaci\'{o}n de Galicia (CESGA) for providing their services for part of the simulations in this work.}}
\\
\vspace{1mm}
\author{Naomi Diz-Rosales$^{1}$, Mar\'{i}a Jos\'{e} Lombard\'{i}a$^{2}$, Domingo Morales$^{3}$
\vspace{0.1 cm}\\
{\small  $^{1}$naomi.diz.rosales@udc.es}\\
{\small $^{2}$maria.jose.lombardia@udc.es}\\
{\small $^{3}$d.morales@umh.es}\\
{\small  $^{1, 2}$Universidade da Coru\~{n}a, CITIC, Spain.}\\
{\small $^{3}$Universidad Miguel Hern\'{a}ndez de Elche, IUICIO, Spain.}\\
{\small 28-04-2024}}
\end{center}
\vspace{3mm}

\appendix

\section{Extending descriptive analysis}\label{app.ext.descr.data}

The domains or spatial areas of interest correspond to the health areas of Castilla y León, that has been one of the most affected communities, with the highest case fatality rates \citep{Lopez2021}.

Nevertheless, until now, studies have only been conducted at the level of autonomous communities, such as \cite{CNE2022}, which evaluate the factors that influenced the greater diffusion of COVID-19 in Spain, being clear that the disease has shown heterogeneous behaviour and that further study is needed.

Therefore, in this study, the target spatial domain or area is defined as Health Area (H.A), a key unit for determining the risk of health collapse during the process of implementing virus containment measures.

The H.A. is a demarcation of each autonomous community, originally defined in Article 56 of Law 14/1986, of 25 April, General Health, as a fundamental structure of the health system as is is responsible for the management of the health service centres of the community in its territorial demarcation and for the health services and programmes that exist there \citep{BOE1986}.

Based on this legislation, the Junta de Castilla y León approved, by means of Decree 108/1991 of 9 May, its Health Organisation, currently structuring its health map around 11 health areas \citep{SG20}.

The H.A. acts as a grouping of Primary Care centres and is used for the disaggregation of statistical data on resources, population and activity of the Primary Care Information System (SIAP) \citep{SG20}. For this reason, in the following subsections, the information provided in Section \ref{main.dataset} is expanded upon in order to help increase understanding of the cases under study.

\subsection{ICU occupancy rate due to COVID-19}\label{app.ext.descr.data.ICU}

Starting with the evolution of occupancy in ICU throughout the period under study, Table \ref{tab:app.ext.descr.data.ICU.CL} provides a basic description of the proportion of people with COVID-19 in ICU per 100,000 inhabitants according to the epidemic period.

\begin{table}[ht!]
\caption{Descriptive of COVID-19 ICU occupancy rate in each EP in Castilla y León.}\label{tab:app.ext.descr.data.ICU.CL}
\renewcommand{\arraystretch}{1.2}
\centering
\begin{tabular}{|r|rrrrrr|}
\hline
Period & \multicolumn{1}{r|}{Min.} & \multicolumn{1}{r|}{1st Qu.} & \multicolumn{1}{r|}{Median} & \multicolumn{1}{r|}{Mean} & \multicolumn{1}{r|}{3rd Qu.} & Max. \\ \hline
2nd EP & 2                         & 9                            & 19                          & 19.9714                   & 30                           & 46   \\ \hline
3rd EP & 2                         & 9                            & 16                          & 18.7737                   & 26                           & 56   \\ \hline
4th EP & 0                         & 6                            & 10                          & 10.6204                   & 14                           & 26   \\ \hline
5th EP & 0                         & 2                            & 4                           & 4.9475                    & 7                            & 20   \\ \hline
6th EP & 0                         & 2                            & 4                           & 5.8295                    & 9                            & 28   \\ \hline
\end{tabular}
\end{table}

It can be seen that the second and third epidemic periods are those with higher ratios of ICU hospitalisations due to COVID-19. These periods are between November 2020 and March 2021. Therefore, vaccination coverage was still minimal, as the campaign began between the end of December 2020 and the beginning of January 2021. The maximum peak of hospitalised patients is recorded during the third epidemic period and 75\% of the observations correspond to occupancy levels classified in the highest risk of collapse category. In addition to lower vaccination coverage in relation to the other periods, and the relaxation of non-pharmacological intervention measures, the second and especially the third period included holiday months and months of high social contact, as well as the incursion of the Alpha variant, which increased both incidence and severity.

The values decrease significantly in the following periods, even when in the fifth period the Delta variant occurs, which, although it is associated with an increase in incidence, does not increase in severity. In addition, it should be considered that in this time range, from spring to summer, there is greater activity in open spaces, which makes contagion more difficult. But, above all, vaccination coverage is higher, with 54.22\% of the population of Castilla y León having received at least one dose at the beginning of the fifth period, and 83.36\% at the beginning of the sixth period \citep{MinisterioSanidad2024}. Thus, in the sixth epidemic period there was a slight upturn compared to the fifth in both regions, which could be caused by the incursion of the Omicron variant that caused an implosion of infections, at a time of year characterised by social contact such as Christmas. However, the increase was not particularly high.

\begin{figure}[htbp]
  \begin{minipage}{0.50\textwidth}
    \centering
    \includegraphics[width=\linewidth]{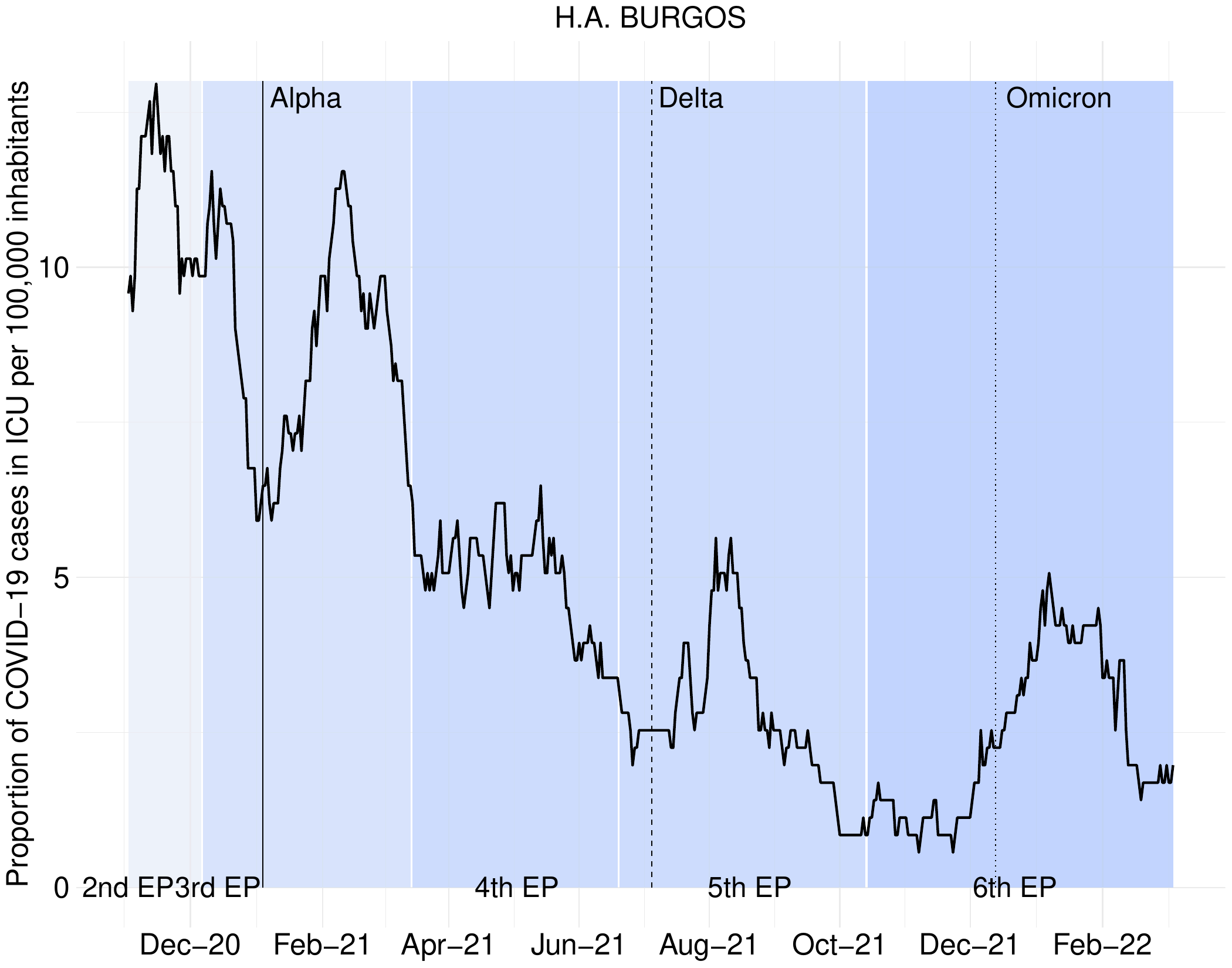}
  \end{minipage}
   \begin{minipage}{0.50\textwidth}
    \centering
    \includegraphics[width=\linewidth]{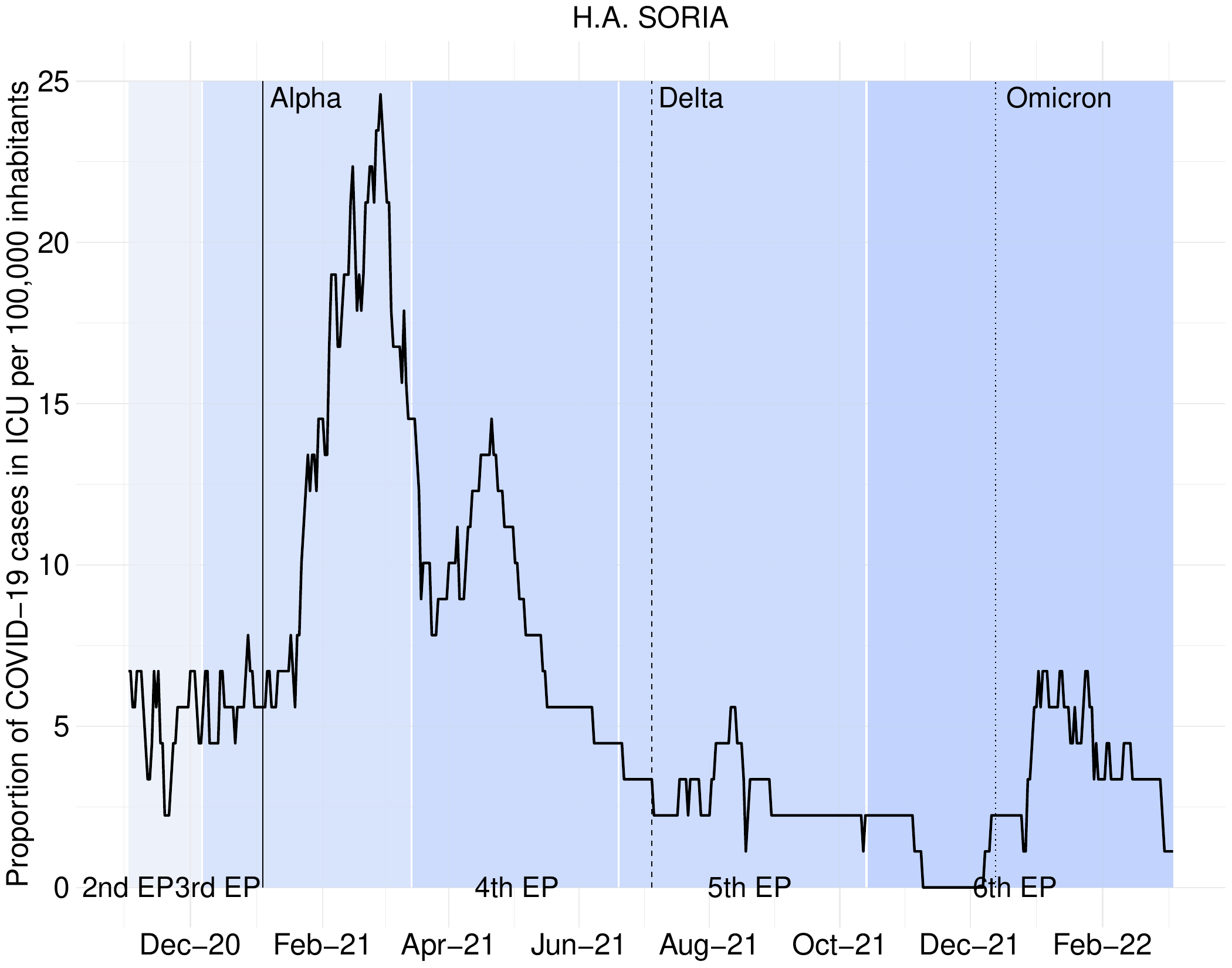}
  \end{minipage}

  \begin{minipage}{0.50\textwidth}
    \centering
    \includegraphics[width=\linewidth]{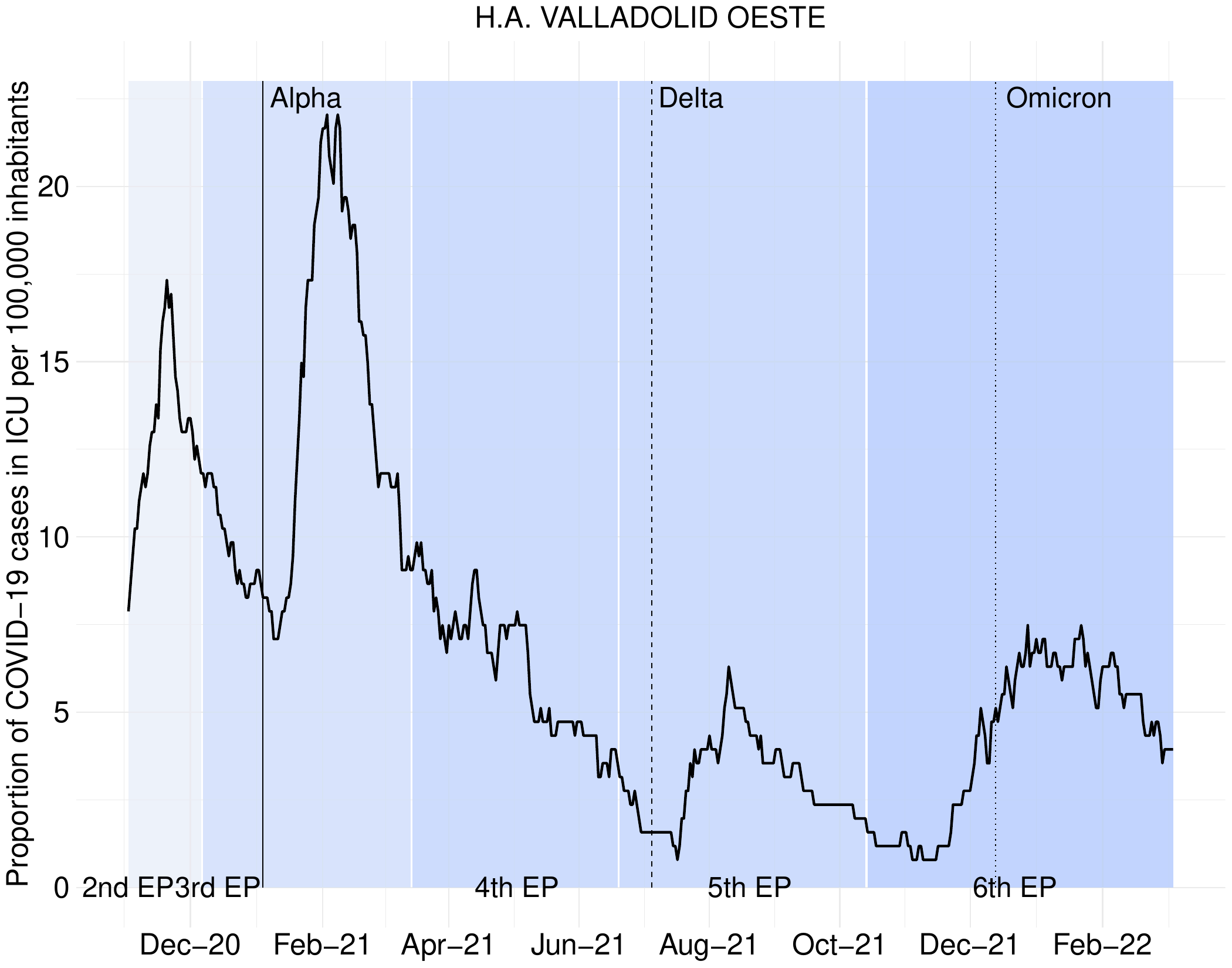}
  \end{minipage}%
  \begin{minipage}{0.50\textwidth}
    \centering
    \includegraphics[width=\linewidth]{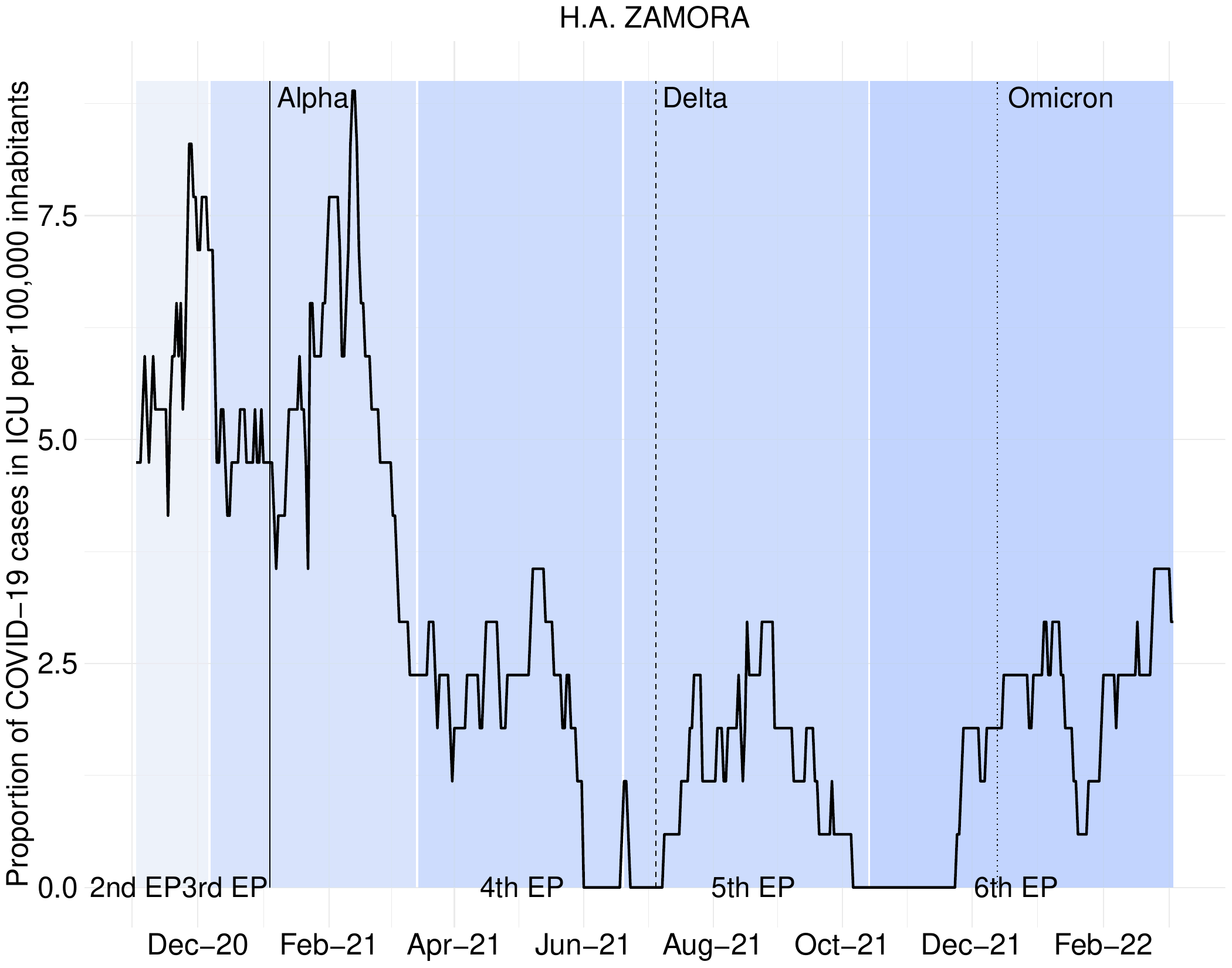}
  \end{minipage}
  \caption{Evolution of the ICU occupancy rate by COVID-19 per 100,000 inhabitants by health area and day in Castilla y León.}
  \label{fig:app.ext.descr.data.ICU.HATIME.I}
\end{figure}

The pattern of ICU admissions due to COVID-19 per 100,000 inhabitants is broken down in Figure \ref{fig:app.ext.descr.data.ICU.HATIME.I} for H.A. and day.

This visualisation allows us to detect both the heterogeneous behaviour between health areas and within the same H.A. as a function of time, with respect to the dynamics of occupancy in ICUs due to COVID-19.
Therefore, it is to be expected that such behaviour will carry over to the auxiliary variables. In the following subsections, each selected variable is analysed one by one.

\newpage
\subsection{Ward occupancy rate due to COVID-19}\label{app.ext.descr.data.WARD}

Beginning with {\it ward.rateL2}, as can be seen in Table \ref{tab:app.ext.descr.data.WARD.CL} the second and third epidemic periods have the highest proportions of ward admissions, with 75\% of the observations in the second period being at maximum risk, and in the third period between medium and maximum risk, according to the indicators of \cite{MinisterioSanidad2022}. As in the case of ICU admissions, from the fourth epidemic period onwards, the rate of hospitalisation in the ward declined significantly, with a slight upturn in the sixth epidemic period, possibly due to the emergence of Omicron and the holiday period.

\begin{table}[ht!]
\caption{Descriptive of the variable \textit{ward.rateL2} per 100,000 inhabitants in each EP in Castilla y León.}\label{tab:app.ext.descr.data.WARD.CL}
\renewcommand{\arraystretch}{1.2}
\centering
\begin{tabular}{|r|rrrrrr|}
\hline
Period & \multicolumn{1}{r|}{Min.} & \multicolumn{1}{r|}{1st Qu.} & \multicolumn{1}{r|}{Median} & \multicolumn{1}{r|}{Mean} & \multicolumn{1}{r|}{3rd Qu.} & Max. \\ \hline
2nd EP & 9.9049                         & 40.6962                            & 52.7695                          & 52.9668                   & 65.2207                           & 96.0480   \\ \hline
3rd EP & 2.6413                         & 18.1116                            & 30.0954                          & 40.0250                   & 58.6510                           & 127.0974   \\ \hline
4th EP & 0.0000                         & 6.3977                            & 9.7489                          & 11.8121                   & 15.7008                           & 45.8316   \\ \hline
5th EP & 0.0000                         & 2.6413                            & 5.1185                           & 8.1857                    & 12.4933                            & 50.7137   \\ \hline
6th EP & 0.0000                        & 4.1504                            & 13.1510                           & 15.4646                    & 24.4182                            & 50.5139   \\ \hline
\end{tabular}
\end{table}

Figure \ref{fig:app.ext.descr.data.WARD.HATIME.I} below shows a breakdown for each H.A. and study day of the occupancy rate with a delay of L2, {\it ward.rateL2}, per 100,000 inhabitants, due to COVID-19.

\begin{figure}[htbp]
  \begin{minipage}{0.50\textwidth}
    \centering
    \includegraphics[width=\linewidth]{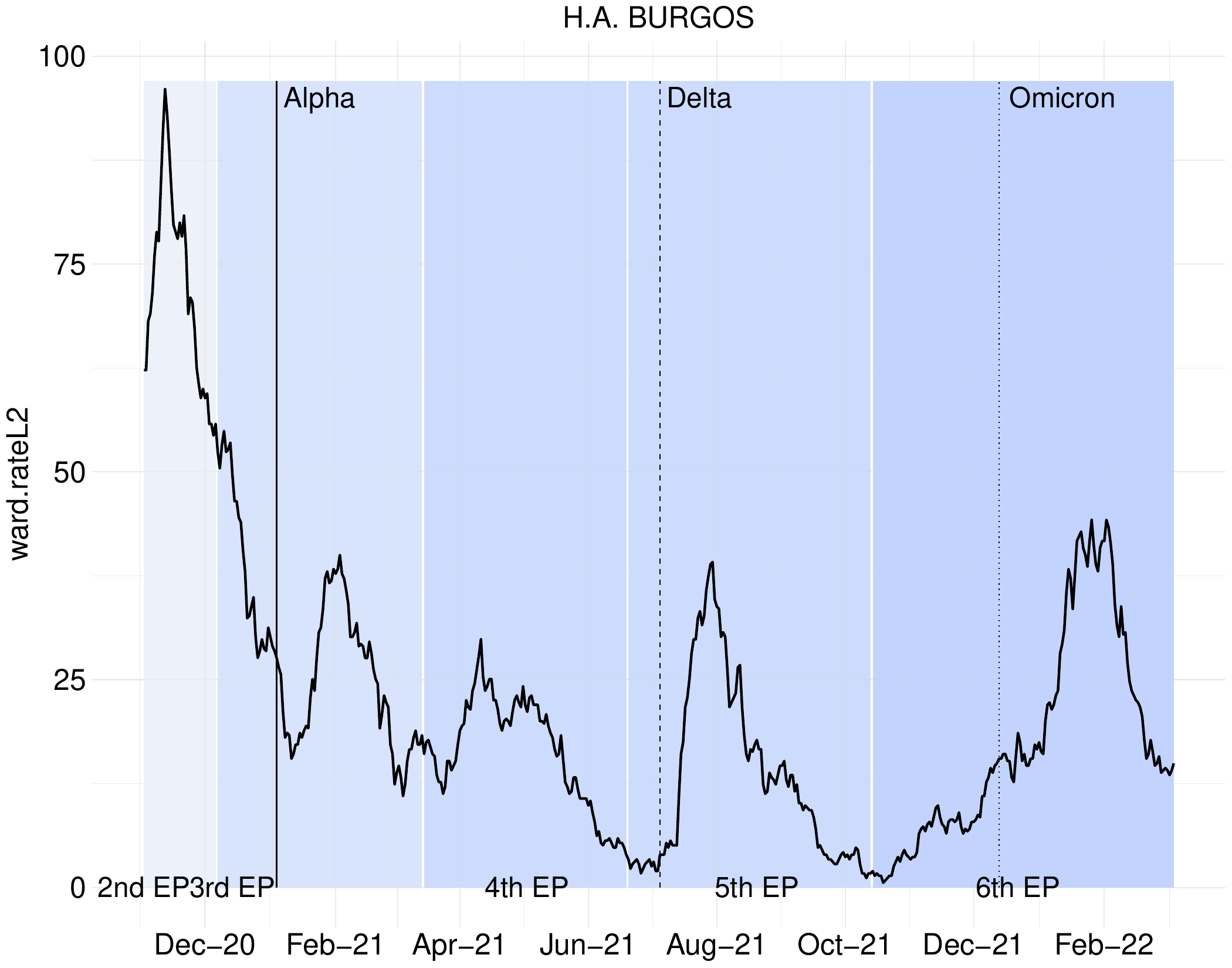}
  \end{minipage}
  \begin{minipage}{0.50\textwidth}
    \centering
    \includegraphics[width=\linewidth]{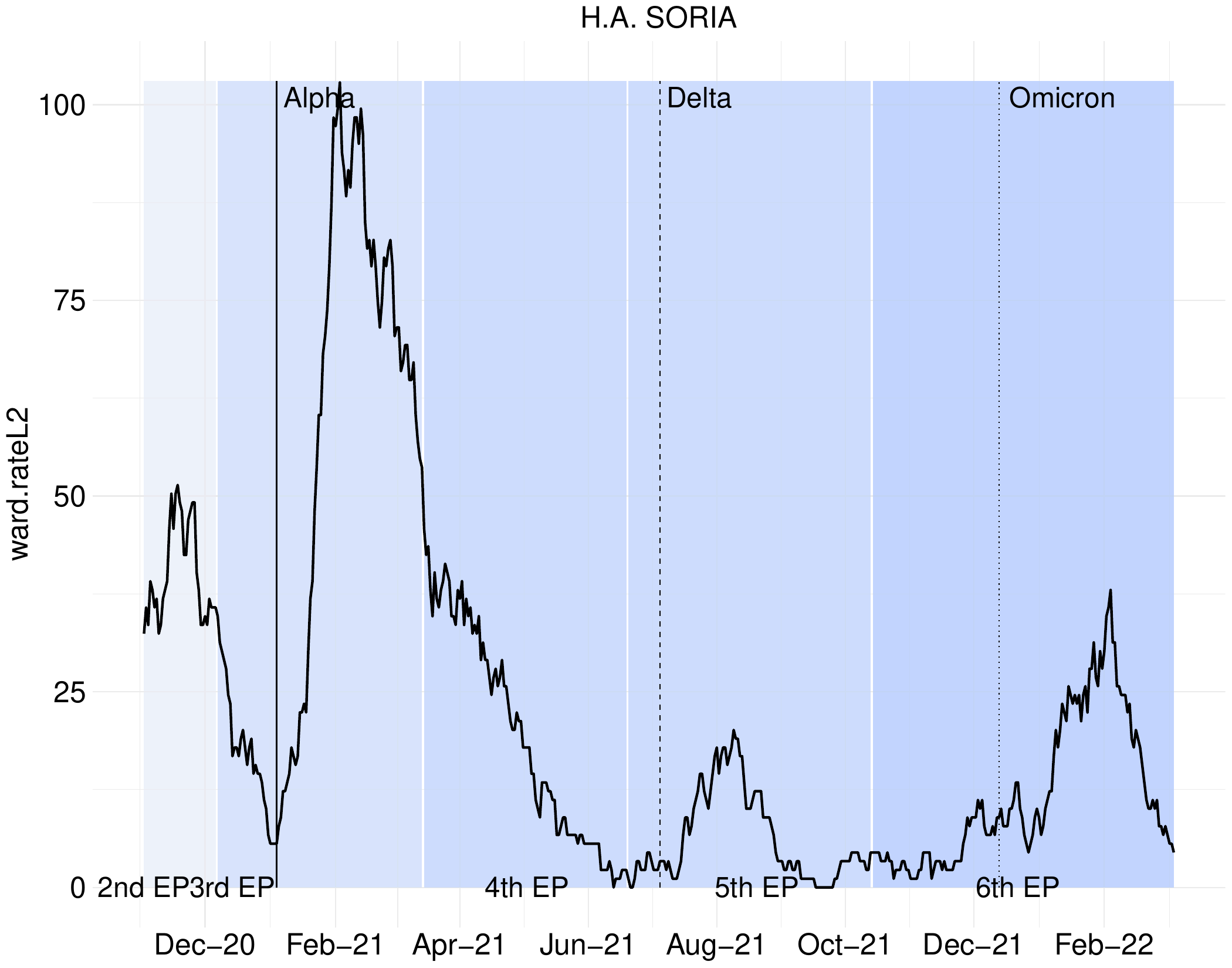}
  \end{minipage}

  \begin{minipage}{0.50\textwidth}
    \centering
    \includegraphics[width=\linewidth]{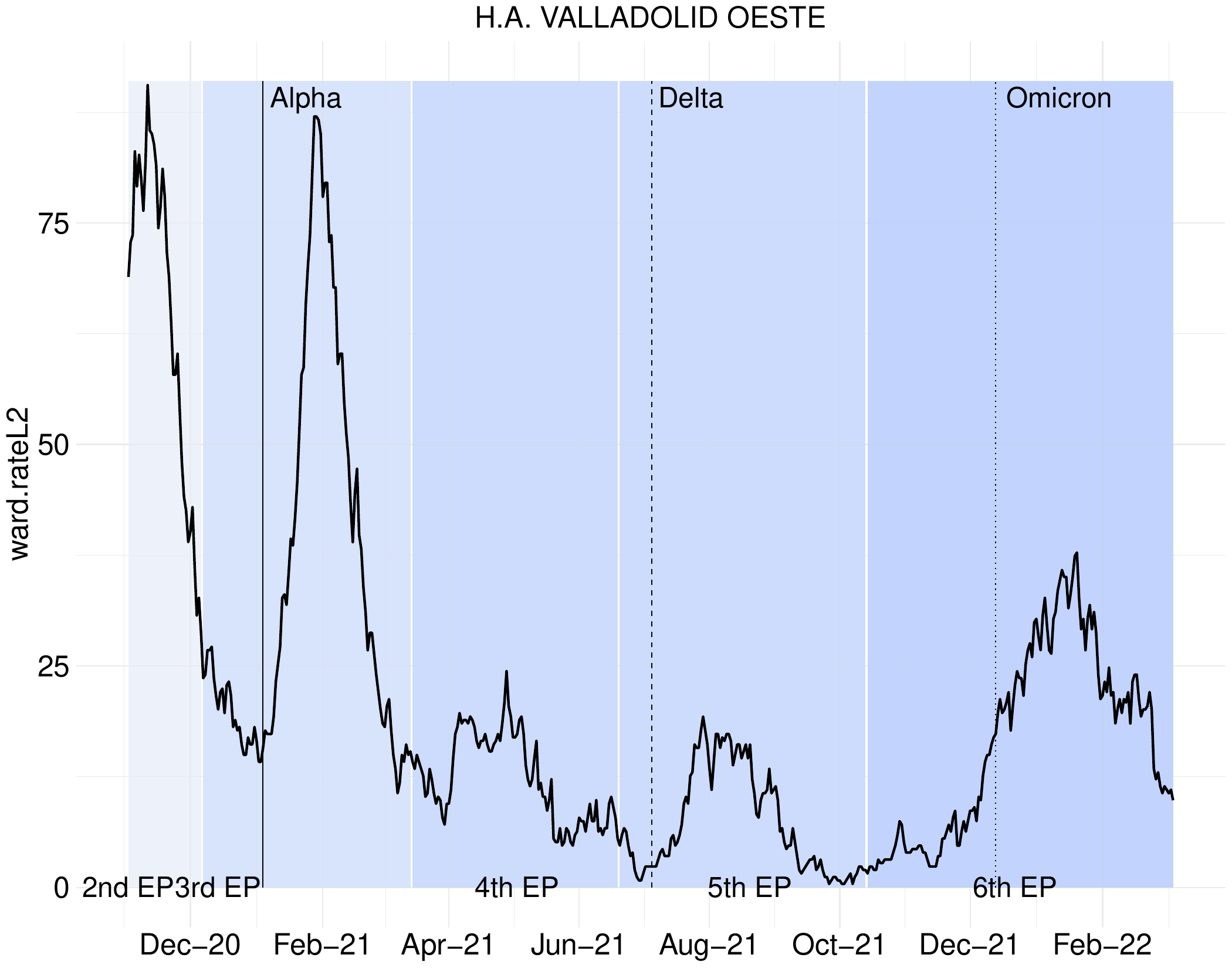}
  \end{minipage}%
  \begin{minipage}{0.50\textwidth}
    \centering
    \includegraphics[width=\linewidth]{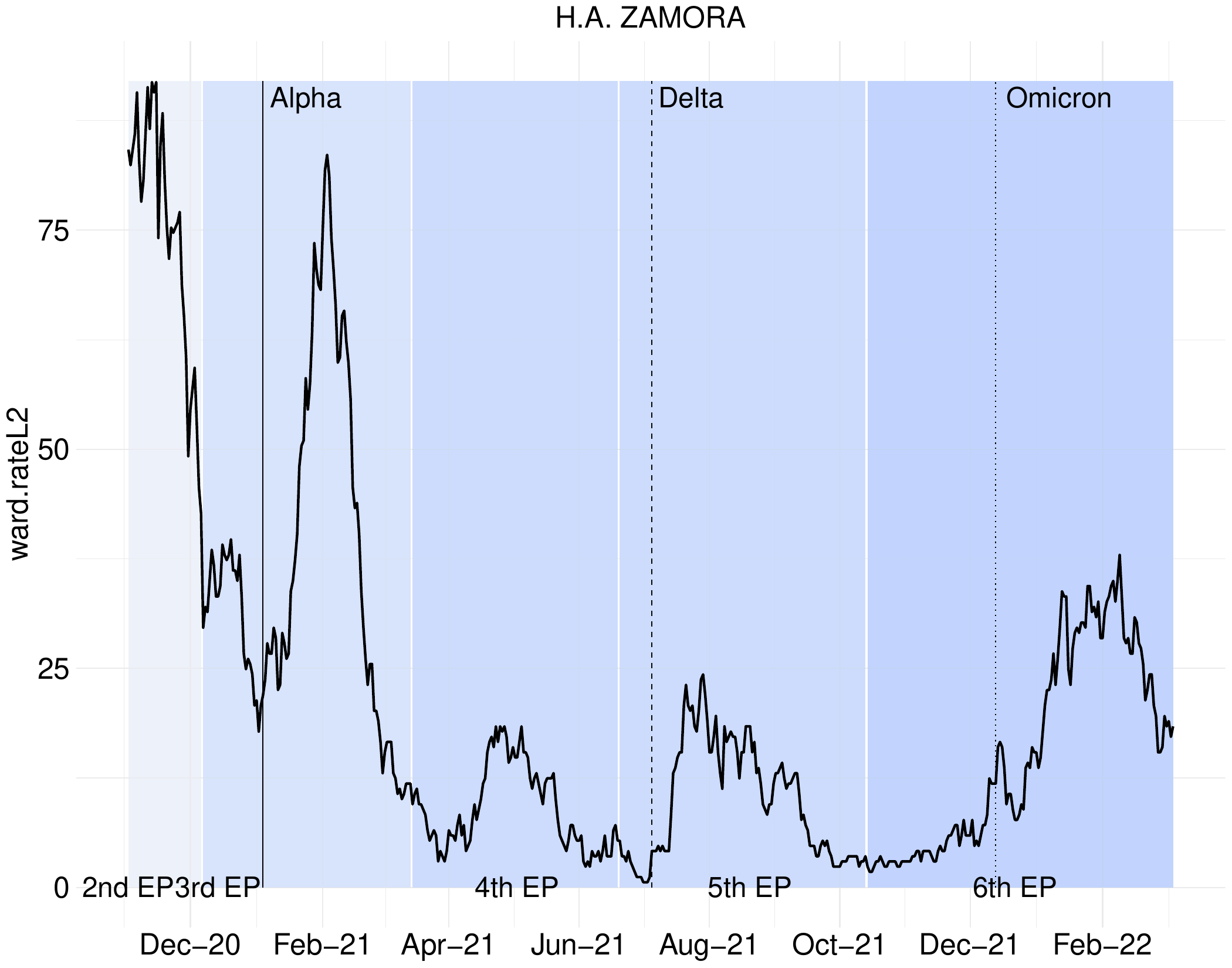}
  \end{minipage}
  \caption{Evolution of the ward occupancy rate by COVID-19 per 100,000 inhabitants by health area and day in Castilla y León.}
  \label{fig:app.ext.descr.data.WARD.HATIME.I}
\end{figure}

\newpage
\subsection{Discharges rate due to COVID-19 by health area}\label{app.ext.descr.data.RD}

The behaviour described for the variable {\it ward.rateL2} is transferred to the variable {\it disch.rate14L3}. As can be seen in Table \ref{tab:app.ext.descr.data.RD.CL}, it shows a higher prevalence in the second and third epidemic period.

This is consistent with the pattern of admissions, since the higher the incidence, the greater the probability of cases requiring hospitalisation, and therefore, this potentially leads to a higher number of discharges.  It can be seen that in the rest of the epidemic periods it decreases, except in the sixth period, when there is an upturn, probably due to the increase in admissions due to the holiday period and the Omicron variant. This increase may also indicate a lower severity of the disease, since if there is an increase in admissions but no increase in discharges, this would imply a higher number of deaths or very long hospital stays.

\begin{table}[ht!]
\caption{Descriptive of the variable \textit{disch.rate14L3} per 1000 inhabitants in each EP in Castilla y León.}\label{tab:app.ext.descr.data.RD.CL}
\renewcommand{\arraystretch}{1.2}
\centering
\begin{tabular}{|l|llllll|}
\hline
Period & \multicolumn{1}{l|}{Min.} & \multicolumn{1}{l|}{1st Qu.} & \multicolumn{1}{l|}{Median} & \multicolumn{1}{l|}{Mean} & \multicolumn{1}{l|}{3rd Qu.} & Max. \\ \hline
2nd EP & 0.2245                         & 0.5925                            & 0.7958                          & 0.7800                   & 0.9765                           & 1.3815   \\ \hline
3rd EP & 0.0330                         & 0.2857                            & 0.5211                          & 0.6026                   & 0.8348                           & 1.8282   \\ \hline
4th EP & 0.0312                         & 0.1127                            & 0.1694                          & 0.2038                   & 0.2650                           & 0.9502   \\ \hline
5th EP & 0.0000                         & 0.0564                            & 0.1047                           & 0.1461                    & 0.2016                            & 0.6950   \\ \hline
6th EP & 0.0000                        & 0.0512                            & 0.1845                           & 0.2486                    & 0.4296                            & 0.8953   \\ \hline
\end{tabular}
\end{table}

The Figure \ref{fig:app.ext.descr.data.RD.HATIME.I} below shows a breakdown for each H.A. in Castilla y León and the day of the variable variable {\it disch.rate14L3} per 1,000 inhabitants.

\begin{figure}[htbp]
  \begin{minipage}{0.50\textwidth}
    \centering
    \includegraphics[width=\linewidth]{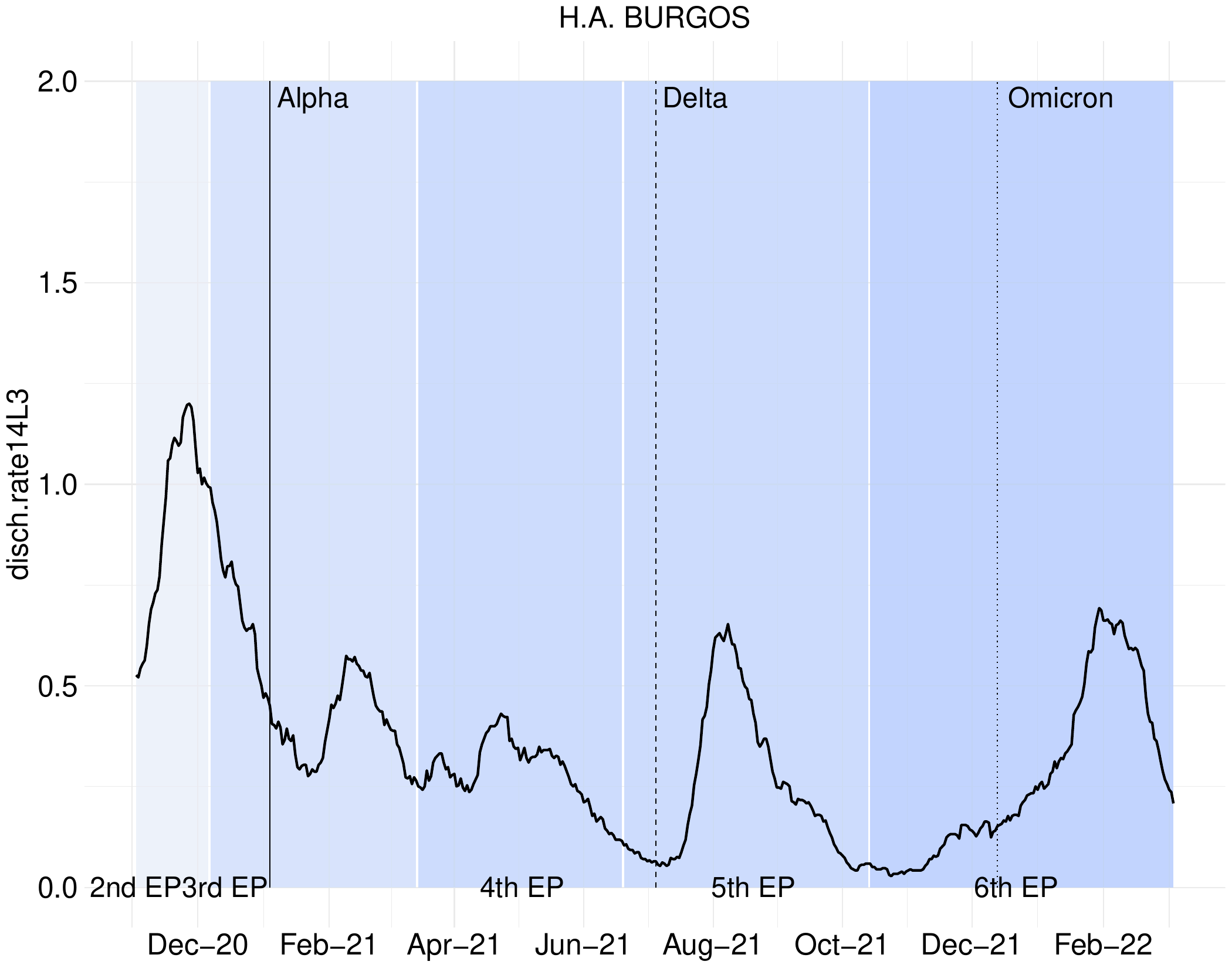}
  \end{minipage}
 \begin{minipage}{0.50\textwidth}
    \centering
    \includegraphics[width=\linewidth]{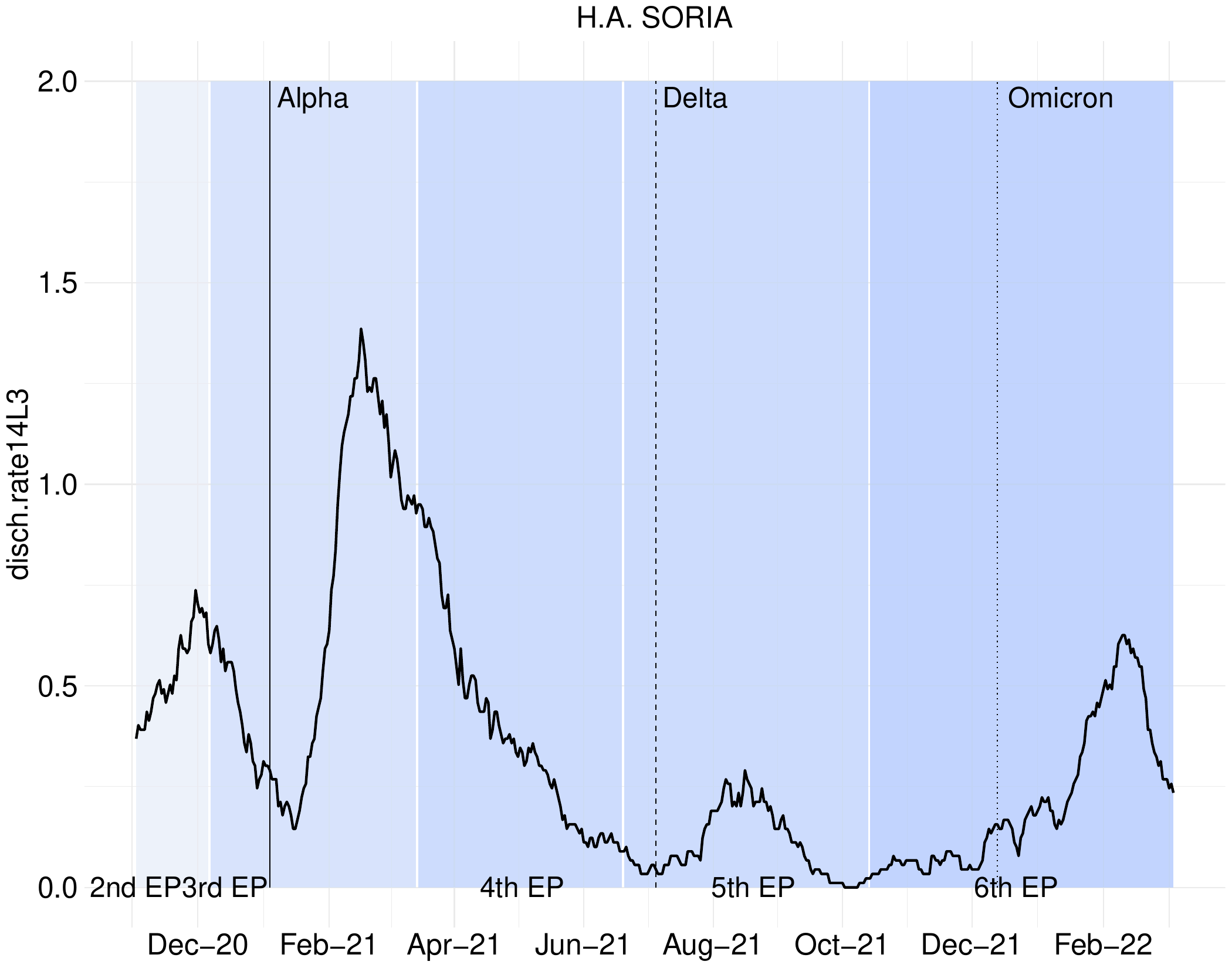}
  \end{minipage}

  \begin{minipage}{0.50\textwidth}
    \centering
    \includegraphics[width=\linewidth]{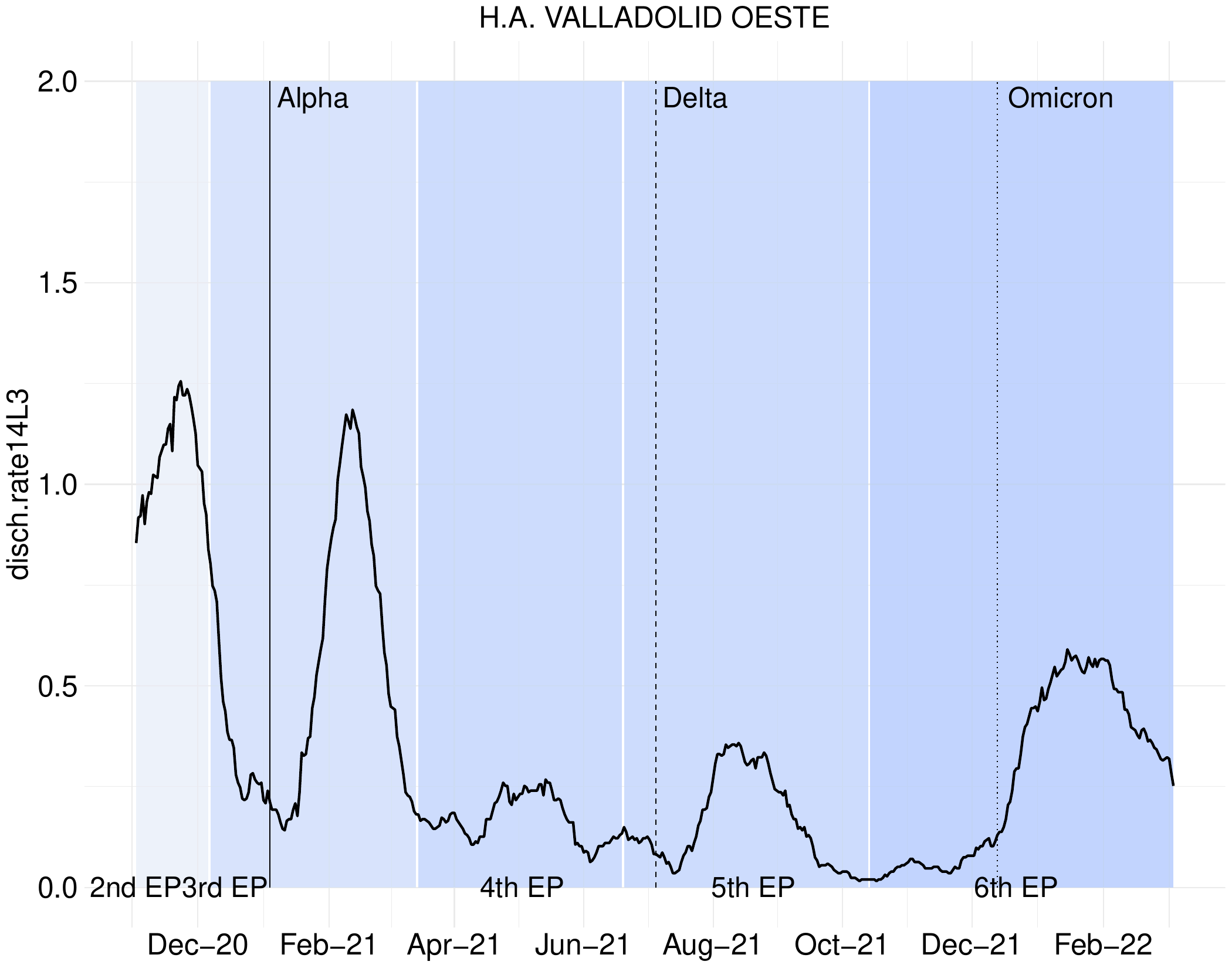}
  \end{minipage}%
  \begin{minipage}{0.50\textwidth}
    \centering
    \includegraphics[width=\linewidth]{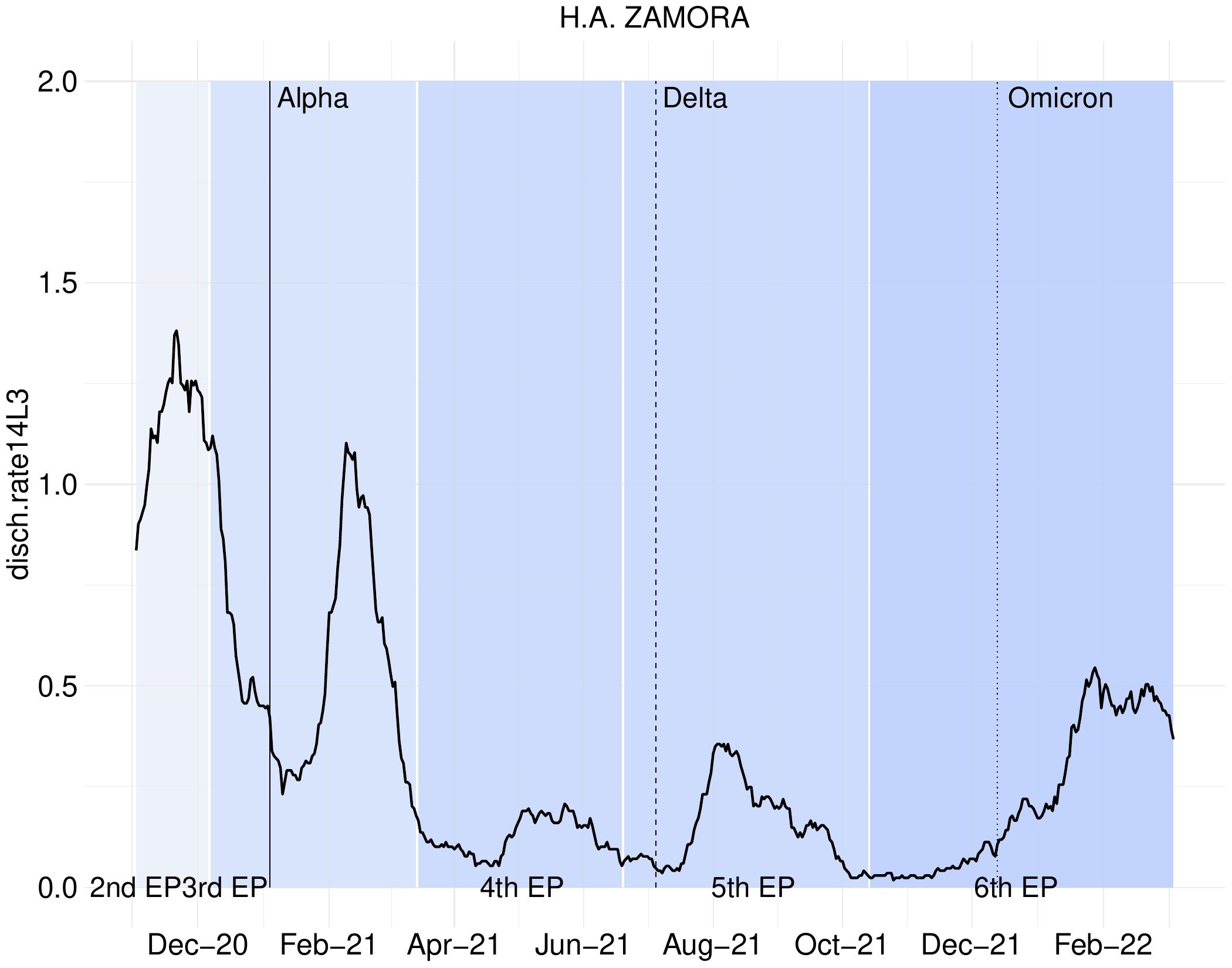}
  \end{minipage}
  \caption{Evolution of the {\it disch.rate14L3} per 1,000 inhabitants by health area (H.A.) and day.}
  \label{fig:app.ext.descr.data.RD.HATIME.I}
\end{figure}

\newpage
\subsection{Acute rate due to COVID-19}\label{app.ext.descr.data.ACUTE}

The variable {\it acute.rate7} evaluates the rate of weekly hospitalisation in ICU in relation to the total number of admissions and shows a global behaviour different from the other auxiliary variables, which can be seen in Table \ref{tab:app.ext.descr.data.ACUTE.CL}. Thus, the maximum averages are obtained from the third epidemic period onwards, reaching a maximum in the fourth period and decreasing progressively. It should be noted that during the second and third periods, although the proportion of patients admitted to the ICU was the highest, the proportion of patients admitted to the ward was also the highest, which could justify the fact that from the fourth period onwards, when the proportion of patients admitted to the ICU decreased drastically, the proportion of patients admitted to the ICU was the highest.
This could justify the fact that from the fourth period onwards, when occupancy on the ward drops drastically, these proportions are higher. Furthermore, although the disease, with the new variants and wide vaccination coverage, is less severe, the duration of admission to ICU is generally of a duration that can reach more than one month, contributing to maintaining the indicator at high values. This would be supported by the lower proportion of discharges detected in the fourth epidemic period.

\begin{table}[ht!]
\caption{Descriptive of the variable \textit{acute.rate7} in each EP in Castilla y León.}\label{tab:app.ext.descr.data.ACUTE.CL}
\renewcommand{\arraystretch}{1.2}
\centering
\begin{tabular}{|l|llllll|}
\hline
Period & \multicolumn{1}{l|}{Min.} & \multicolumn{1}{l|}{1st Qu.} & \multicolumn{1}{l|}{Median} & \multicolumn{1}{l|}{Mean} & \multicolumn{1}{l|}{3rd Qu.} & Max. \\ \hline
2nd EP & 4.7059                         & 11.0487                            & 13.8462                          & 14.8220                   & 18.5696                           & 31.0345   \\ \hline
3rd EP & 4.0351                         & 14.1108                            & 19.2545 & 21.0367 & 26.0870                           & 53.9823   \\ \hline
4th EP & 0.0000                         & 24.8172                            & 31.0127                          & 32.4958                   & 40.2987                           & 75.6757   \\ \hline
5th EP & 0.0000                         & 15.7582                            & 23.2505                          & 27.6888                    & 36.9389                            & 100.0000   \\ \hline
6th EP & 0.0000                         & 9.2712                            & 15.5982                           & 17.0443                    & 21.8750                            & 100.0000   \\ \hline
\end{tabular}
\end{table}

\begin{figure}[htbp]
  \begin{minipage}{0.50\textwidth}
    \centering
    \includegraphics[width=\linewidth]{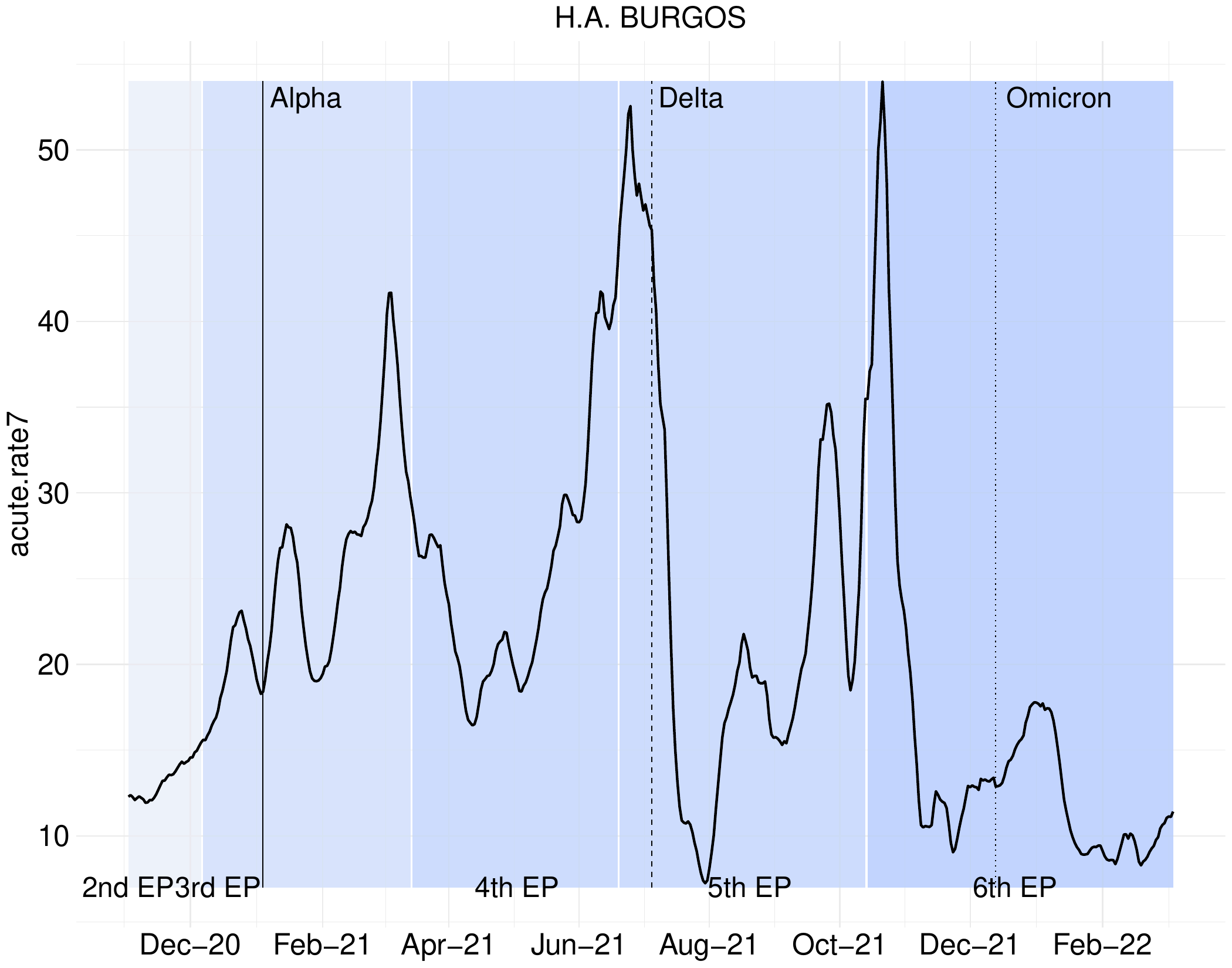}
  \end{minipage}
 \begin{minipage}{0.50\textwidth}
    \centering
    \includegraphics[width=\linewidth]{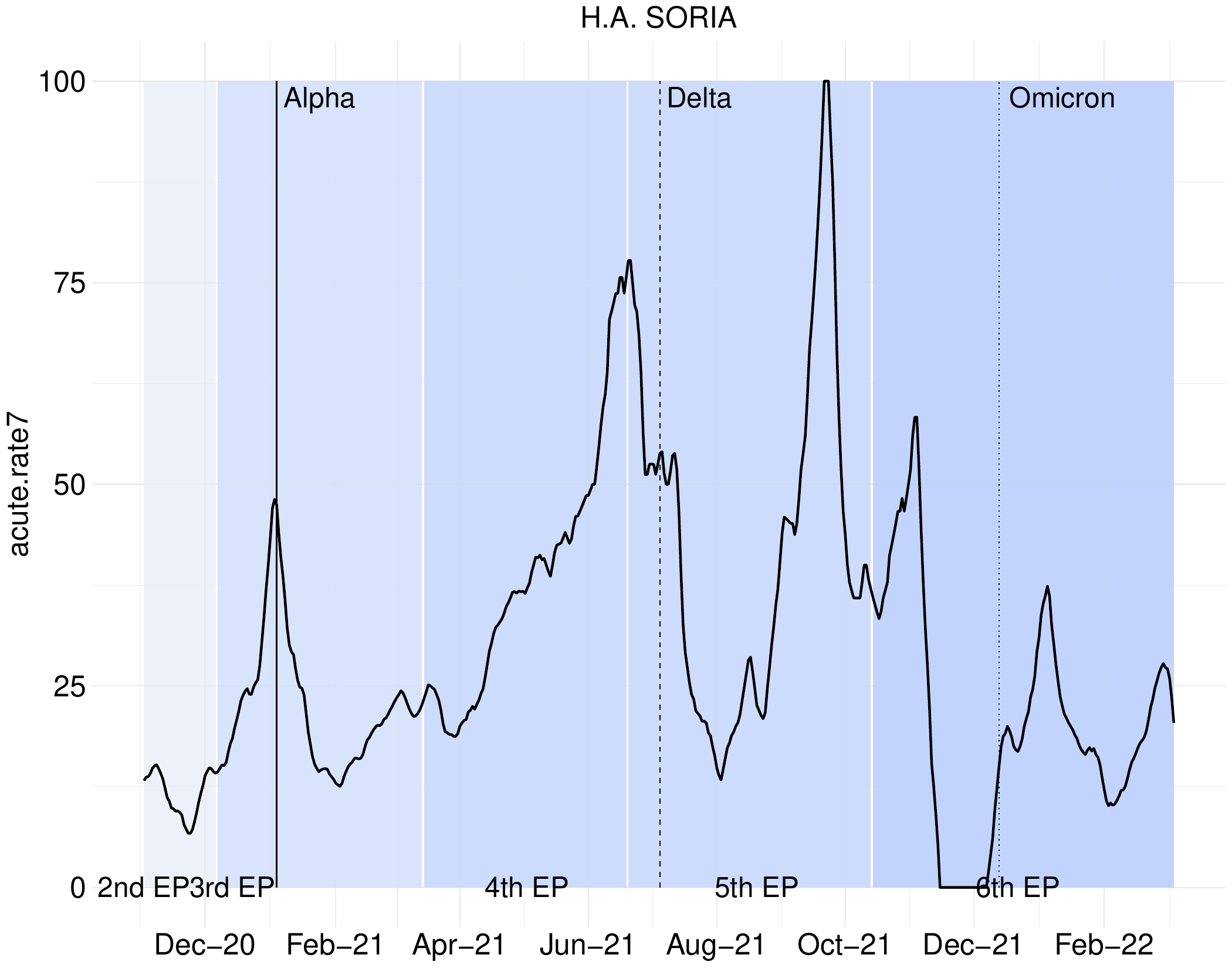}
  \end{minipage}

  \begin{minipage}{0.50\textwidth}
    \centering
    \includegraphics[width=\linewidth]{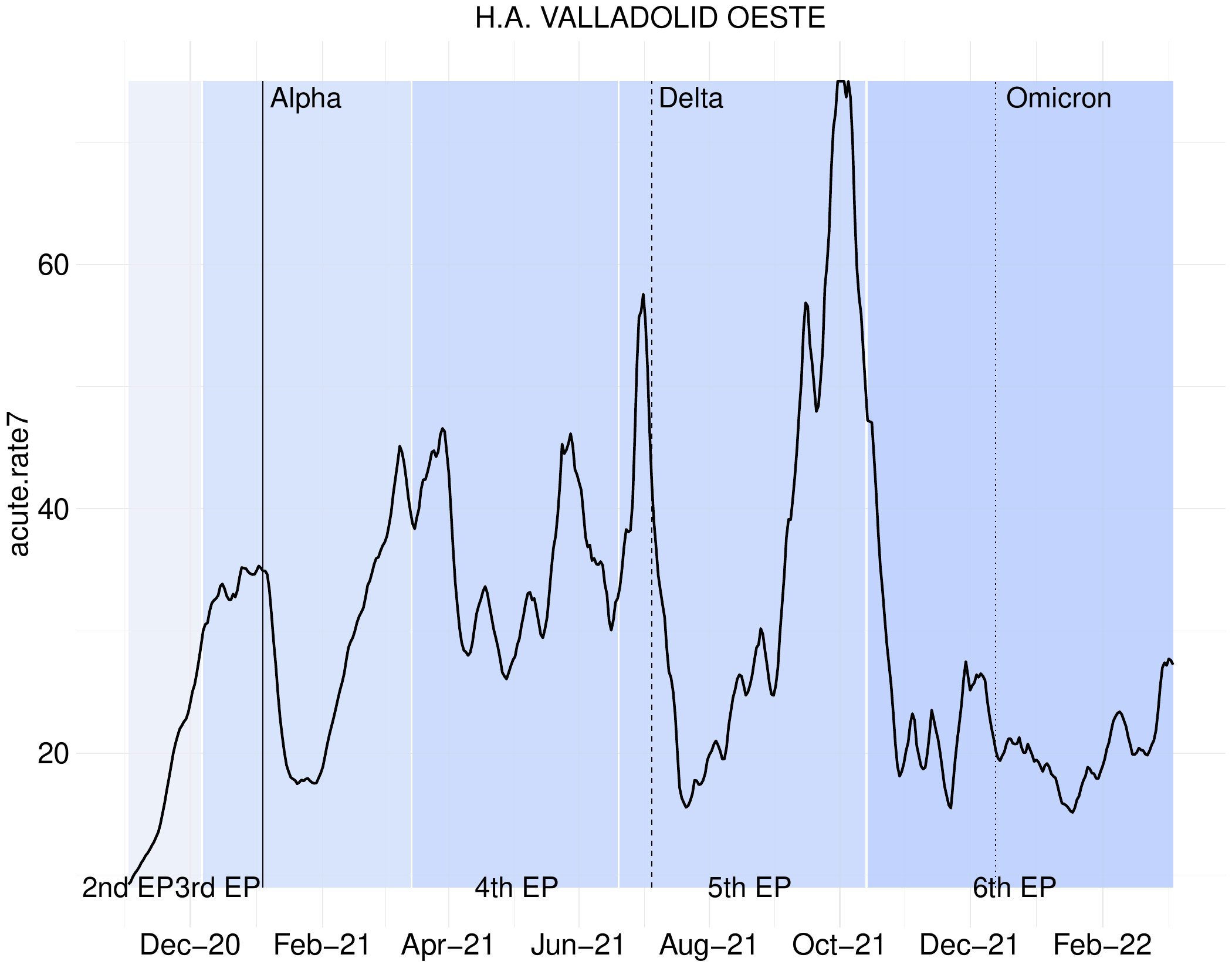}
  \end{minipage}%
  \begin{minipage}{0.50\textwidth}
    \centering
    \includegraphics[width=\linewidth]{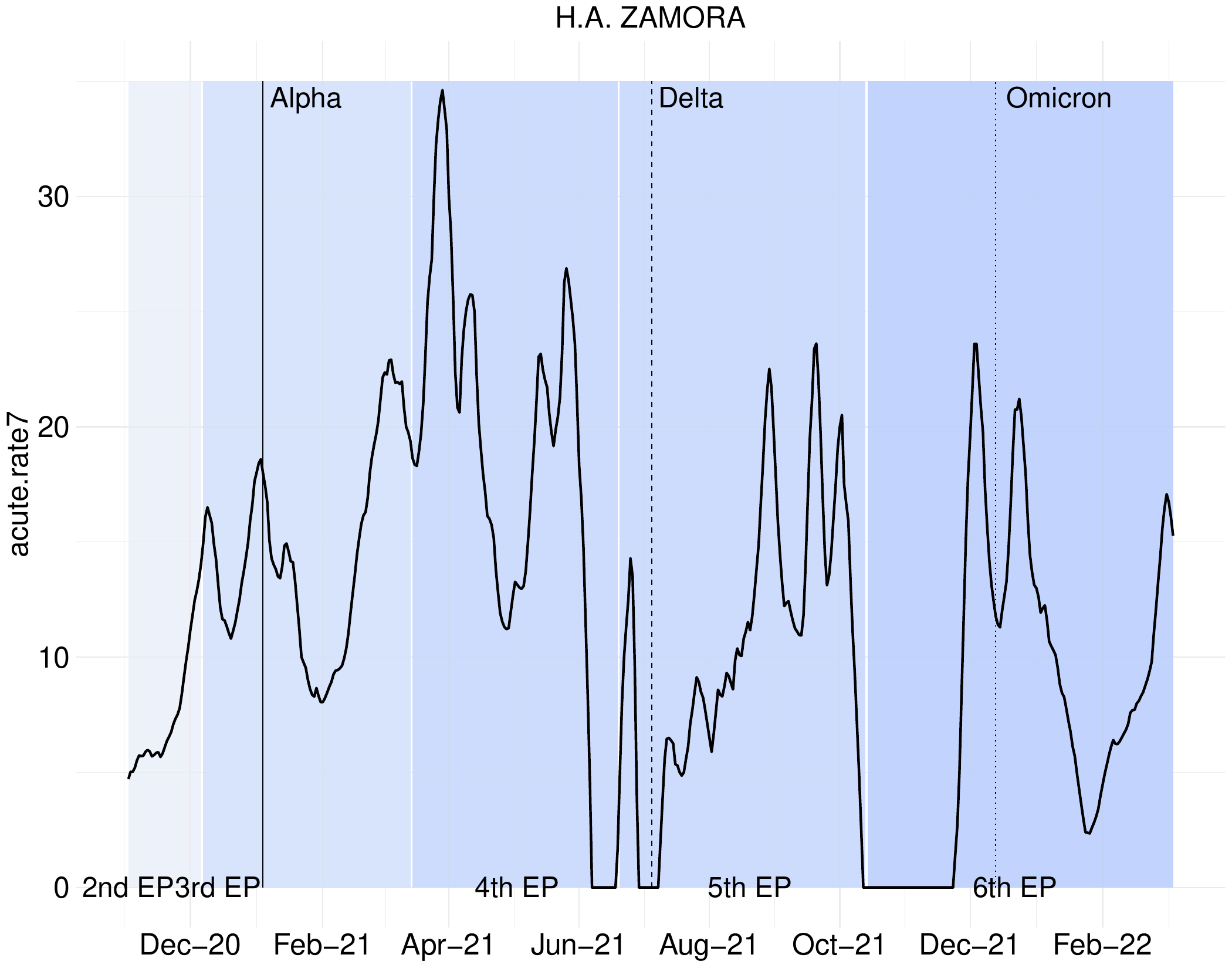}
  \end{minipage}
  \caption{Evolution of the {\it acute.rate7} (\%) by health area (H.A.) and day in Castilla y León.}
  \label{fig:app.ext.descr.data.ACUTE.HATIME.I}
\end{figure}

The Figure \ref{fig:app.ext.descr.data.ACUTE.HATIME.I} shows a breakdown for health area and day of the variable {\it acute.rate7} in \%.

\newpage
\subsection{Slopes of auxiliary variables by health area}\label{app.ext.descr.data.SLOPES}

\begin{figure}[htbp]
\centering
  \begin{minipage}{0.45\textwidth}
    \centering
    \includegraphics[width=\linewidth]{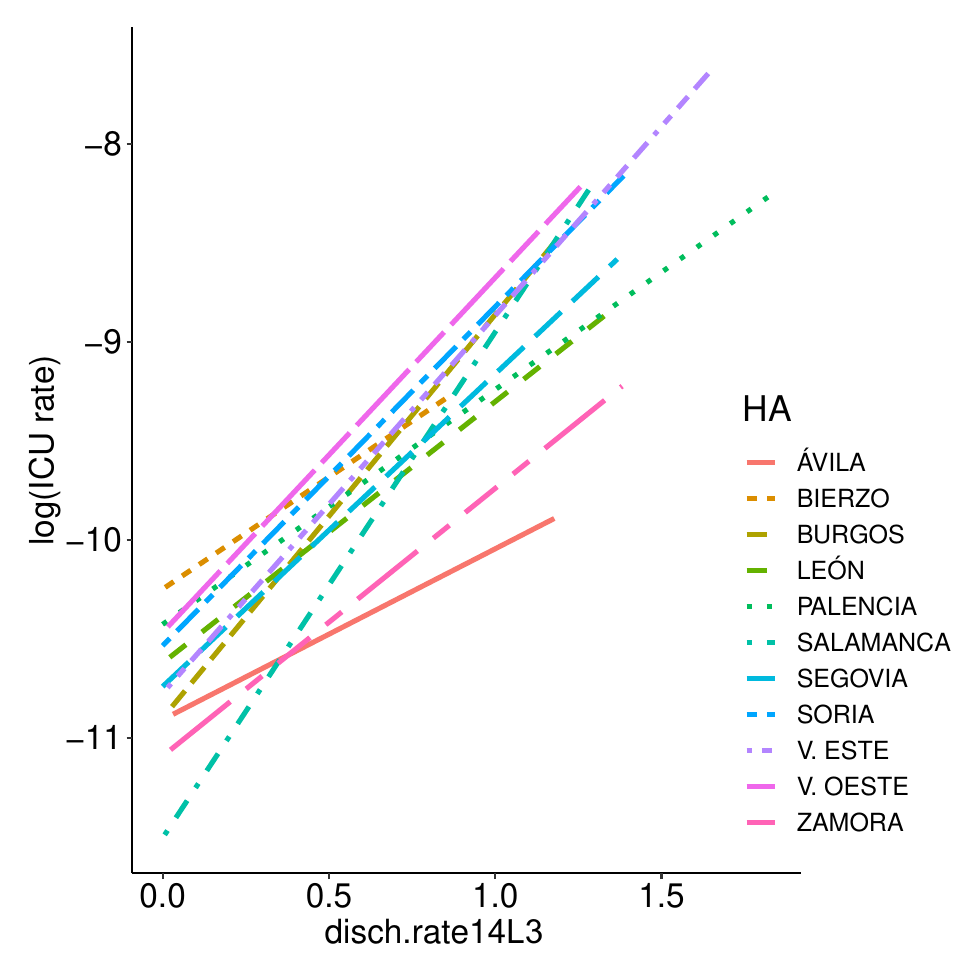}
  \end{minipage}%
  \begin{minipage}{0.45\textwidth}
    \centering
    \includegraphics[width=\linewidth]{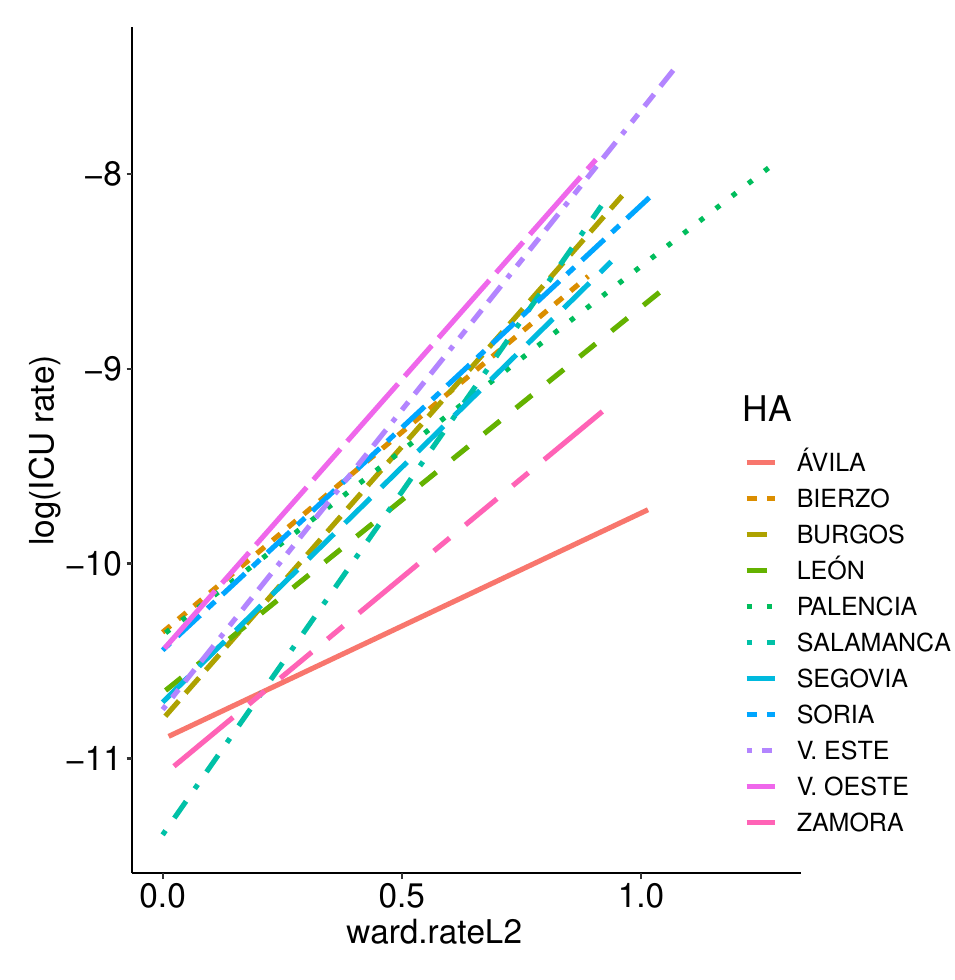}
  \end{minipage}%

  \begin{minipage}{0.45\textwidth}
    \centering
    \includegraphics[width=\linewidth]{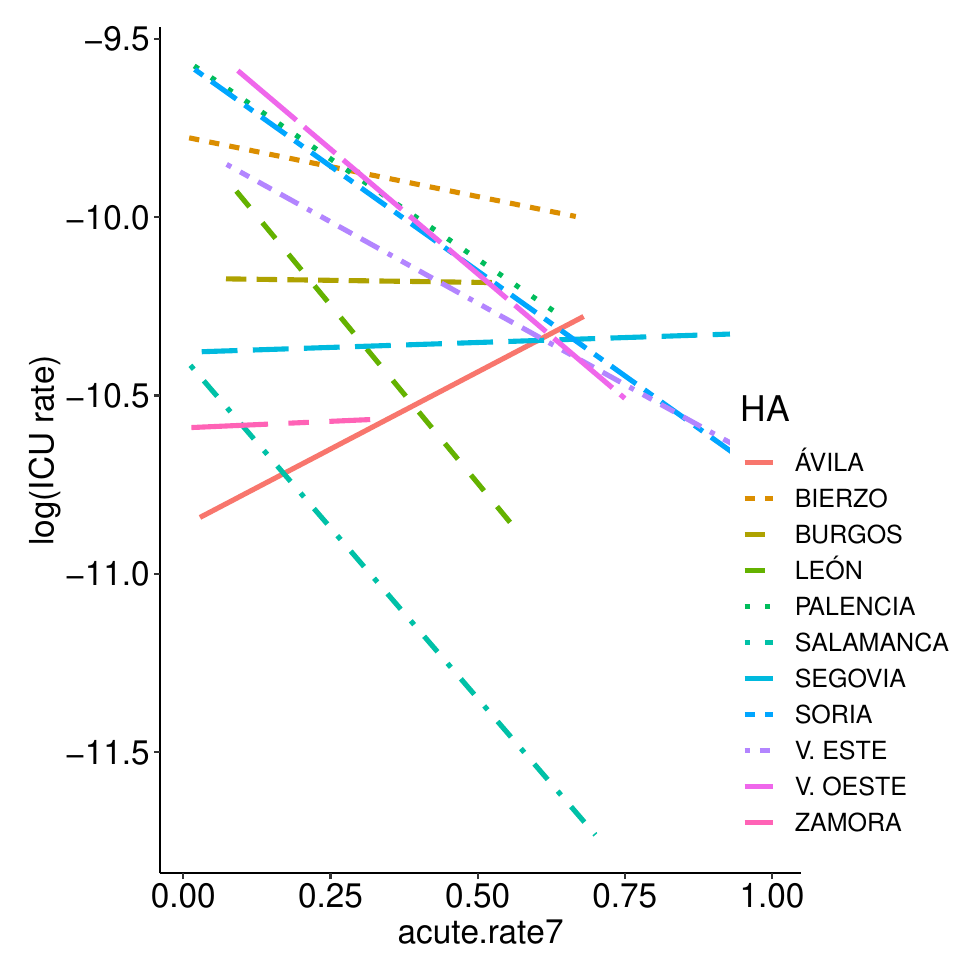}
  \end{minipage}
  \caption{Relationship of the auxiliary variables with the target variable according to the Health Area (H.A.).}
  \label{fig:app.ext.descr.data.SLOPES}
\end{figure}

The diversity of patterns by H.A. and day contemplated both in the analysis of ICU admission and in the evolution of each of the auxiliary variables suggests the need to assess whether this behaviour carries over to the relationship of the auxiliary variables with the target variable.
In this subsection, Figure \ref{fig:app.ext.descr.data.SLOPES} shows the relationship between the logarithm of the proportion of ICU occupancy due to COVID-19 by H.A. and day and each of the ancillary variables selected in the final model, {\it ward.rateL2}, {\it disch.rate14L3} and {\it acute.rate7}.

In view of the results, it can be seen that the variables with the most heterogeneous relationship to the target variable according to health area {\it acute.rate7}  and {\it discharges.rate14L3}.

\newpage
\section{Further results from application to real data}\label{app.further.ap.data}

\begin{figure}[h!]
  \begin{minipage}{0.50\textwidth}
    \centering
    \includegraphics[width=\linewidth]{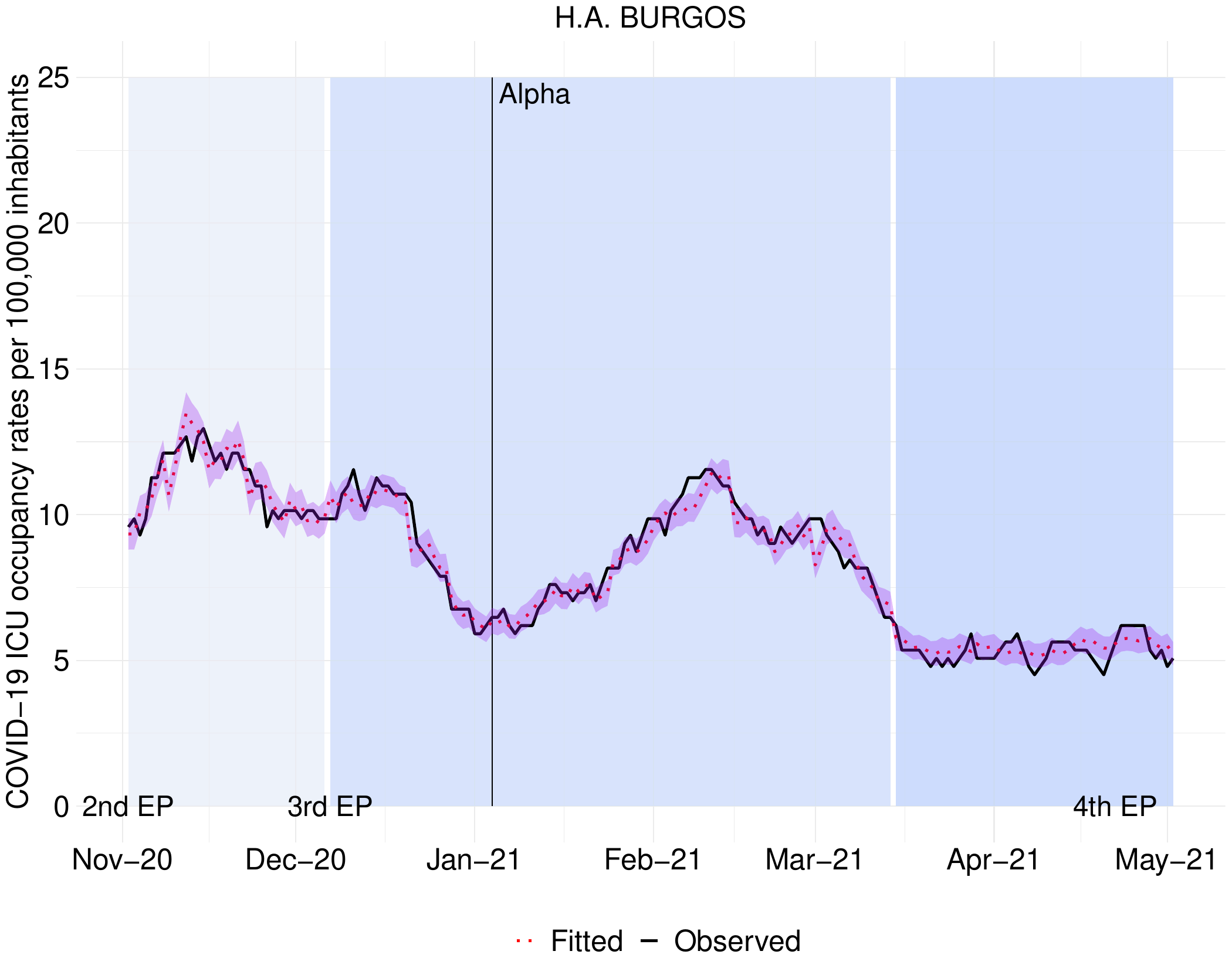}
  \end{minipage}
  \begin{minipage}{0.50\textwidth}
    \centering
    \includegraphics[width=\linewidth]{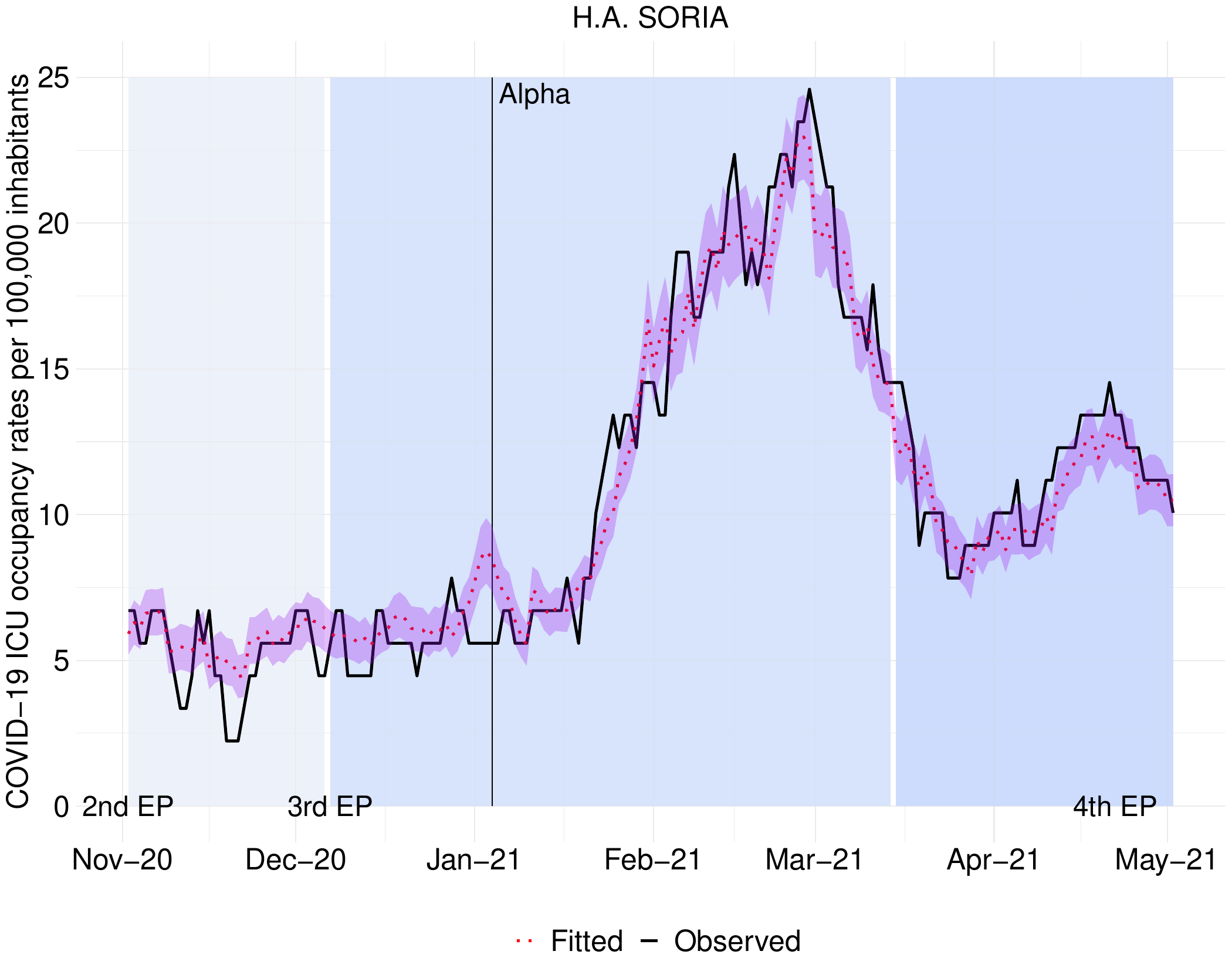}
  \end{minipage}

  \begin{minipage}{0.50\textwidth}
    \centering
    \includegraphics[width=\linewidth]{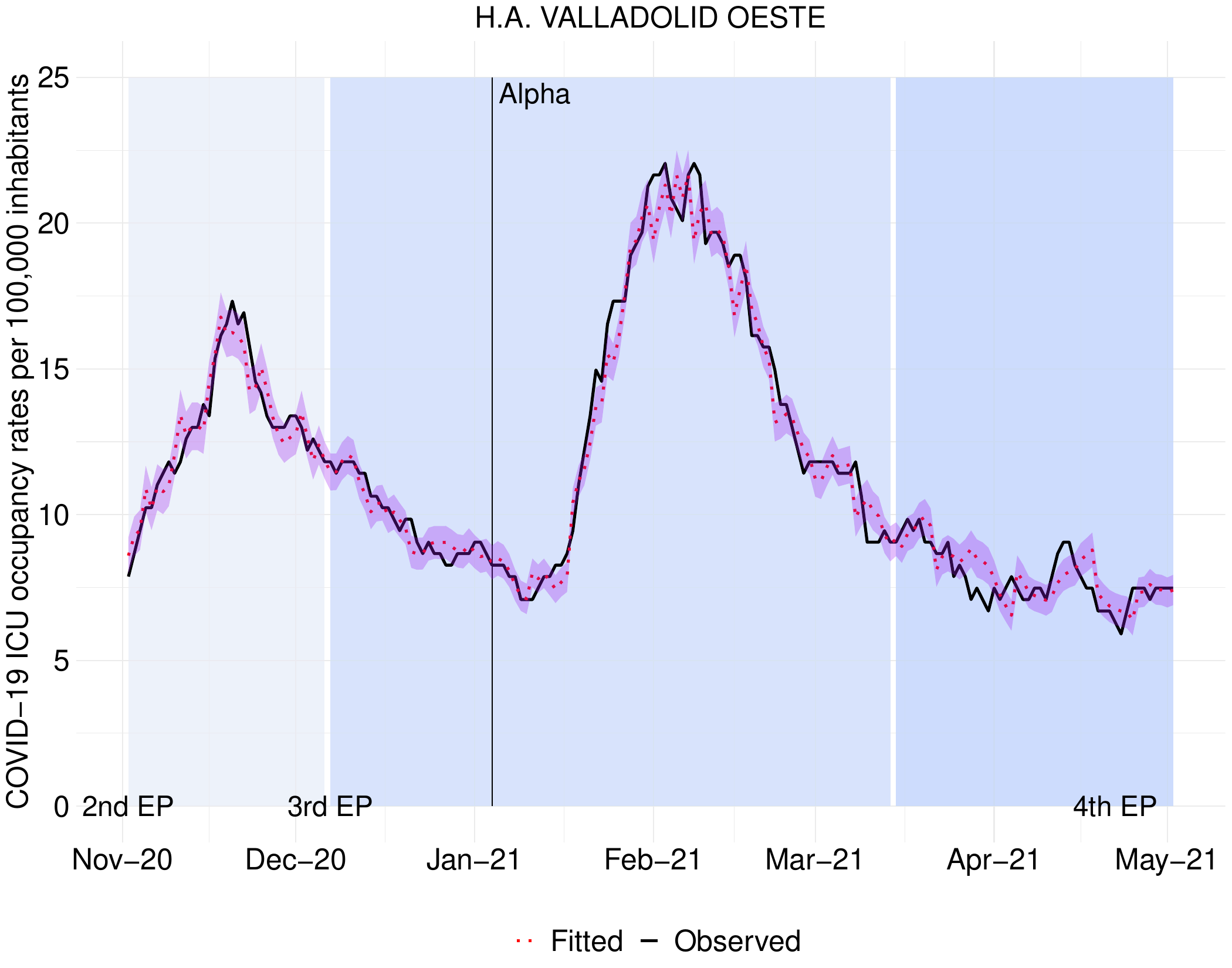}
  \end{minipage}%
  \begin{minipage}{0.50\textwidth}
    \centering
    \includegraphics[width=\linewidth]{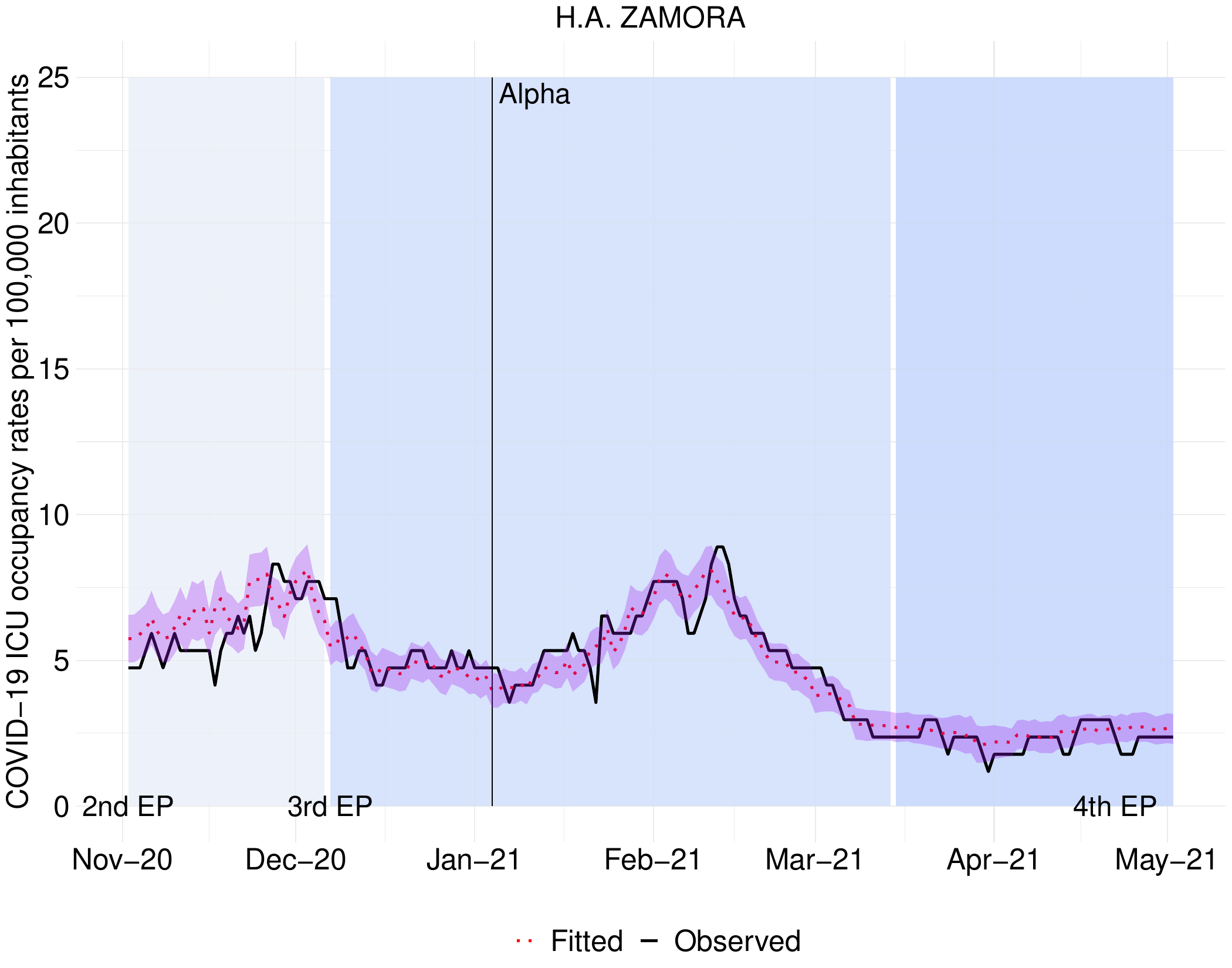}
  \end{minipage}
  \caption{Observed vs estimated ICU occupancy rate per 100,000 inhabitants.}
  \label{fig:app.further.ap.data.FIT.I}
\end{figure}

\begin{figure}[h!]
  \begin{minipage}{0.50\textwidth}
    \centering
    \includegraphics[width=\linewidth]{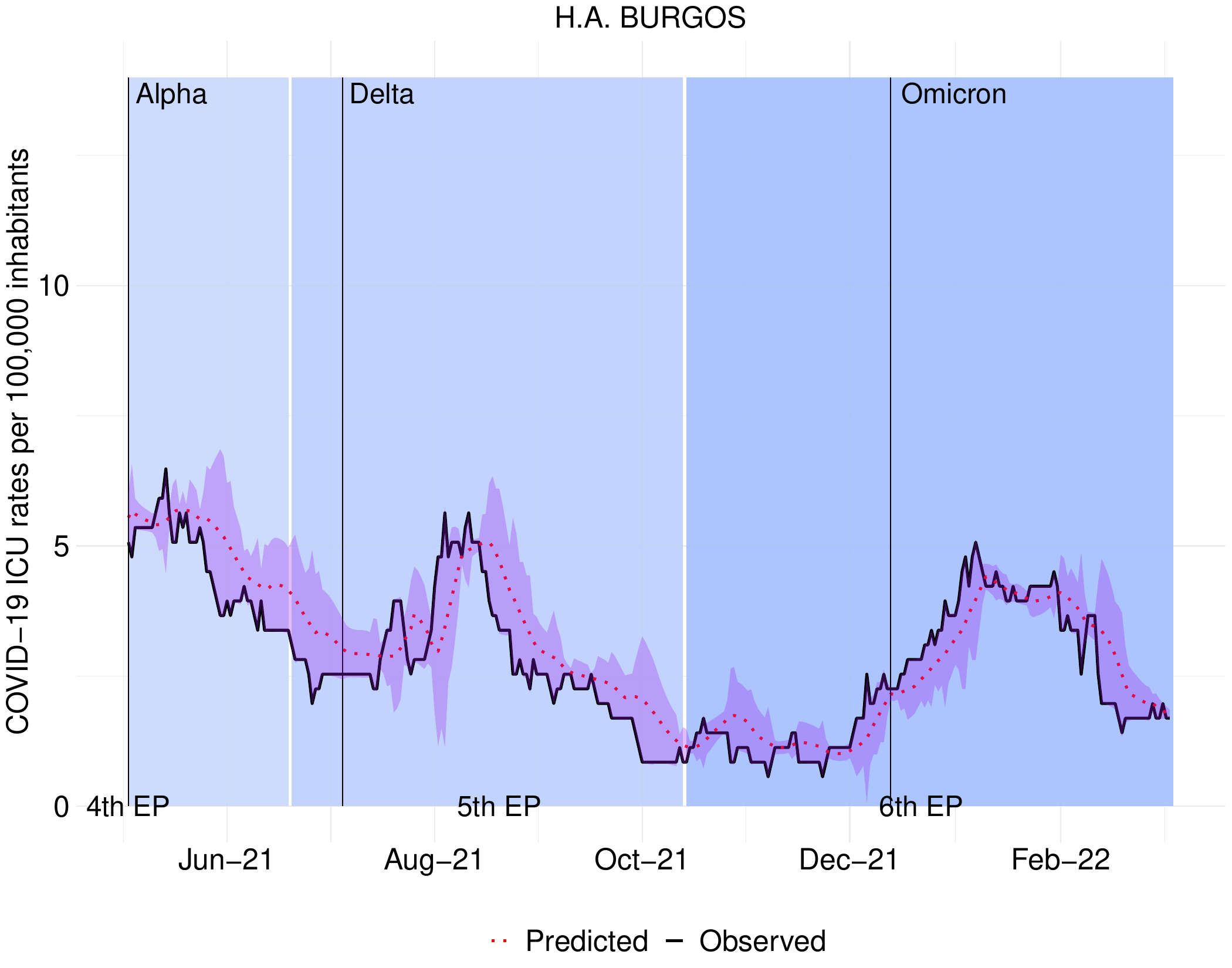}
  \end{minipage}
\begin{minipage}{0.50\textwidth}
    \centering
    \includegraphics[width=\linewidth]{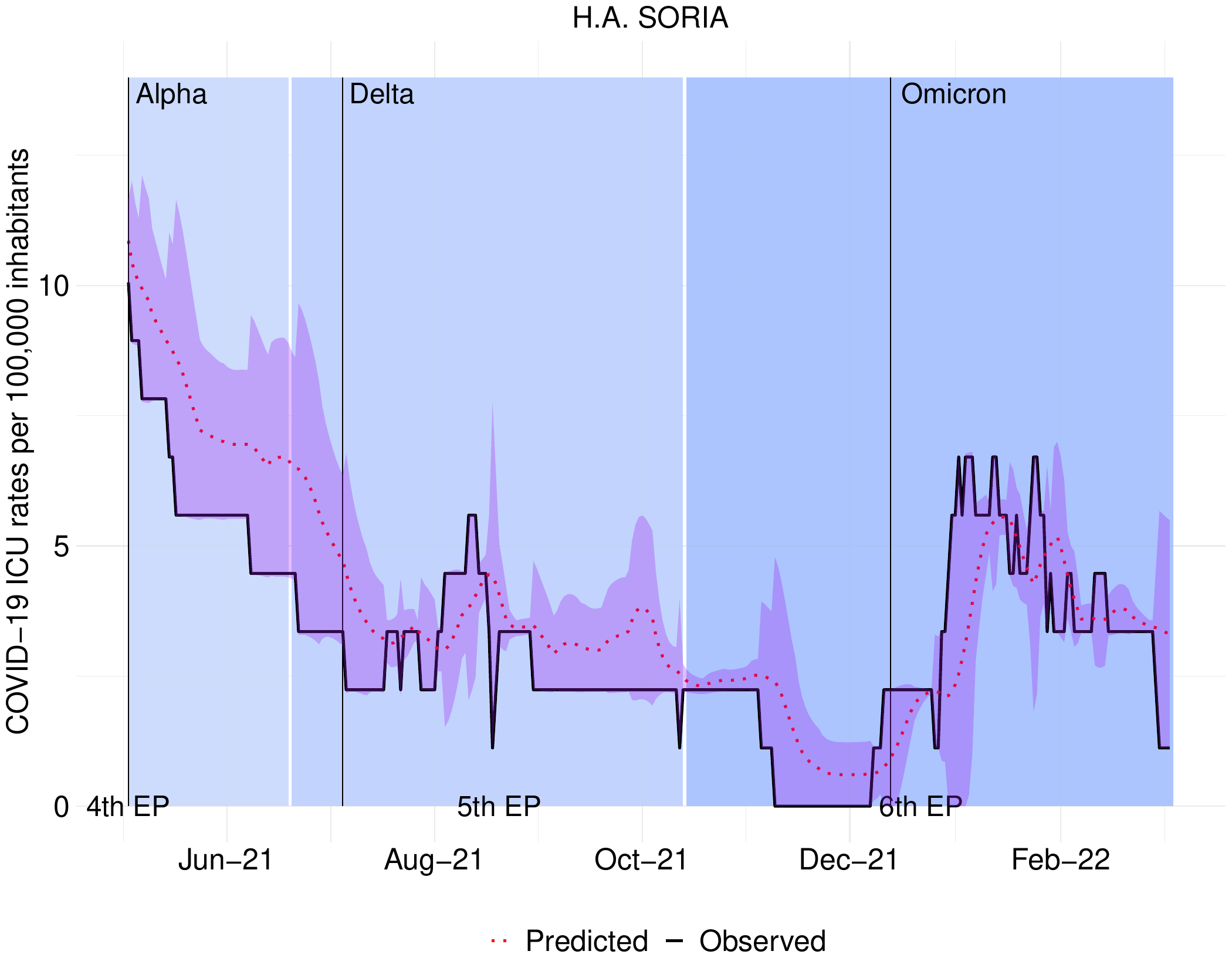}
  \end{minipage}

  \begin{minipage}{0.50\textwidth}
    \centering
    \includegraphics[width=\linewidth]{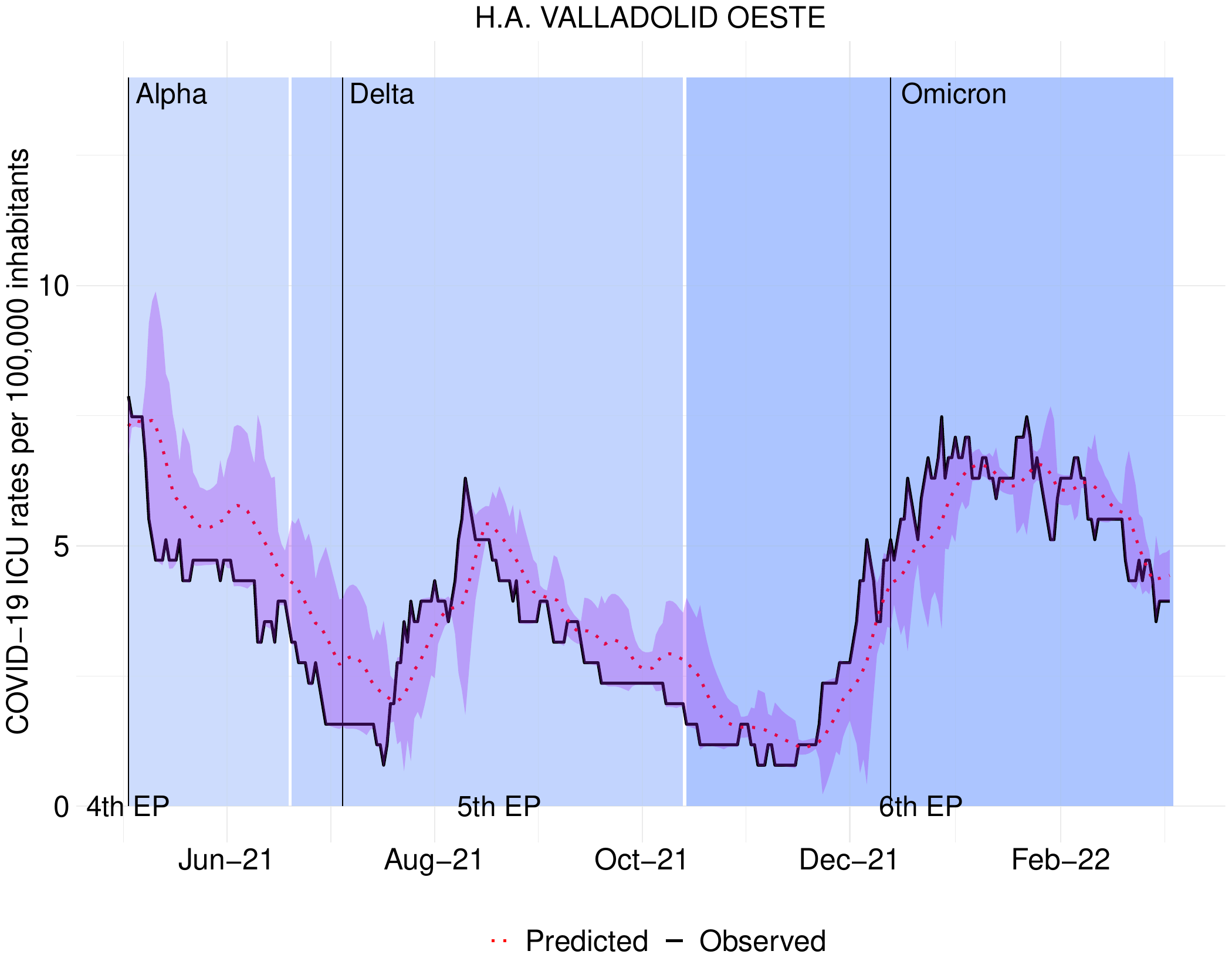}
  \end{minipage}%
  \begin{minipage}{0.50\textwidth}
    \centering
    \includegraphics[width=\linewidth]{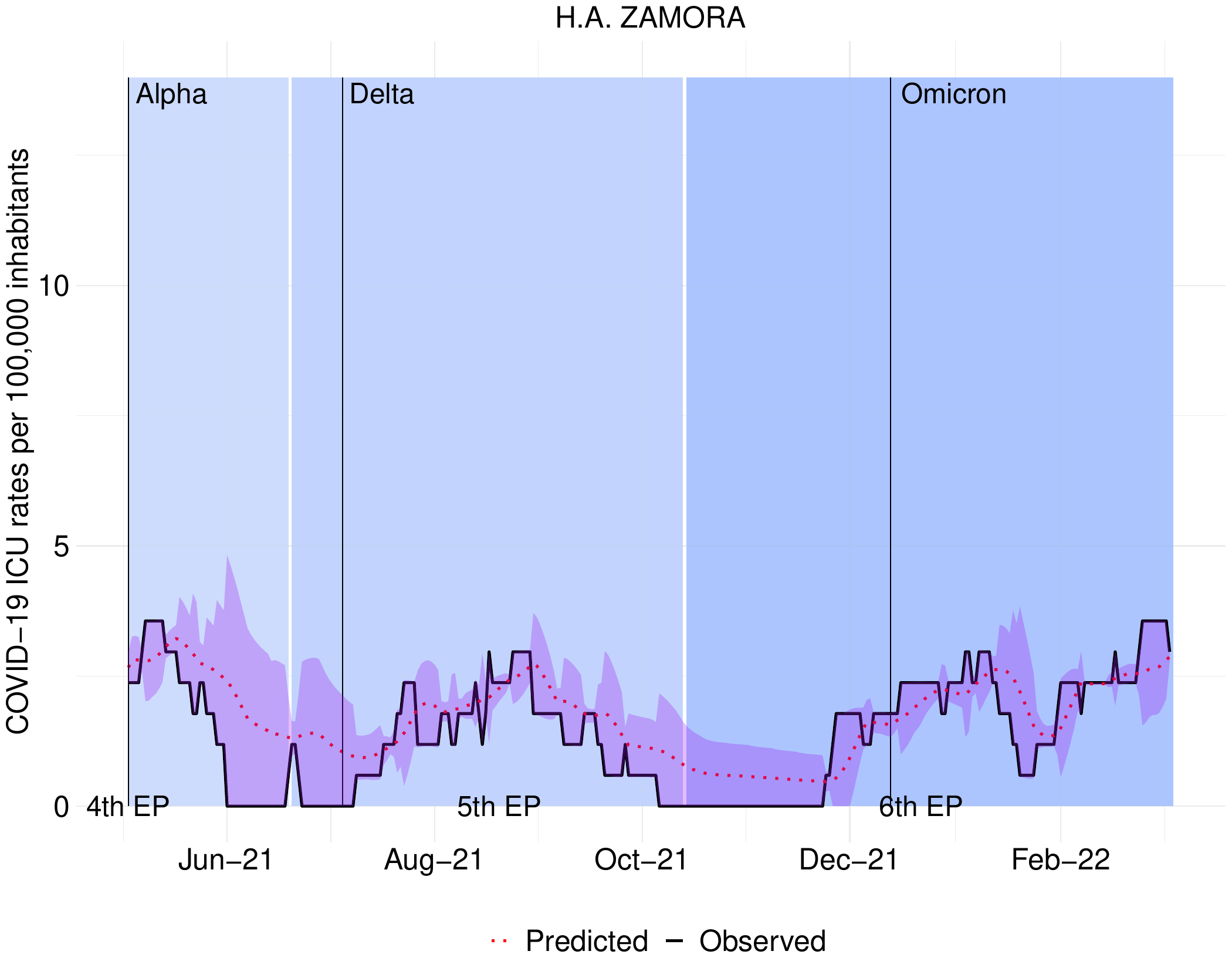}
  \end{minipage}
  \caption{Observed vs predicted ICU occupancy rate per 100,000 inhabitants with actual RMSE.}
\label{fig:app.further.ap.data.PRED.I}
\end{figure}

\begin{figure}[h!]
  \begin{minipage}{0.50\textwidth}
    \centering
    \includegraphics[width=\linewidth]{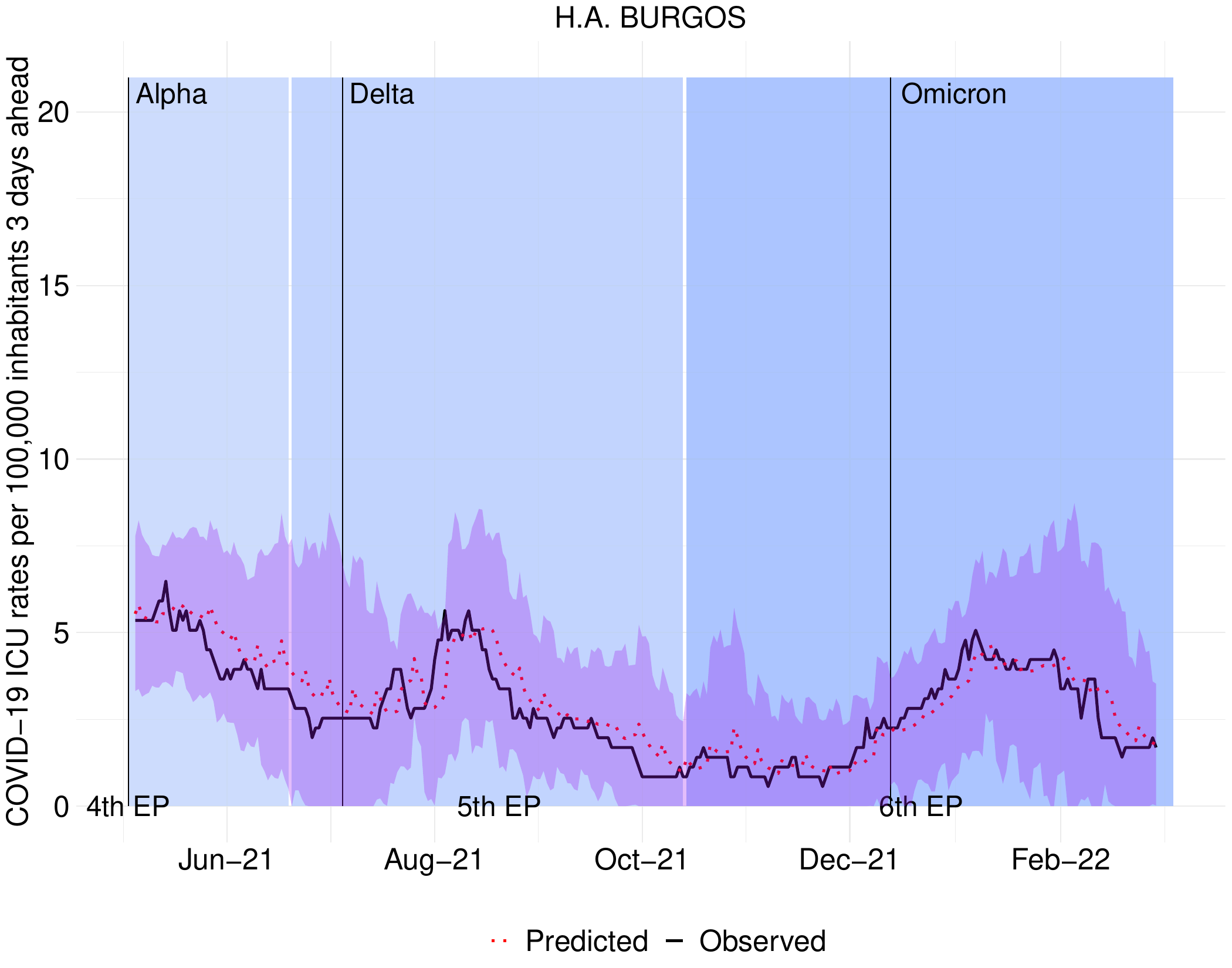}
  \end{minipage}
  \begin{minipage}{0.50\textwidth}
    \centering
    \includegraphics[width=\linewidth]{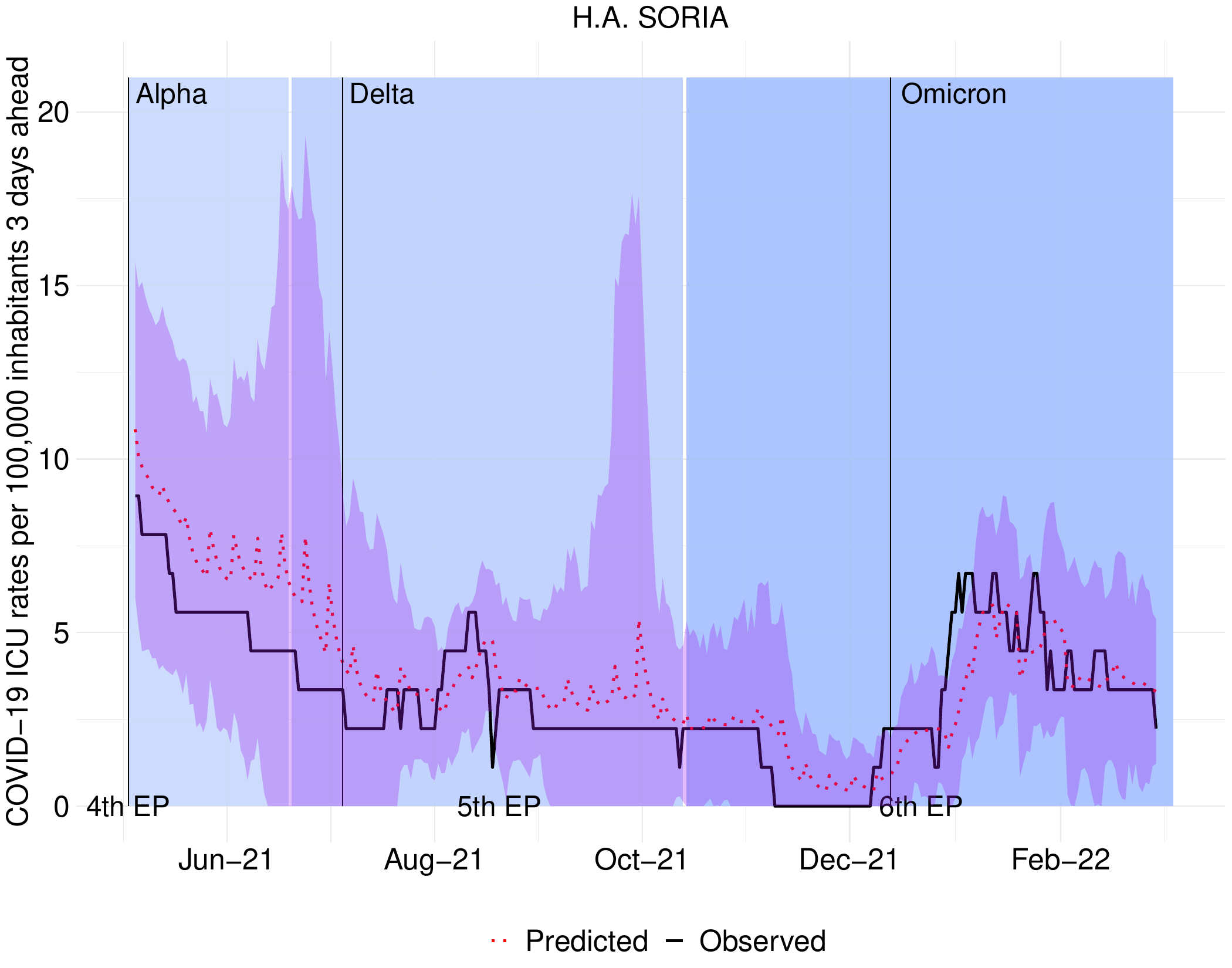}
  \end{minipage}

  \begin{minipage}{0.50\textwidth}
    \centering
    \includegraphics[width=\linewidth]{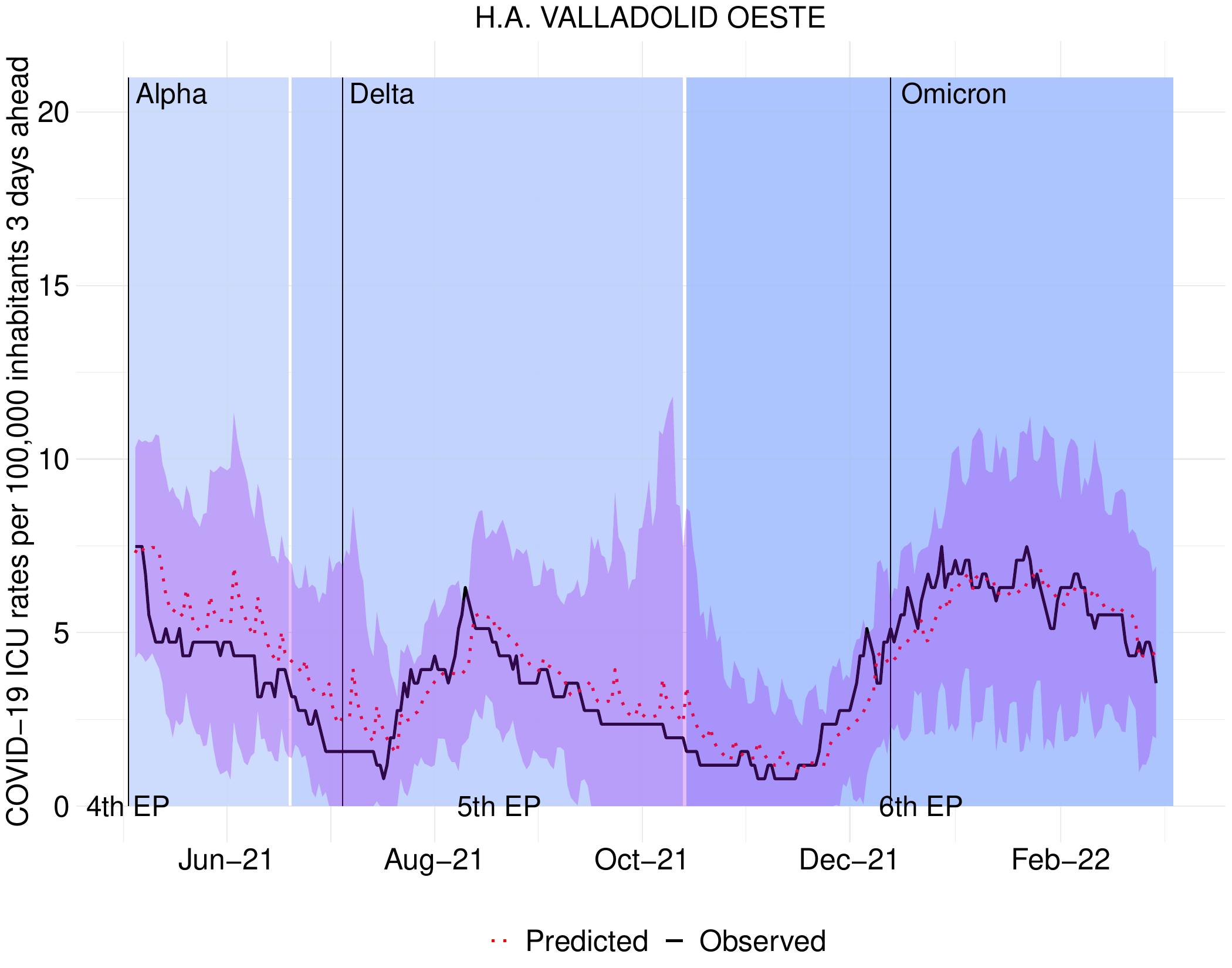}
  \end{minipage}%
  \begin{minipage}{0.50\textwidth}
    \centering
    \includegraphics[width=\linewidth]{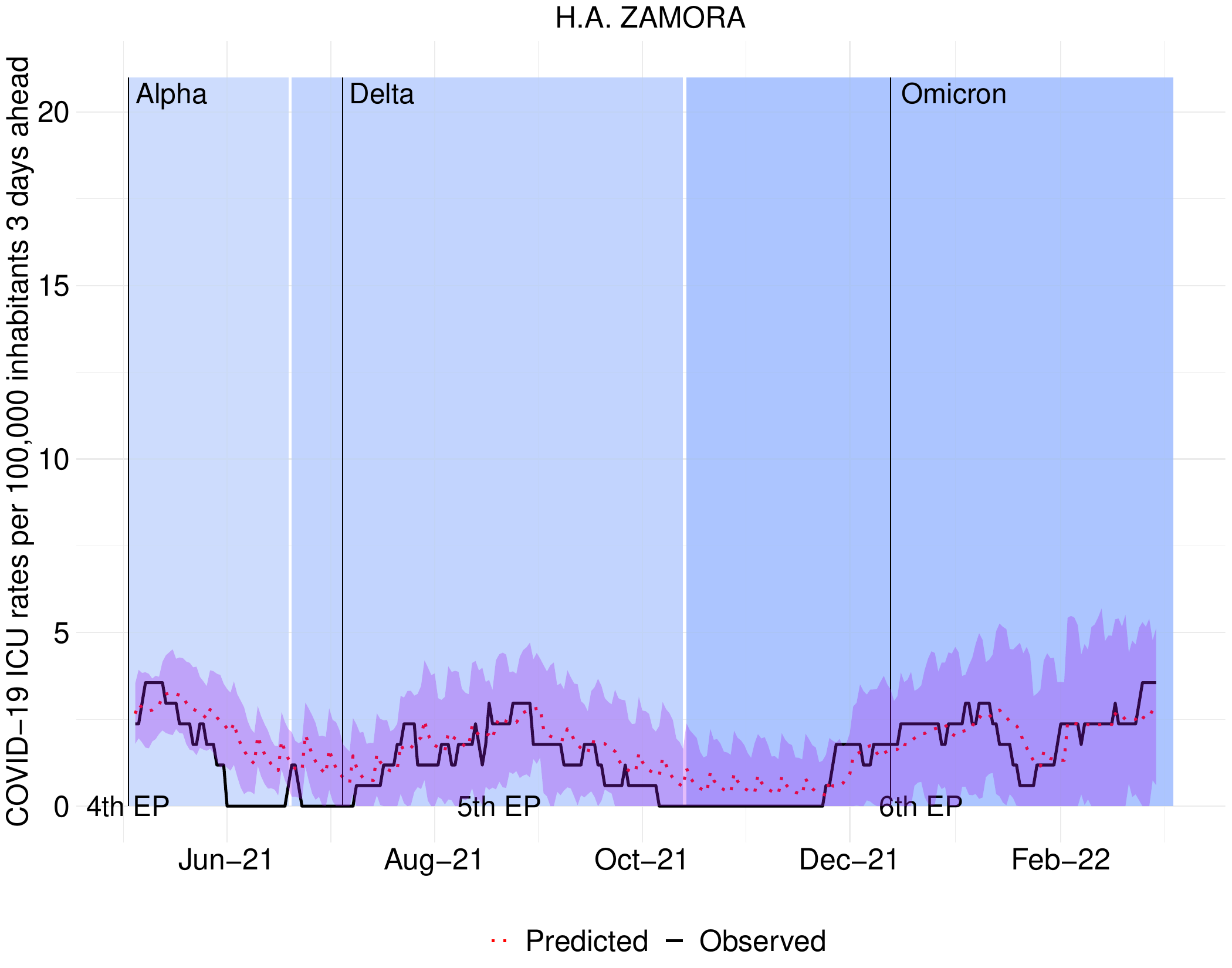}
  \end{minipage}
  \caption{Observed versus predicted three-day ICU occupancy rate per 100,000 population with simulated RMSE.}
  \label{fig:app.further.ap.data.PREDSIM.H3.I}
\end{figure}

\begin{figure}[htbp]
  \begin{minipage}{0.50\textwidth}
    \centering
    \includegraphics[width=\linewidth]{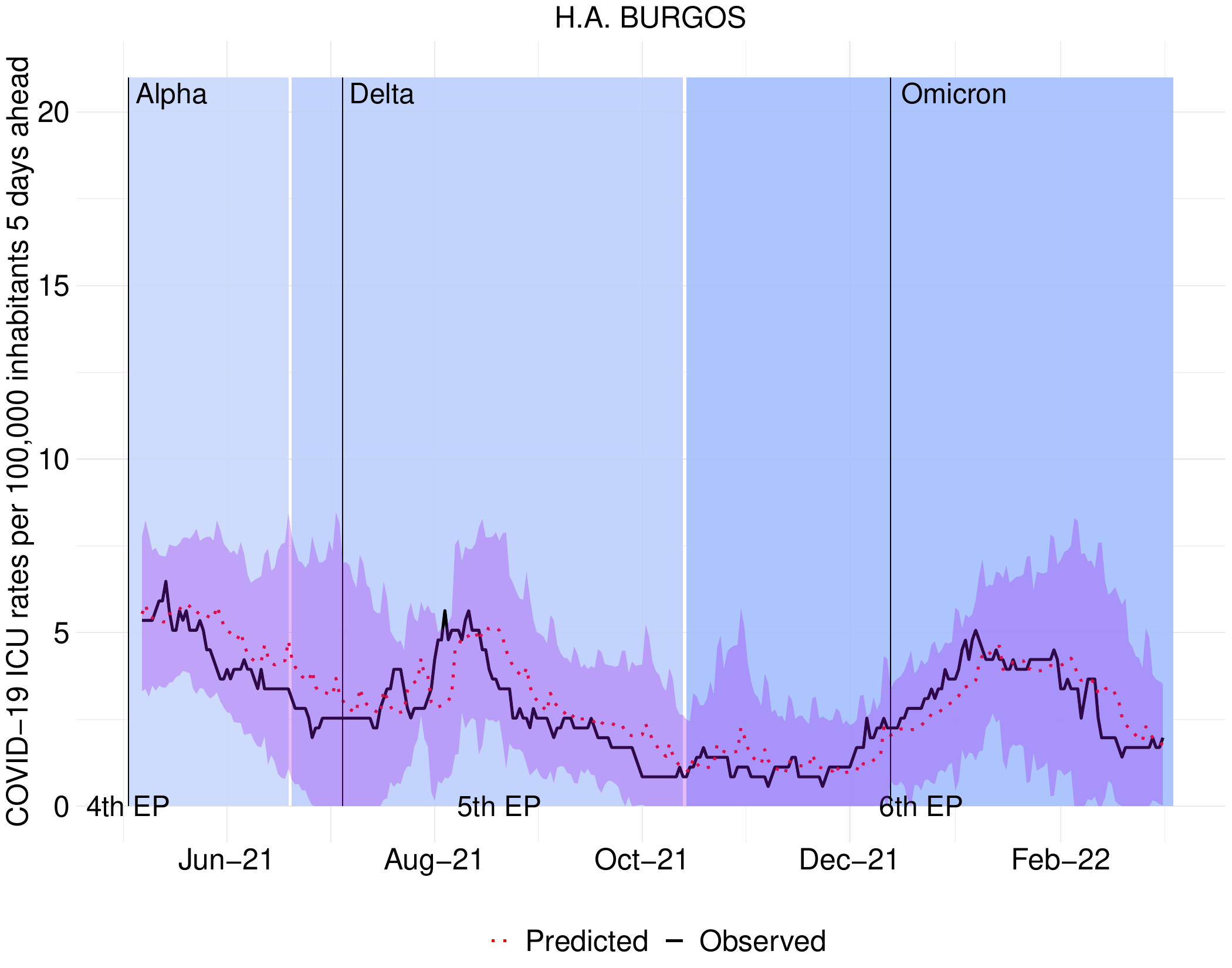}
  \end{minipage}
  \begin{minipage}{0.50\textwidth}
    \centering
    \includegraphics[width=\linewidth]{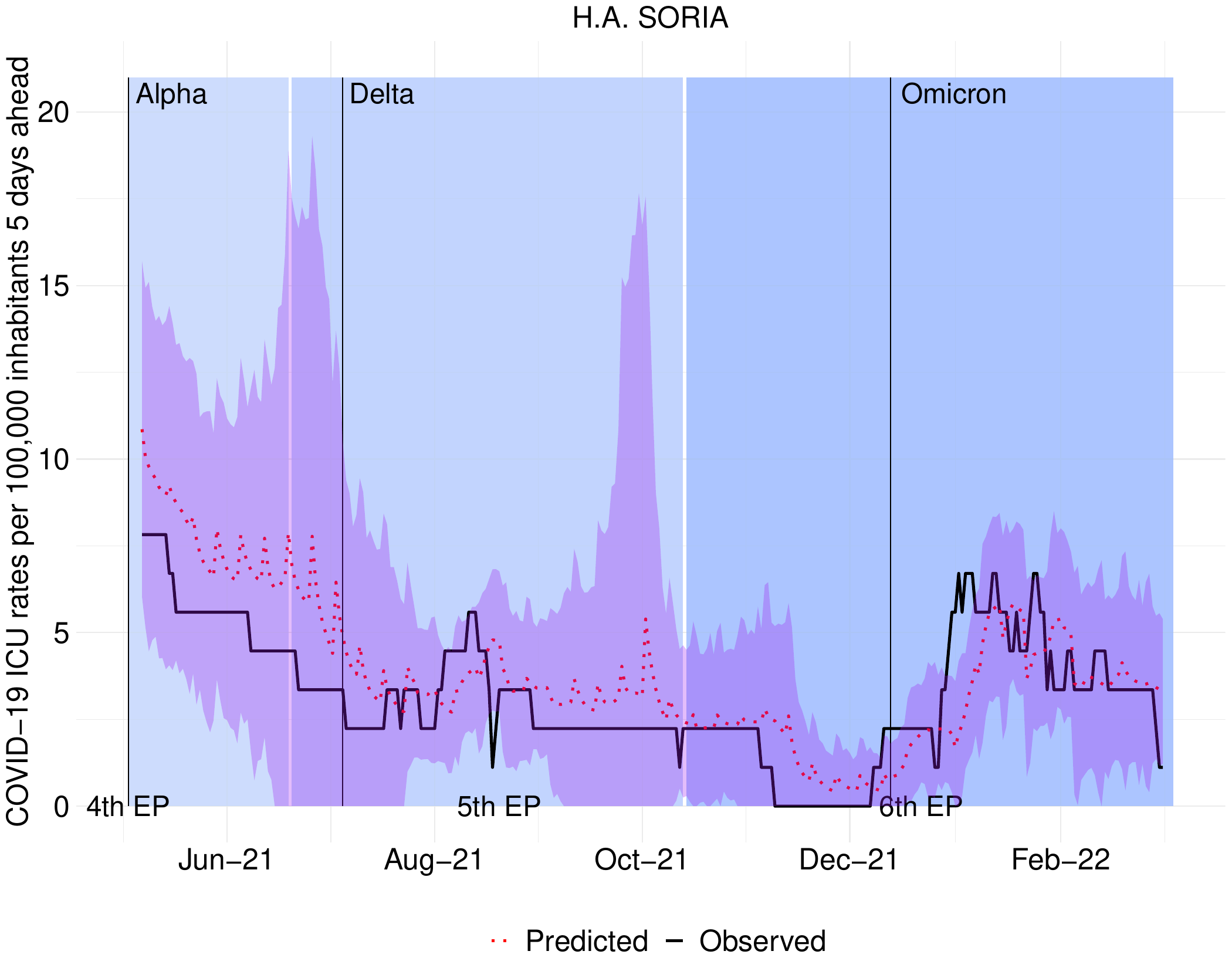}
  \end{minipage}

  \begin{minipage}{0.50\textwidth}
    \centering
    \includegraphics[width=\linewidth]{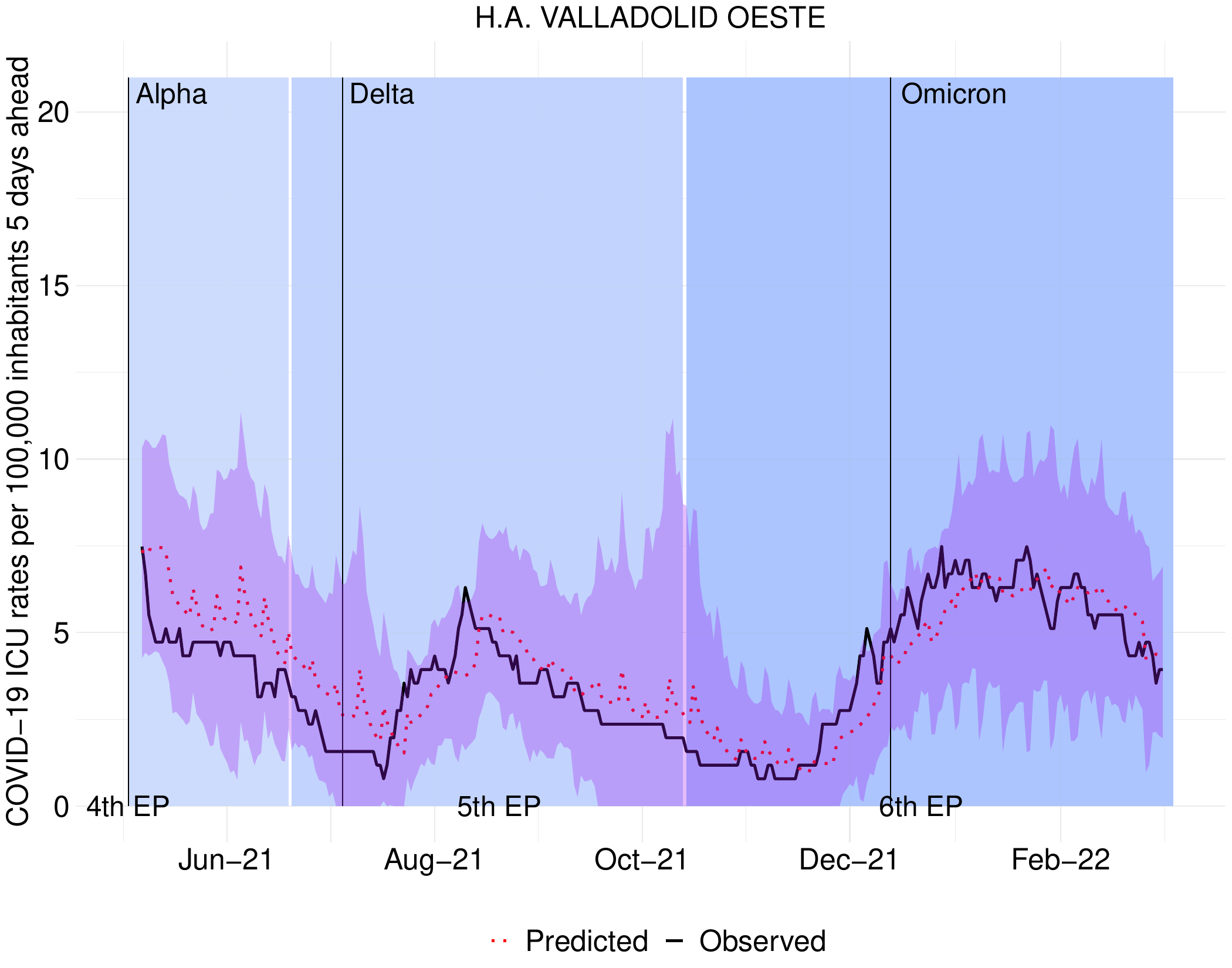}
  \end{minipage}%
  \begin{minipage}{0.50\textwidth}
    \centering
    \includegraphics[width=\linewidth]{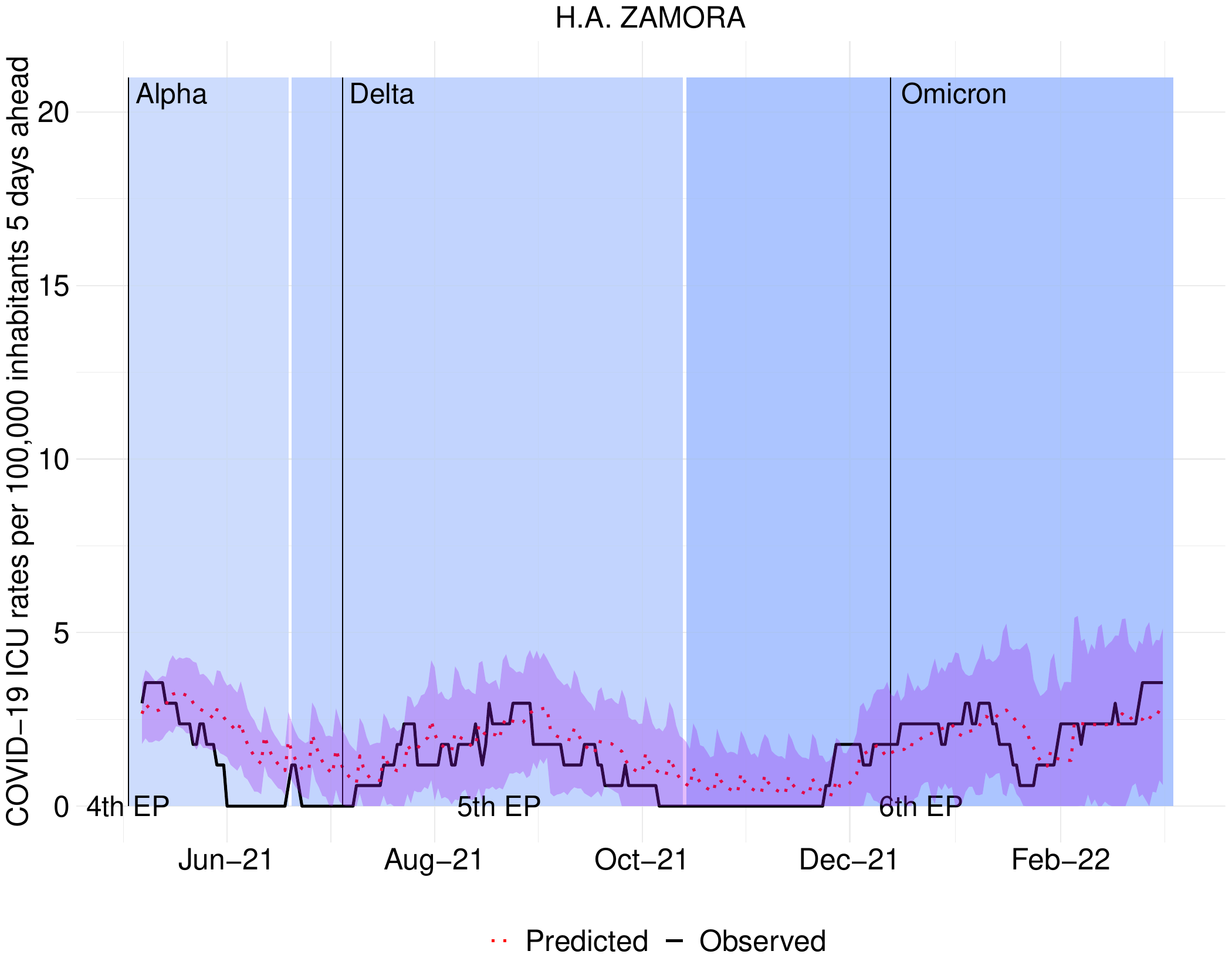}
  \end{minipage}
  \caption{Observed versus predicted five-day ICU occupancy rate per 100,000 population with simulated RMSE.}
  \label{fig:app.further.ap.data.PREDSIM.H5.I}
\end{figure}

\begin{figure}[htbp]
  \begin{minipage}{0.50\textwidth}
    \centering
    \includegraphics[width=\linewidth]{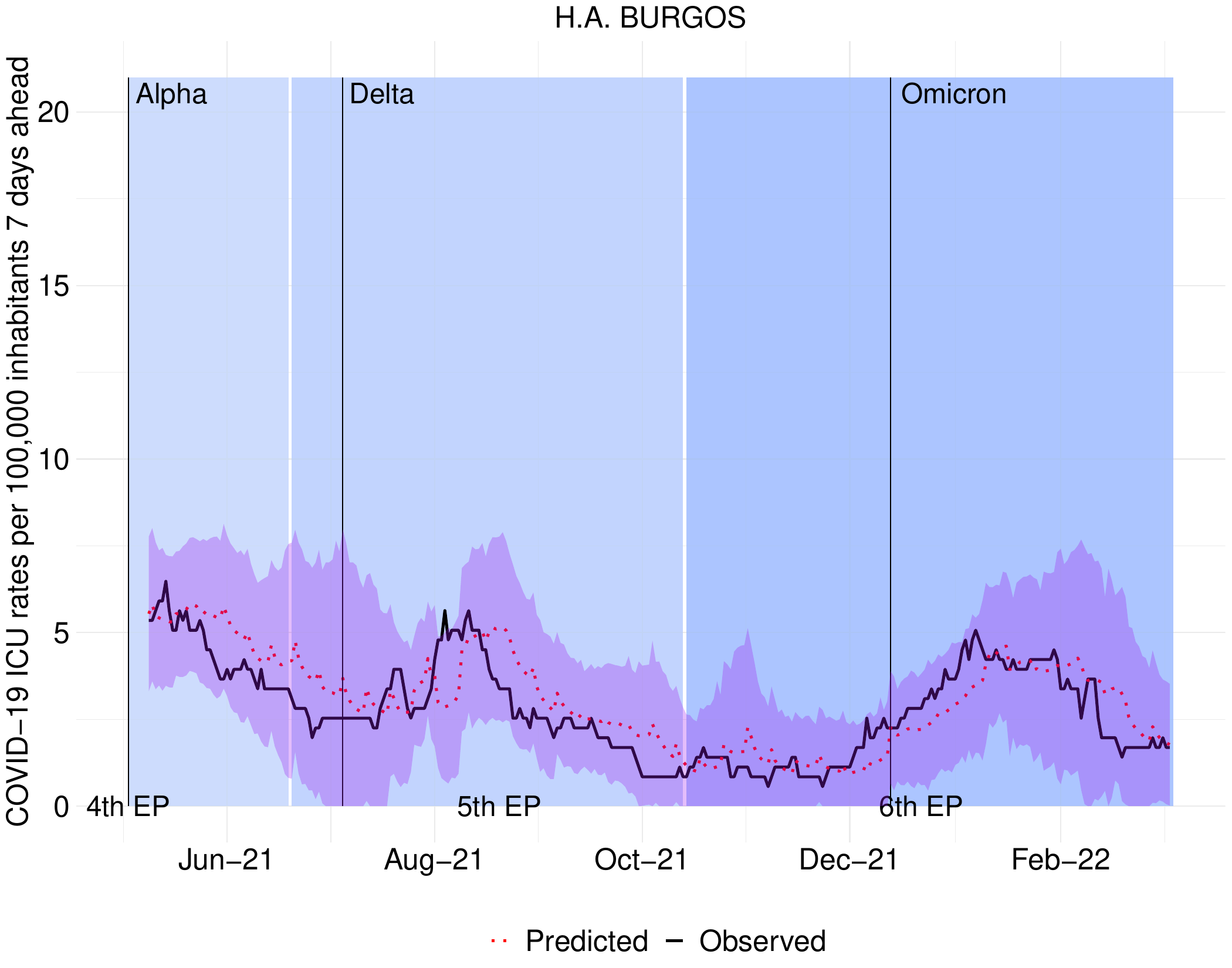}
  \end{minipage}
  \begin{minipage}{0.50\textwidth}
    \centering
    \includegraphics[width=\linewidth]{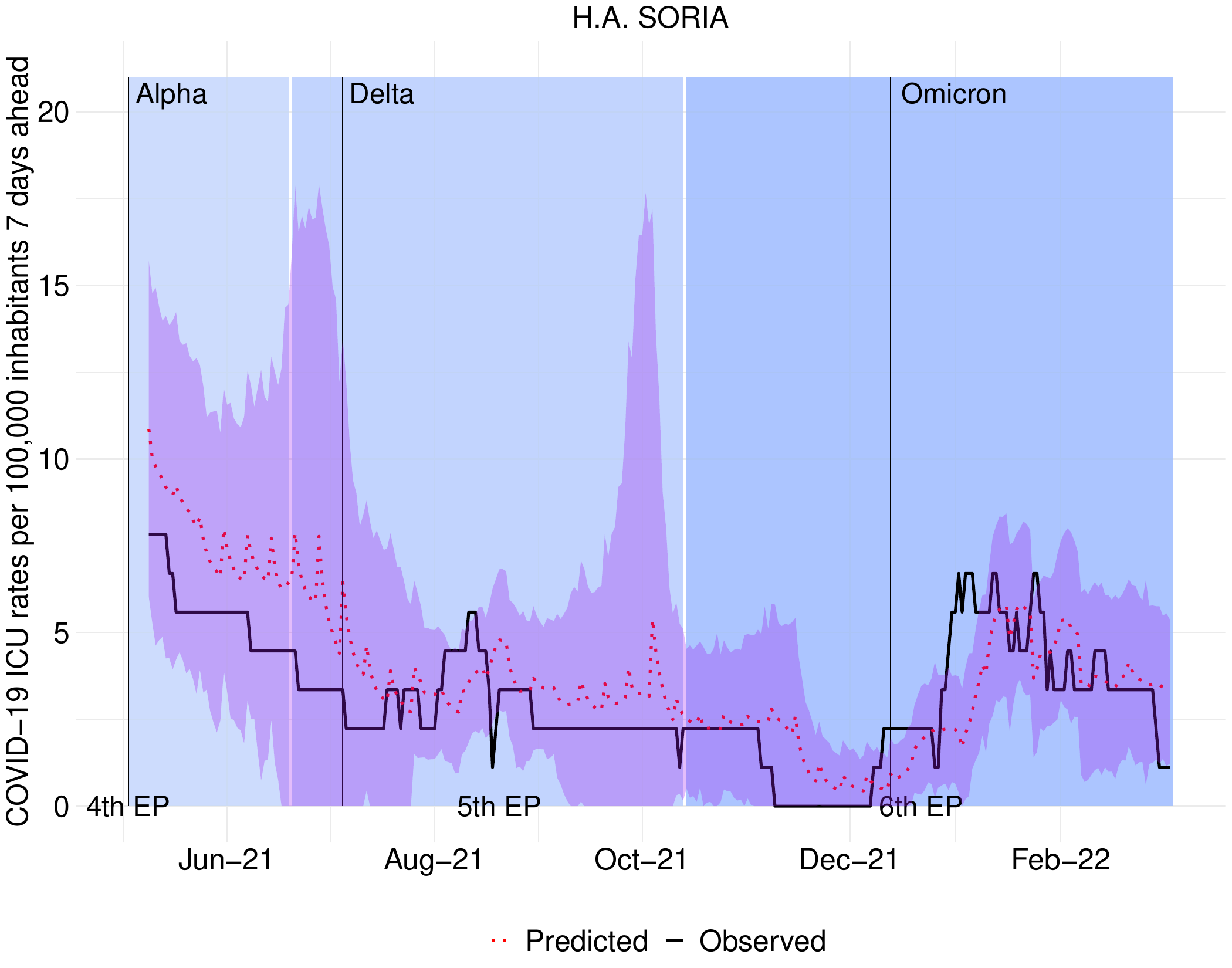}
  \end{minipage}

  \begin{minipage}{0.50\textwidth}
    \centering
    \includegraphics[width=\linewidth]{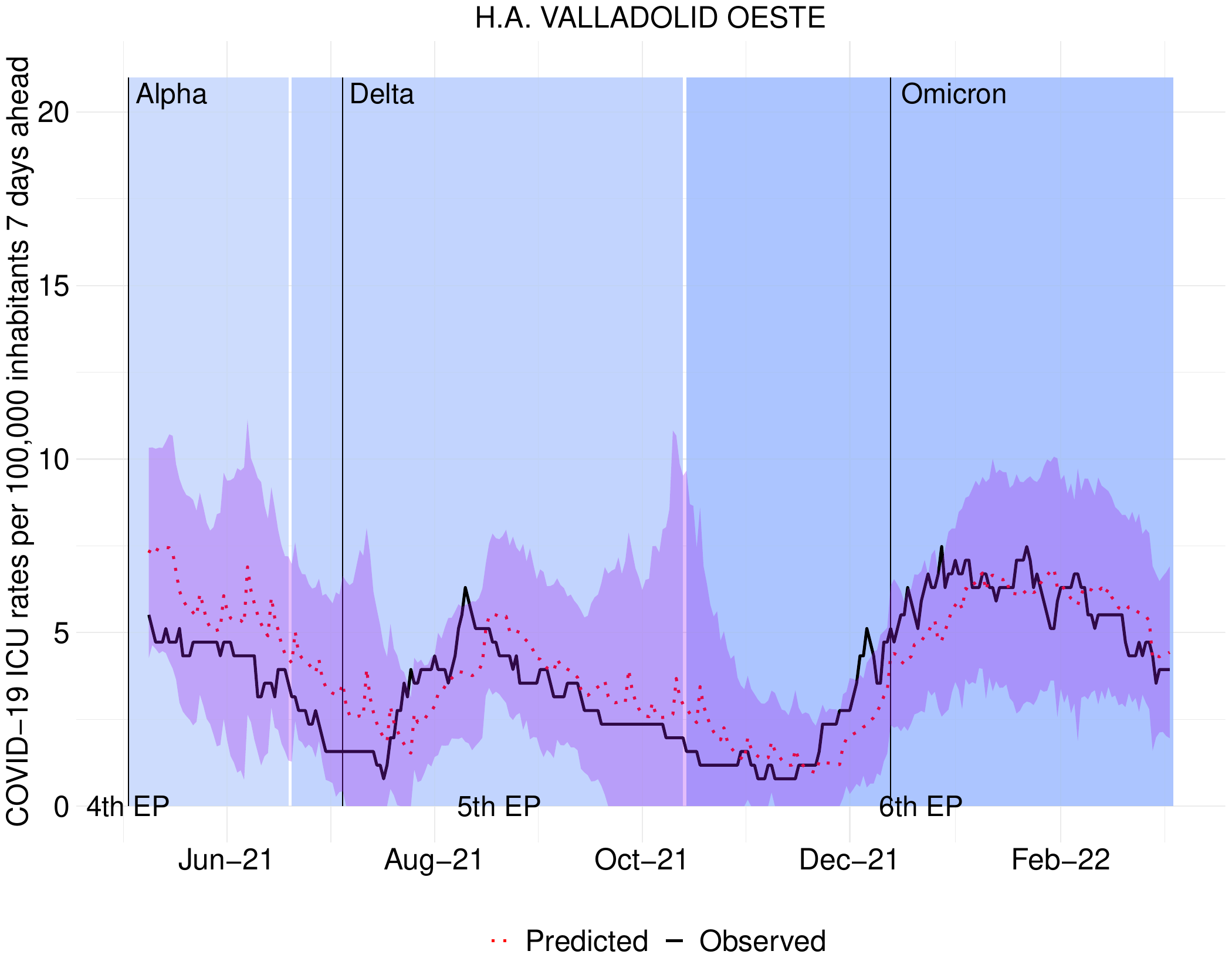}
  \end{minipage}%
  \begin{minipage}{0.50\textwidth}
    \centering
    \includegraphics[width=\linewidth]{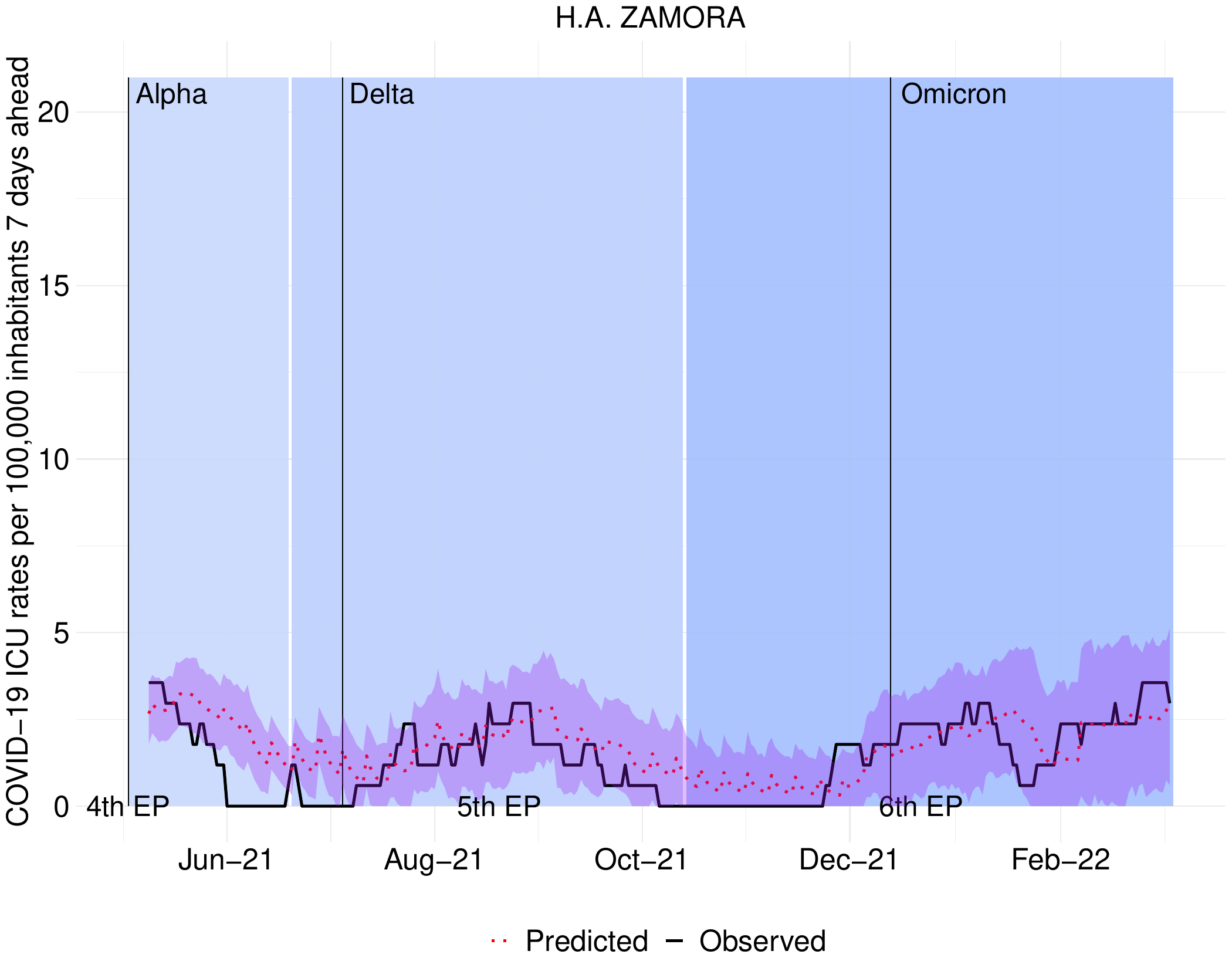}
  \end{minipage}
  \caption{Observed versus predicted seven-day ICU occupancy rate per 100,000 population with simulated RMSE.}
  \label{fig:app.further.ap.data.PREDSIM.H7.I}
\end{figure}

This section contains an extension of the results obtained after applying the ARRCP model to the data for Castilla y León Section \ref{main.app}. Specifically, Figure \ref{fig:app.further.ap.data.FIT.I} is provided to evaluate the quality of the adjustment by H.A and day, Figure \ref{fig:app.further.ap.data.PRED.I} to evaluate the quality of the prediction in the future through the real RMSE recorded, and finally, Figure \ref{fig:app.further.ap.data.PREDSIM.H3.I}, Figure \ref{fig:app.further.ap.data.PREDSIM.H5.I}, and Figure \ref{fig:app.further.ap.data.PREDSIM.H7.I} provide the predictions 3, 5 and 7 days ahead with the RMSE simulated at the time of prediction.

\subsection{Assessment of the fit of the ARRCP model}\label{app.further.ap.data.FIT}

\newpage
\subsection{Actual assessment of the forward prediction error of the ARRCP model}\label{app.further.ap.data.PRED}

\newpage
\subsection{Simulated assessment of the forward prediction error of the ARRCP model.}\label{app.further.ap.data.PREDSIM}

\newpage

\section*{References}

\end{document}